\documentclass[12pt]{article}
\usepackage{geometry}                
\usepackage{vmargin,bm,dsfont,amsmath,algorithmicx,multirow}
\usepackage[linesnumbered,ruled]{algorithm2e}
\setmarginsrb{0.6in}{1in}{0.6in}{1in}{0pt}{0mm}{10pt}{0.5in}
\usepackage{url}
\usepackage{graphicx}
\usepackage{amssymb}
\usepackage{authblk}
\usepackage{amsthm}
\usepackage{epstopdf}
\usepackage{url}
\usepackage{mathtools}
\usepackage{dsfont,color}
\usepackage{color,xcolor,cleveref}

\usepackage{subfig,animate}
 \newcommand{\h}[1]{\mathbf{#1}}
\usepackage{algpseudocode}

\DeclareMathOperator*{\argmin}{arg\,min}

\title{Joint 3D Localization and Classification of Space Debris using a Multispectral Rotating Point Spread Function}

\author[1,*]{Chao Wang}
\author[1]{Grey Ballard}
\author[1]{Robert Plemmons}
\author[2,3]{Sudhakar Prasad}

\affil[1]{{\small Department of Computer Science, Wake Forest University, Winston-Salem, NC 27109 }}
\affil[2]{{\small Department of Physics and Astronomy, The University of New Mexico, Albuquerque, NM 87131}}
\affil[3]{{\small School of Physics and Astronomy, University of Minnesota, Minneapolis, MN 55455}}
\affil[*]{{\small Corresponding author: chaowang.hk@gmail.com}}
\begin{document}
\maketitle
\abstract{We consider the problem of joint three-dimensional (3D) localization and material classification of unresolved space debris using a multispectral rotating point spread function (RPSF). The use of RPSF allows one to estimate the 3D locations of point sources from their rotated images acquired by a single 2D sensor array, since the amount of rotation of each source image about its $x, y$ location depends on its axial distance $ z$. Using multi-spectral images, with one RPSF per spectral band, we are able not only to localize the 3D positions of the space debris but also classify their material composition. We propose a three-stage method for achieving joint localization and classification. In Stage 1, we adopt an optimization scheme for localization in which the spectral signature of each material is assumed to be  uniform, which significantly improves efficiency and yields better localization results than possible with a single spectral band. In Stage 2, we estimate the spectral signature and refine the localization result via an alternating approach. We process classification in the final stage.  Both Poisson noise and Gaussian noise models are considered, and the implementation of each is discussed.  Numerical tests using multispectral data from NASA show the  efficiency of our three-stage approach and illustrate the improvement of point source localization and spectral classification from using multiple bands over a single band.   } 
\section{Introduction}

Image data taken with a specially designed point spread function (PSF) that encodes, via a simple rotation \cite{DH2008pavani,prasad2013rotating,3dsml2017review,tetrapods2014shechtman}, the source distance can be employed to acquire a three dimensional (3D) field of unresolved sources like space debris \cite{Internal_report2016prasad,AMOS2018novel}.
By employing a spiral phase mask with the phase winding number changing in regular integer steps from one annular zone to the next and the zone radii having a square root dependence on their index \cite{prasad2013rotating}, one can create a PSF that rotates uniformly with changing source distance while largely maintaining its shape and size. Specifically, the off-center, shape-preserving PSF is rotated relative to a reference PSF corresponding to the point source being at the plane of best focus, with the rotation angle being approximately proportional to the source misfocus from the plane of best focus.

Hyperspectral and multispectral imaging techniques constitute another important topic in optics and remote sensing problems. Multiple images are collected in contiguous spectral bands for a wide range of applications from space situational awareness \cite{Nagy2015deblurring,hege2004hyperspectral,mult_space2006AMOS,zhang2008tensor},  agriculture \cite{datt2003preprocessing,agri_hype2001} to military detection \cite{mili2002detection,stein2002anomaly}.  One can classify the material composition of an object using detailed spectral information.  

Combining the techniques of PSF engineering and multispectral imaging is a promising direction.  In \cite{zhao2013deblurring}, a joint deblurring and sparse unmixing is shown to efficiently analyze the image objects using hyperspectral image data.   In \cite{Nagy2015deblurring}, the authors consider the different blurring effects in different spectral wavelengths and address the space unmixing problem with estimated PSF in each band. Another direction is PSF engineering with multispectral/multicolor imaging.   \cite{gahlmann2013quantitative} extends 3D localization using double helix PSF in multispectral imaging problem but its considerations are limited to detecting point sources with the spectral intensity mainly distributed in two bands, i.e., red (440 nm) and green (550 nm). 
This modification can be performed by adding a spectrally-sensitive, phase-modulating
element to the imaging path which
is positioned in a plane conjugate to the back focal plane of the imaging device.  
  Fluorescent labels encode point sources into red and green classes for biological contextual information. Furthermore, via chromatic dispersion, the multiple colors are encoded in the PSF \cite{DL_localization,multicolour_loc}. Then the spectral information can be detected even in a single 2D snapshot.  However, the spectral energy of each targeted object is required to be peaked in a characteristic narrow band, which is not true for space debris, a fact that precludes their classification in this way. For a description of the NASA/DOD space object material spectral signature  library we are using see \cite{NASApaper}.

We discuss here the problem of 3D localization and classification of closely spaced point sources from simulated noisy image data obtained by using Prasad's rotating-PSF imager \cite{prasad2013rotating} in multiple bands. For space debris, this would require an active laser illumination system and collection of light scattered by the debris in multiple narrow bands. The localization problem is discretized on a cubical lattice where the coordinates and values of its nonzero entries represent the 3D locations and fluxes of the sources, respectively. The flux value for one point source in the $i$-th band is the $i$-th entry in its corresponding  spectral signature. 
		Finding the locations and spectral signatures of a few point sources on a large lattice is  a large-scale sparse 3D inverse problem. Using both Poisson and Gaussian  statistical noise models, we describe the results of simulations and propose  a three-stage method to extract both the 3D location coordinates and spectral signatures of individual debris particles from  rotating-PSF imagery, thus achieving both 3D localization and classification. The Poisson noise case involves an EMCCD sensor operating in the photon-counting (PC) regime \cite{daigle2010} while the Gaussian noise case involves conventional CCD sensors operating at high per-mode photon numbers
		 and large read-out noise.  We show different implementations for these two noise models and compare our results to those obtained when using only a single band. The joint approach obtains more accurate results than  one that first localizes the sources with single-band data and then classifies them with multispectral image data.
We use the terms ``band'', ``channel'', and ``spectral wavelength'' interchangeably, depending on the context.

The rest of the paper is organized as follows. \Cref{sec:Physic_model} describes the physics model for the single-lobe rotating PSF for both single and multiple bands.  In \Cref{sec:single_band}, we review a non-convex optimization method to solve the point source localization problem for a single band image using the Poisson noise model. In \Cref{sec:3Stages}, we introduce multispectral rotating PSF imaging and propose a three-stage method, which we analyze for the Poisson noise case.  We adapt this method For the Gaussian noise case in \Cref{sec:Gaussian}. We discuss numerical experiments, including comparisons with the single band model, in \Cref{sec:numerical}, and present some concluding remarks in \Cref{sec:conclusions}.

\section{Physics Models for RPSFs at Single and Multiple Spectral Wavelengths}\label{sec:Physic_model}

In this section, we review the single-lobe rotating PSF forward model \cite{prasad2013rotating} for single and multiple bands. If the number of point sources is $M$, then the 2D observed image $G$ in a single band has the form,
\begin{equation}
	G (\h s) = \mathcal{N}\left(\sum_{j=1}^M \mathcal{H}_{j}(\h s - \h s_j)f_j+ b\right), 
	\label{equ:forward_model}
\end{equation}
where  $\mathcal{N}$ is the noise operator and $b$ is the uniform background value. Here  
$\mathcal{H}_j(\h s - \h s_j)$ is the rotating PSF for the $j$-th point source of flux $f_j$. 
The depth information $z_j$  in the 3D position coordinates $(x_j, y_j, z_j)$ is encoded in  $\mathcal{H}_j$, and $\h s = (x,y)$ is  a scaled version of the image-plane position vector, $\mathbf{r}$, namely 
\begin{equation}\label{equ:s}
	\mathbf{s} =  \frac{R}{\lambda z_I}\mathbf{r}
\end{equation}
where $z_I$ is the distance between the image plane and the lens. $\h s_j=(x_j,y_j)$ is the similarly scaled version of the corresponding 2D image-plane location, $\mathbf{r}_j$, of the $j$th source. The imaging wavelength is denoted by $\lambda$ and the radius of the pupil is $R$. 
Following the Fourier optics model, the  incoherent PSF for a clear aperture containing a phase mask with optical phase retardation, $\psi$, is given by 
\begin{equation}
	\label{equ:H_j}
\mathcal{H}_{j}(\mathbf{s} ) = 
{1\over \pi}\left|\int P(\mathbf{u} )\mathrm{exp} \left[ \iota( 2\pi \mathbf{u}\cdot\mathbf{s} +  \zeta_j u^2 - \psi(\mathbf{u}))  \right] d \mathbf{u} \right|^2,
\end{equation}
where $\iota = \sqrt{-1} $  and the depth information is encoded in
the defocus parameter 
\begin{equation}\label{equ:zeta}
	\zeta_j = \frac{\pi  (l_0-z_j) R^2}{\lambda l_0 z_j}.
\end{equation}
 Here, $l_0$ denotes the distance between the lens and the best focus point in the object space.  The indicator function for the pupil of radius $R$ is represented as $P(\mathbf{u} )$. 
 The normalized pupil-plane position vector $\boldsymbol{u}$ is given by dividing the actual pupil-plane position vector, $\boldsymbol{\rho}$, by the pupil radius, $\mathbf{u} = \frac{\boldsymbol{\rho}}{R}$. 
For the single-lobe rotating PSF, $\psi(\mathbf{u}) $ is chosen to be the spiral phase
 profile defined as $$\psi(\mathbf{u}) = l\phi_{\mathbf{u}}, \ \ \text{for } \sqrt{\frac{l-1}{L}}\leq u\leq \sqrt{\frac{l}{L}}, \ l = 1,\cdot\cdot\cdot, L, $$  in which $L$ is the number of concentric annular zones in the phase mask and $\mathbf{u}$ is in the  polar coordinate $(u, \phi_{\mathbf{u}}) $. 
In the following, we directly focus on estimation of the location in the $x$-$y$ plane, $(x_j, y_j)$, as well as the defocus parameter $\zeta_j$. From \eqref{equ:zeta}, we directly obtain the depth as
\begin{equation}\label{equ:depth}
	z_j = \frac{\pi l_0 R^2}{\lambda l_0 \zeta_j + \pi R^2}. 
\end{equation}
Since the $(x,y)$ location of the $j$-th point source in the object space is related to its $(x,y)$ location in the image plane by the simple magnification factor $-z_j/z_I$, its 3D location in the object space is determined in terms of its image plane coordinates as
\begin{equation}
	\left(-\frac{z_j x_j}{ z_I}, -\frac{z_j y_j}{z_I},  z_j  \right),
\end{equation}
in which $z_j$ is calculated from the source defocus parameter, $\zeta_j$, using \eqref{equ:depth}.

In the multispectral problem, we have a sequence of 2D observed images $\{G^{(i)}, \quad i = 1, 2, \dots, K\}$ corresponding to  $K$ spectra.
We need to solve for the 3D locations of the point sources and their flux values for each spectral wavelength, namely
\begin{equation*}
	\{(x_j,y_j,\zeta_j,\mathbf{f}_j), \quad j = 1, 2, \dots, M \},
\end{equation*}
where 
$\mathbf{f}_j = (f^{(1)}_j, f^{(2)}_j, \dots, f^{(K)}_j)$ and $\zeta_j$ is determined from Eq.~(\ref{equ:zeta}) in which $\lambda$ is set equal to 400 nm. Each point source has a value for its flux in each spectral band, and the vector of those values defines its spectral signature. Thus the forward model is 
\begin{equation}
	G^{(i)}(\h s^{(i)}) = \mathcal{N}\left(\sum_{j=1}^M \mathcal{H}^{(i)}_{j}(\h s^{(i)} - \h s^{(i)}_j)f^{(i)}_j + b\right), 
	\label{equ:forward_model_multi}
\end{equation}
where $\mathcal{H}^{(i)}_j$ is defined in terms of $\zeta_j^{(i)}$. Here the symbols $\h s^{(i)}$, $\h s^{(i)}_j$, and $\zeta_j^{(i)}$ generalize the corresponding single-wavelength symbols to multiple wavelengths, namely
$$\h s^{(i)} = \frac{R}{\lambda^{(i)} z_I}\h s,\ \h s^{(i)}_j = \frac{R}{\lambda^{(i)} z_I}\h r_j,\ {\rm and}\ \zeta_j^{(i)} = \frac{\pi (l_0-z_j) R^2}{\lambda^{(i)} l_0 z_j} $$
and the spatially uniform background flux $b$ is assumed for simplicity to be the same for each spectral wavelength. 

Note that the depth information $z$  is transformed into a defocus parameter called $\zeta$. For a given defocused source, since $\boldsymbol{s}$ and $\zeta$ both scale inversely with the wavelength, its PSF at a longer wavelength is less rotated but spatially more extended than that at a shorter wavelength. Thus, for example, for the RGB (red green blue) wavelenths of $\lambda = 660, \, 550,$ and 440 nm,  if we denote the $\zeta$ parameter of the source for the blue wavelength as $\zeta_B$, then the same parameter for the green and red wavelengths will be equal to $\zeta_G=\frac{4}{5}\zeta_B$ and $\zeta_R=\frac{2}{3} \zeta_B$, respectively, corresponding to progressively smaller rotations from the blue to the red wavelengths. Correspondingly, the spatial footprints of the PSF blur functions, on the other hand, get increasingly extended, 
a fact that we incorporate in our simulations, in effect, by changing the physical dimensions of the image plane  in proportion to the wavelength.

\section{Numerical Localization Schemes for a Single Wavelength}\label{sec:single_band}
Here, we  review the forward model for the single wavelength problem covered in \cite{Rice2016generalized,kl_nc}. 
In order to estimate the 3D locations of the point sources, we assume their distribution is approximated by a discrete lattice  $\mathcal{X}\in \mathds{R}^{m \times n \times d}$. 
The indices of the nonzero entries of $\mathcal{X}$ are the 3-dimensional  locations of the point sources and the values at these entries   correspond to the fluxes,  i.e., the energy emitted by the illuminated point source. The 2D observed image $G \in \mathds{R}^{m \times n}$ can be represented as 
\begin{equation*}
	G = \mathcal{N}\left(\mathcal{T}(\mathcal{A} \ast \mathcal{X} )+b \mathbb{1}  \right),
\end{equation*}
where
 $b$ is background signal value, $\mathbb{1}$ is a matrix of $1$s of size the same as the size of $G$, and 
$\mathcal{N}$ is the noise operator. Here   $\mathcal{A} \ast \mathcal{X}$ is the convolution of $\mathcal{X}$ with the 3D PSF $\mathcal{A}$, in which the latter is a 3D tensor that comprises a sequence of images with respect to point sources at different depths. The $j$-th slice is the image corresponding to a point source at the origin in the $(x,y)$ plane and at the  defocus parameter $\zeta_j$ which encodes the depth information. 
The dictionary $\mathcal{A}$ {is constructed} by sampling depths at regular intervals in the range, $\zeta_j\in [-\pi L, \ \pi L]$  over which the PSF performs one complete rotation about the geometric image center before it largely break apart.  Here $\mathcal{T}$ is an operator for extracting the last slice of the cube $\mathcal{A} \ast \mathcal{X}$ since the observed information is a 2D snapshot.

Here we consider $\mathcal{N}$ to be the Poisson noise operator, which is data-dependent.  
Determining the source locations and fluxes is a large-scale sparse 3D inverse problem.
In \cite{kl_nc}, Kullback-Leibler (KL) divergence \cite{poisson_formula} is used as the data-fitting term,

\begin{equation}\label{equ:data_fitting_D}
D_s(\mathcal X, \mathcal A, G) = \left\langle \mathbb{1}, \mathcal{T}(\mathcal{A} \ast \mathcal{X})- G \log(\mathcal{T}(\mathcal{A} \ast \mathcal{X})+b \mathbb{1}) \right\rangle,
\end{equation}
to which is added a nonconvex regularization term to enforce sparsity, and the minimization problem for 3D localization for a single wavelength amounts to
\begin{equation}
\min\limits_{\mathcal{X}\geq 0 }\left\{ D_s(\mathcal X, \mathcal A, G) + \mu\sum_{p,q,r = 1}^{m,n,d} \frac{|\mathcal{X}_{pqr}|}{a+|\mathcal{X}_{pqr}|}\right\}, \label{equ:min_fun_poisson}
\end{equation}
where $a$ is fixed and determines the degree of non-convexity and $\mu $ is the regularization parameter.  
The minimization problem \eqref{equ:min_fun_poisson} is solved by the iteratively reweighted $\ell_1$ algorithm \cite{IRL1_2015} with the iterative scheme being 

\begin{equation}
	\begin{cases}
w^k_{pqr}  &=   \frac{a \mu } { \left(a + \hat{\mathcal{X}}_{pqr}^k  \right)^2 }, \quad \forall p,q,r; \\
        \hat{\mathcal{X}}^k  &=   \arg\min\limits_{\mathcal{X}\geq 0  } \left\{  D_s(\mathcal X, \mathcal A, G)+ \sum\limits_{p,q,r=1}^{m,n,d} w^k_{pqr}  |\mathcal{X}_{pqr} | \right\}.
        \end{cases}
        \label{equ:kl_nc_outer}
    \end{equation} 
    In the $\mathcal{X}$-subproblem of \eqref{equ:kl_nc_outer}, the alternating direction method of multipliers (ADMM) \cite{ADMM2011boyd} is used by introducing two auxiliary variables, namely $\mathcal{U}_0$ and $\mathcal{U}_1$.  Then the augmented Lagrangian function  $\mathcal{L}_s(\mathcal{U}_0, \mathcal{U}_1, \mathcal{X}, \eta_0, \eta_1 )$ is 
\begin{equation*}
\begin{split}
	& \langle \mathbb{1}, \mathcal{T} \mathcal{U}_0 - G \log (\mathcal{T}\mathcal{U}_0 + b \mathbb{1}) \rangle  + \sum_{p,q,r=1}^{m,n,d} w^k_{pqr}  |(\mathcal{U}_1)_{pqr} | \\
	 +& \frac{\beta_0}{2}\|\mathcal{U}_0-\mathcal{A} \ast \mathcal{X} - \eta_0 \|^2 + \frac{\beta_1 }{2} \|\mathcal{U}_1-\mathcal{X}-\eta_1\|^2+ I_+(\mathcal{U}_1 ),
	\end{split}
\end{equation*}
where $\eta_0, \eta_1 \in \mathds{R}^{m\times n \times d} $ are the Lagrange multipliers and $\beta_0, \beta_1 >0$ and $\|\mathcal X\|$ is equal to the $\ell_2$ norm of the vectorized $\mathcal{X}$. Here,
$I_+(\mathcal{X})$ 
enforces the nonnegative constraint,
    \begin{equation*}
    	I_+(\mathcal{X})=
    	\begin{cases}
    		0, & \mathcal{X} \geq 0,\\
 \infty, & \text{otherwise}.     	
 \end{cases}
    \end{equation*}

Therefore,  the iterative scheme for the $\mathcal{X}$-subproblem of \eqref{equ:kl_nc_outer} is 
\begin{subequations}\label{equ:kl_nc_sub}
	\begin{align}
		\mathcal{U}_0^{t+1} &= \argmin\limits_{\mathcal{U}_0} \mathcal{L}_s(\mathcal{U}_0, \mathcal{U}_1^t, \mathcal{X}^t, \eta_0^t, \eta_1^t ) \label{equ:U_0} \\
		\mathcal{U}_1^{t+1} &= \argmin\limits_{\mathcal{U}_1 \geq 0}\mathcal{L}_s(\mathcal{U}_0^{t+1}, \mathcal{U}_1, \mathcal{X}^t, \eta_0^t, \eta_1^t )\label{equ:U_1} \\
		\mathcal{X}^{t+1} &= \argmin\limits_{\mathcal{X}} \mathcal{L}_s(\mathcal{U}_0^{t+1}, \mathcal{U}_1^{t+1}, \mathcal{X}, \eta_0^t, \eta_1^t )\label{equ:X} \\
		\eta_0^{t+1} & = \eta_0^{t} - {\rho (\mathcal{U}_0^{t+1} -\mathcal{A} \ast \mathcal{X}^{t+1})} \\
		\eta_1^{t+1} & = \eta_1^{t} - {\rho( \mathcal{U}_1^{t+1} - \mathcal{X}^{t+1})},
	\end{align}
\end{subequations} 
where the superscript $t$ indexes the inner loop iterations, as opposed to the superscript $k$ used for outer iterations in \eqref{equ:kl_nc_outer}. 
Here $\rho\in \left(0, \frac{1+\sqrt{5}}{2}\right)$ is the dual steplength. We choose $\rho=1.618$ in our numerical tests \cite{kl_nc}. 
By using \cite[Proposition 4.1]{kl_nc}, we can solve the $\mathcal{U}_0$-subproblem \eqref{equ:U_0} by
\begin{equation*}
    	\left(\mathcal{U}_0^{t+1}\right)_{pqr} = 
    	\begin{cases} 
    		\frac{- (\xi_1^t)_{pqr} + \sqrt{(\xi_1^t)_{pqr}^2+4 \beta_0  G_{pq}   }}{2 \beta_0  }, & \text{if } r =  d, \\
    		(\xi_0^t)_{pqr}, & \text{otherwise}.
    	\end{cases}\label{equ:u_closed}
    \end{equation*}
    where $\xi_1^t =1 - \beta_0 b - \beta_0 \xi_0^t,$ and $ \xi_0^t = \mathcal{A} \ast \mathcal{X}^{t} +\eta_0^{t}. $
The $\mathcal{U}_1$-subproblem \eqref{equ:U_1} is solved by soft-thresholding under nonnegative constraint. So  the closed-form solution is given by 
\begin{equation}\label{equ:U1_closed}
	\left(\mathcal{U}_1^{t+1}\right)_{pqr} = \max \left\{\mathcal{X}^t+\eta_1^t - w^k_{pqr}/\beta_1,\  0 \right\}.
\end{equation}

The $\mathcal{X}$-subproblem \eqref{equ:X} is a least squares problem. We turn the convolution into  componentwise multiplication by using the Fourier transform. 
Then the closed-form solution for $\mathcal{X}^{t+1}$ reads as
\begin{equation}
	 \mathcal{F}^{-1}\left\{\Omega_s \cdot \left(\overline{\mathcal{F}\{\mathcal{A}\}} \cdot \mathcal{F}\{\mathcal{U}_0^{t+1}-\eta_0^{t}\} +\frac{\beta_1}{\beta_0} \mathcal{F}\{\mathcal{U}_1^{t+1}-\eta_1^{t}\}\right)\right\},
	\label{equ:fft}
	\end{equation}
	where $\Omega_s= \left(|\mathcal{F}\{\mathcal{A}\}|^2+\frac{\beta_1}{\beta_0} \right)^{-1}. $  Here $\left|\mathcal{X}\right|^2$ and $\overline{\mathcal{X}}$ are the componentwise operations of the square of the absolute value of $\mathcal{X}$, and complex conjugate of $\mathcal{X}$, respectively.
	By assuming the boundary condition to be periodic, we use the 3D fast Fourier transformation (3D FFT) to compute \eqref{equ:fft} efficiently.   
For real data, the point sources may be not on a grid, which means the discrete model may not be accurate. In order to avoid missing point sources, the regularization parameter $\mu$ is kept small, which can potentially lead to over-fitting.  
Our optimization solution generally contains tightly clustered point sources, so we need to regard any such cluster of point sources as a single point source.
The same phenomenon has been observed in \cite{FALCON2014,Rice2016generalized,clustered2012fasterstorm}. 
We apply a post-processing approach following
\cite{clustered2012fasterstorm}.  The method is based on the well-defined tolerance distance for recognizing tightly clustered neighbors as single point sources. 
We compute the centroid of each such cluster, which we regard as the location of a single point source. 
	 
\section{Numerical Scheme for Joint Localization and Classification using Multiple Bands} \label{sec:3Stages}
We now propose a three-stage method for multispectral rPSF imaging. Our aim is to localize in 3D and to classify the materials of unresolved space debris (i.e., point sources) based on multispectral information. 

Denote the observed image for the $i$-th spectral band as $G^{(i)}$. We need to restore the 3D tensor $\mathcal{X}^{(i)}$ from the observed image. Note that the positions of the non-zero entries of $\{\mathcal{X}^{(i)}\}$ corresponding to the source locations are the same for all $i = 1, \dots, K. $ The values of nonzero entries in each spectral band correspond to the fluxes of the point sources in that band. 
Since the PSF differs across spectral wavelengths, our dictionaries corresponding to different spectral bands are also different. The discretized forward model becomes  
\begin{equation}
	G^{(i)} \approx \mathcal{N}\left(\mathcal{T}(\mathcal{A}^{(i)} \ast \mathcal{X}^{(i)}  ) + b\mathbb{1}  \right), \quad i = 1, 2, \dots, K,
	\label{equ:dependent}
\end{equation}
where $\mathcal{A}^{(i)}$ is the 3D tensor corresponding to the $i$-th spectral band.
In this section, we consider the  Poisson noise case for imaging space debris \cite{daigle2010}. 
The Gaussian noise case will be introduced in \Cref{sec:Gaussian}. 

\subsection{Stage 1: 3D Localization Using Multiple Bands}

In the first stage, we consider 3D localization via all the spectral images. 
One approach is to restore $\mathcal{X}^{(i)}$ by each $G^{(i)}$, separately, based on the algorithm in \Cref{sec:single_band}. 
However, the separate single-wavelength problems are coupled by the fact that the nonzero pattern must be consistent across $\{\mathcal{X}^{(i)}\}$, as it corresponds to the true physical locations of the point sources.
Enforcing this constraint while solving all single-wavelength problems would be computationally expensive.

Another reasonable approach is to perform the localization using a single wavelength. 
Shorter wavelengths lead to smaller side length of the PSF image, so choosing the image corresponding to the smallest available wavelength might be a good choice. 
However, this approach ignores the observed information from all other spectral bands, which we will see leads to suboptimal results.   

Here, we try to utilize all the information of spectral images while maintaining computational efficiency.
We make the simplifying assumption that for each point source the intensities in the various spectral bands are all equivalent. 
That is to say, we do the 3D localization based on assuming a  uniform spectral signature for each point source. 
We correct this assumption in Stage 2, where we more accurately estimate the intensities of the point sources in the different spectral bands.
Therefore, in Stage 1, we assume that $\mathcal{X}^{(i)}$ are the same for all $i$, and the discretized forward model becomes 
\begin{equation*}
	G^{(i)} \approx \mathcal{N}\left(\mathcal{T}(\mathcal{A}^{(i)} \ast \mathcal{X} ) + b \mathbb{1}  \right).
\end{equation*}

Our numerical scheme attempts to minimize the sum of the errors across bands, and we adapt the KL-NC model in \cite{kl_nc} from single wavelength to multispectral  images. The optimization  \eqref{equ:min_fun_poisson} becomes 
\begin{equation}
	 	\min\limits_{\mathcal{X}\geq 0 }\left\{  \sum\limits_{i=1}^K D_s(\mathcal X, \mathcal A^{(i)}, G^{(i)}) + \mu\sum_{p,q,r = 1}^{m,n,d} \frac{|\mathcal{X}_{pqr}|}{a+|\mathcal{X}_{pqr}|}\right\}, \label{equ:min_fun_muti}
	 \end{equation}
	 where $K$ is the number of bands.
Similarly, we use the iteratively reweighted $\ell_1$ algorithm \cite{IRL1_2015} to solve \eqref{equ:min_fun_muti}:
\begin{equation}\label{equ:outer_mult}
	\begin{cases}
w^k_{pqr} &=\frac{a \mu } { \left(a + \hat{\mathcal{X}}_{pqr}^k  \right)^2 }, \quad \forall p,q,r; \\
        \hat{\mathcal{X}}^k &= \arg\min\limits_{\mathcal{X}\geq 0  } \left\{ \sum\limits_{i=1}^K D_s(\mathcal X, \mathcal A^{(i)}, G^{(i)})+ \sum\limits_{p,q,r=1}^{m,n,d} w^k_{pqr}  |\mathcal{X}_{pqr} | \right\}.
        \end{cases}
    \end{equation}
    Here the subproblem of \eqref{equ:outer_mult} is solved by ADMM.  
We introduce $(K+1)$ auxiliary variables, namely $\mathcal{U}_{0}^{(1)}, \mathcal{U}_{0}^{(2)}, \dots, \mathcal{U}_{0}^{(K)}$ and $\mathcal{U}_1,$ and we have $(K+1)$ Lagrange multipliers. 
Then the augmented Lagrangian function  $\mathcal{L}_m(\mathcal{U}_0, \mathcal{U}_1, \mathcal{X}, \eta_0, \eta_1 )$ is 
\begin{equation*}
\begin{split}
	& \sum_{i=1}^{K}\langle \mathbb{1}, \mathcal{T} \mathcal{U}_0^{(i)} - G^{(i)} \log (\mathcal{T}\mathcal{U}_0^{(i)} + b \mathbb{1}) \rangle  + \sum_{p,q,r=1}^{m,n,d} w^k_{pqr}  |(\mathcal{U}_1)_{pqr} | \\
	 +&  \frac{\beta_0}{2}\sum_{i=1}^{K}\|\mathcal{U}_0-\mathcal{A}^{(i)} \ast \mathcal{X} - \eta_0 \|^2 + \frac{\beta_1 }{2} \|\mathcal{U}_1-\mathcal{X}-\eta_1\|^2+   I_+(\mathcal{U}_1 ).
	\end{split}
\end{equation*}
The iterative scheme is 
\begin{equation}\label{equ:outer}
	\begin{cases}
		\mathcal{U}_{0}^{t+1} &= \argmin\limits_{\mathcal{U}_{0}} \mathcal{L}_m(\mathcal{U}_0, \mathcal{U}_1^t, \mathcal{X}^t, \eta_0^t, \eta_1^t ) \\ 
		\mathcal{U}_1^{t+1} &= \argmin\limits_{\mathcal{U}_1 \geq 0} \mathcal{L}_m(\mathcal{U}_0^{t+1}, \mathcal{U}_1, \mathcal{X}^t, \eta_0^t, \eta_1^t ) \\
		\mathcal{X}^{t+1} &= \argmin\limits_{\mathcal{X}} \mathcal{L}_m(\mathcal{U}_0^{t+1}, \mathcal{U}_1^t, \mathcal{X}, \eta_0^t, \eta_1^t )\\
		(\eta_0^{(i)})^{t+1} & = (\eta_0^{(i)})^{t} - {\rho ((\mathcal{U}_0^{(i)})^{t+1} -\mathcal{A}^{(i)} \ast \mathcal{X}^{t+1})} \\ &\quad i = 1, 2, \dots, K\\
		\eta_1^{t+1} & = \eta_1^{t} - {\rho( \mathcal{U}_1^{t+1} - \mathcal{X}^{t+1})},
	\end{cases}
\end{equation}

The $\mathcal{U}_1$-subproblem is the same as \eqref{equ:U_1} in the single band image, so it has the same closed-form solution \eqref{equ:U1_closed}. 

The $\mathcal{X}$-subproblem  is a least squares problem. Compared to \eqref{equ:X}, the first part of objective function is the summation.  Similarly, we rewrite the convolution into componentwise multiplication by using the Fourier transform. The closed-form solution of $\mathcal{X}^{t+1}$ reads as
\begin{equation}
\begin{split}
 \mathcal{F}^{-1}
	\left\{\Omega_m\left(\sum_{i=1}^K\overline{\mathcal{F}\{\mathcal{A}^{(i)}\}} \cdot \mathcal{F}\left\{\left(\mathcal{U}_0^{(i)}\right)^{t+1}-\left(\eta_0^{(i)}\right)^{t}\right \} \right.\right. \\
	\left.\left. +\frac{\beta_1}{\beta_0} \mathcal{F}\{\mathcal{U}_1^{t+1}-\eta_1^{t}\}\right)  \right\},
	\label{equ:fft}
	\end{split}
	\end{equation}
	where $\Omega_m= \left(\sum\limits_{i=1}^K\left|\mathcal{F}\{\mathcal{A}^{(i)}\}\right|^2+\frac{\beta_1}{\beta_0} \right)^{-1}$.
	As in the single band case, we use the 3D FFT to compute \eqref{equ:fft} efficiently.
	
The solution of the $\mathcal{U}_0$-subproblem is the same as \eqref{equ:U_0} for each $\mathcal{U}^{(i)}_0$.  Therefore, we need Proposition 1 in \cite{kl_closedform2009total}. Then the closed-form solution of $\mathcal{U}_0^{(i)}$ is 
    
    \begin{equation*}
    	(\mathcal{U}^{(i)})^{t+1}_{pqr} = 
    	\begin{cases} 
    		\frac{-\Big(\xi^{(i)}_1\Big)^{t}_{pqr} + \sqrt{\left\{\Big(\xi^{(i)}_1\Big)^{t}_{pqr}\right\}^2+4 \beta_0  G_{pq}^{(i)}   }}{2 \beta_0  }, & \text{if } r =  d, \\
    		\Big(\xi^{(i)}_0\Big)^{t}_{pqr}, & \text{otherwise}.
    	\end{cases}\label{equ:u_closed}
    \end{equation*}
    where $\Big(\xi^{(i)}_1\Big)^{t} =1 - \beta_0 b - \beta_0 \Big(\xi^{(i)}_0\Big)^{t},$ and $ \Big(\xi^{(i)}_0\Big)^{t} = \mathcal{A}^{(i)} \ast \mathcal{X}^{t} +(\eta_0^{(i)})^{t}. $
    
Note that $\{\mathcal{U}_0^{(i)}\}$ and $\eta_0^{(i)}$ are a sequence of 3D tensors. In our implementation, we store them in two 4D tensors
and  use component-wise operators for 4D tensors in the updating scheme for the $\mathcal{U}_0$ and $\eta_0$ subproblems. 
In addition, both $\{\eta_0^{(i)}\}$ and $\eta_1$ can be stored in the Fourier domain to reduce per-iteration computations. 

{\bf Remarks:} Similar to the single band case, we need to do post-processing to remove some clustered false positives. 
We compute the centroid for each clustered point source and then represent the clustered set by the centroid \cite{kl_nc}. 
We subsequently perform false positive removal, as described in the following section.

\subsection{Stage 2: Alternating approach for estimating spectral signatures and removing false positives}
In the 1st stage, we obtain the locations of the point sources 
$$
(x_j, y_j, z_j), \ j = 1, \dots, M,
$$
where $M$ is the number of point sources. 
In the 2nd stage, our goal is to more accurately estimate the spectral signature of each point source. 
Here the vectorized PSF image corresponding to the $j$-th point source and the $i$-th band is denoted as $\mathbf{h}_j^{(i)}$, which is normalized, i.e., $\h 1^T \h h_j^{(i)}=1$, where $\h 1$ is the vector of all 1's. 
We let $L$ be the total number of pixels in the vectorized data array, so $L=mn. $ 
Then we define a system PSF matrix $H^{(i)}$ for the $i$-th band with 
\begin{equation*}
	H^{(i)} = \Big[\mathbf{h}_1^{(i)}, \mathbf{h}_2^{(i)}, \dots, \mathbf{h}_M^{(i)}\Big] \in \mathds{R}^{L\times M}. 
\end{equation*}

The vectorized observed image in the $i$-th band is denoted by  $$\mathbf{g}^{(i)} = \mathcal{N}(H^{(i)}\mathbf{f}^{(i)}+ b \h 1) \in \mathds{R}^{ L \times 1}, $$ 
where $\mathbf{f}^{(i)}\in \mathds{R}^{M}$ is the flux vector for all estimated point sources in the $i$-th band.  In second stage, we follow \cite{kl_nc} to estimate the flux information in each band via the iterative algorithm:

\begin{equation}
	(\mathbf{f}^{(i)})^{k+1} = \mathbf{f}^{(i)}_G + \mathcal{K}^{(i)}\left((\mathbf{f}^{(i)})^{k}\right),  \quad k = 1, 2, \dots
	\label{iterative_scheme_flux}
\end{equation}
in which $\mathbf{f}^{(i)}_G = \left(H^{(i)}\right)^{+}(\mathbf{g}^{(i)}-b \mathbf{1} )$,
$\left(H^{(i)}\right)^{+}$ being the pseudo inverse of $H^{(i)}$, is the solution 
corresponding to the Gaussian noise model and $$\mathcal{K}^{(i)}(\mathbf{v}) := \sum\limits_{p = 1}^L\frac{\mathbf{e}_p^T\left( H^{(i)} \mathbf{v} + b \, 1-\mathbf{g} \right)\mathbf{e}_p^TH^{(i)} \mathbf{v}}{\mathbf{e}_p^T (H^{(i)} \mathbf{v} + b \mathbf{1})} \left(H^{(i)}\right)^+ \mathbf{e}_p, $$
where 
 $\mathbf{e}_p$ is the $p$-th canonical basis unit vector.  

The iterative scheme can be implemented as a fixed point iteration away from the starting Gaussian noise solution $\mathbf{v}_G$ to the final Poisson noise solution,  a process that shows the connection between these two noise models. 

In multispectral  images, this iterative scheme is processed for each $\mathbf{f}^{(i)}$ separately. Then the estimated spectral signature for the $j$-th point source is 
\begin{equation*}
	\h f_{j} = \Big(f^{(1)}_{j}, f^{(2)}_j, \dots,  f^{(K)}_j\Big). 
\end{equation*}
 Unlike the single band image case, we next utilize the estimated spectral information to refine the localization and spectrum estimation. We do these processes in an alternating fashion for further removal of false positives in the estimated point sources. In \cite{kl_nc}, we observe that the flux value in false positives is usually lower than the value of most true positives.  However, it is impractical to set up the threshold with a single band as it requires some information on the flux value in ground truth. In multispectral  images, with more information on fluxes in each band, we can estimate the threshold.  

Here we propose two criteria for identifying the false positives. The first is the positivity of the fluxes. Note that the iteration scheme \eqref{iterative_scheme_flux} does not require a positivity  constraint. In the numerical tests, we observe negative flux in some bands for some of the false positives. 
This can be explained as the effect of false positives compensating for an overestimate  of some true positive flux values.  
The second criterion involves the summation of the spectral values in each point source. 
We know the flux varies for different materials in each band and we may get a large flux value for false positives when considering only one band. 
However, because the PSF images are different in different bands, it is very seldom that we have large flux values in every band for false positives. 
Thus, the probability of having a large flux value in every band for false positives is small. In other words, multispectral imaging data contain more information to detect the false positives.  
Moreover, the ground-truth spectrum is normalized for each point source. 
The summation of the spectra of ground truths is in a certain range, while we expect, as we confirmed in our numerical tests, false positives fail to reach this range of energy.  
In fact, we observed false positives typically to have much lower summation values
 
Here, we can regard the point source a as false positive if its corresponding estimated spectral signature $\h f$ satisfies 
 \begin{equation*}
 	 \h 1^T \h f  \leq\gamma \max_{k} \h 1^T \h f_k,
 \end{equation*}
 where the threshold \begin{equation}\label{equ:gamma}
 	\gamma < \min\limits_{k \neq l}\frac{\h 1^T \h f^\ast_k}{\h 1^T \h f^\ast_l },
 \end{equation}
    and   $ \h f^\ast_k$ denotes the true spectral signature for the $k$-th material. The value $\min\limits_{k \neq l}  \frac{\h 1^T \h f^\ast_k}{\h 1^T \h f^\ast_l }$ is computed from  the spectral library \cite{NASA_DATA}  and the choice of $\gamma$ is robust in not being tightly linked to the ground truth; see also \Cref{sec:thr_s2}. 
 The algorithm for our 2nd stage is summarized as \Cref{alg:s2}.

\begin{algorithm}[htp]
\begin{algorithmic}[1]
\Require Estimated  $\left\{(\hat{x}_j, \hat{y}_j, \hat{\zeta}_j) \right\}$ and dictionaries $\mathcal{A}^{(i)}, i = 1, 2, \dots, K $.
\Ensure The locations and the estimated spectrum for each estimated point source.
    \State Construct the matrices $H^{(i)}, i = 1, 2, \dots, K$;
    \State Do the post-processing by removing the clustered false positives to get the locations of estimated point sources   $\left\{(\hat{x}_j, \hat{y}_j, \hat{z}_j)\right\}$;
    \State Estimate spectra $\{\hat{\h f_j}\}$ of point sources by the iterative scheme  \eqref{iterative_scheme_flux};
    \State Identify the false positives by negative flux values and threshold;
    \State Remove the corresponding columns of false positives in $H^{(i)}$ and go to Step 1 until there are none detected in Step 4.   
\end{algorithmic}
\caption{Alternating approach in Stage 2. }
\label{alg:s2}
\end{algorithm}

\subsection{Stage 3: Multispectral Classification of Point Sources} 

In the last stage, we  classify  the materials of space debris based on the estimated spectra in the 2nd stage. There are many methods for classification or clustering developed for hyperspectral images. See e.g., \cite{bioucas2008hyperspectral,mianji2011svm,iordache2011sparse}, which determine the underlying materials in each pixel. Here, for the multispectral rotating PSF image, we detect the material for each point source instead of for each pixel in the observed image. The measured spectrum at each point source is assumed to be a linear combination of spectral signatures (called endmembers), so we may formulate the linear spectral mixture model as follows:
\begin{equation*}
	F = X M + E,
\end{equation*}
 where the estimated spectra is
\begin{equation*}
	F: = \left[\h f^{(1)},  \h f^{(2)},  \dots, \h f^{(K)}\right] \in \mathds{R}^{M \times K}.
\end{equation*}
 Here, $M\in \mathds{R}^{N\times K}$ is a spectral library containing  
spectral signatures of $N$ endmembers with $K$ spectral bands
  and $E$ is the computational error from the former two stages. Here we need to solve $X\in\mathds{R}^{M\times N}$ whose entries are the coefficients for the linear combination of different endmembers. Here the summation of entries in each row of $X$ is one. We assume each space debris contains only a few kinds of materials, so $X$ is a sparse and nonnegative matrix.  
We formulate the classification problem as a least-squares minimization problem.
In order to determine the matrix $X$, we add the abundance nonnegativity constraint $X \geq 0$ (i.e., each entry is nonnegative) and the abundance sum-to-one constraint $$\sum\limits_{q=1}^N X_{jq}  = 1, \text{for } j = 1, \dots, M. $$ 

Therefore, the minimization for spectral unmixing classification is 
\begin{equation}\label{equ:classification}
\min\limits_{X\in \mathds{R}^{M \times N}} \|XM- F\|_F \ \text{s.t.} \ X \geq 0,  \sum\limits_{q=1}^N X_{jq}  = 1,  \ j = 1, \dots, M.
\end{equation}

Note that we do not need to involve a total variation (TV) term for requiring the sparsity of the gradient in $X$ as in \cite{Nagy2015deblurring,li2012compressive,li2012coupled,zhao2013deblurring}. 
This is because the objects we classify are not pixels in the image but point sources of space debris which are independent of one another. 
The minimization is solved by an Interior Point Least Squares solver \cite{clas_alg2,clas_alg1}. 
The proposed algorithm for all the stages is summarized as \Cref{alg:whole_process}. 

\begin{algorithm}[btp]
\begin{algorithmic}[1]
\Require $\mathcal{X}^0\in \mathds{R}^{m\times n \times d}$ and $\left\{G^{(i)}\in  \mathds{R}^{m \times n}, i = 1, 2, \dots, K\right\}$.
\Ensure The location and the labels of materials for each point source with the corresponding spectral signature  $\left\{(\hat{x}_j, \hat{y}_j, \hat{\zeta}_j, \hat{\mathbf{f}_i}, \text{label}_j), \ j  = 1, 2, \dots, M \right\}$.
    \State Solve the nonconvex optimization problem by \Cref{equ:outer} and obtain the minimizer $\hat{\mathcal{X}}$ by assuming uniform spectral signature for each point source;
    \State Do the post-processing by removing the clustered false positives to get the locations of estimated point sources   $\left\{(\hat{x}_j, \hat{y}_j, \hat{\zeta}_j)\right\}$;
    \State Estimate spectrum $\hat{\mathbf{f}}$ of point sources by the iterative scheme  \eqref{iterative_scheme_flux} and further remove the false positive by two criterions via \Cref{alg:s2} to get $F$. 
    \State Classify the point sources in Stage 3 by \eqref{equ:classification}.  
\end{algorithmic}
\caption{Three-stage method for multispectral imaging via rPSFs. }
\label{alg:whole_process}
\end{algorithm}

\section{Gaussian Noise Case}\label{sec:Gaussian}
In this subsection, we briefly discuss cases where the point source images are contaminated by additive Gaussian noise, rather than signal-dependent Poisson noise.  The noisy  forward model now becomes: 
\begin{equation}
	G^{(i)} \approx \mathcal{T}(\mathcal{A}^{(i)} \ast \mathcal{X}^{(i)})  + b\mathbb{1}  + N^{(i)},
	\label{equ:dependent}
\end{equation}
 where $N^{(i)}$ is additive Gaussian noise with zero mean and standard deviation equal to 10\% of the highest pixel value in the $i$-th band of the original image $ \mathcal{T}(\mathcal{A}^{(i)} \ast \mathcal{X}^{(i)}  )$. 

In Stage 1, we use the $\ell_1$ norm as the regularization term. Then the optimization model is 
\begin{equation}
	 	\min\limits_{\mathcal{X}\geq 0 }\left\{  \sum\limits_{i=1}^K \frac{1}{2} \left\| \mathcal{T}(\mathcal{A}^{(i)} \ast \mathcal{X}) +b\mathbb{1} - G^{(i)}  \right\|^2_F + \mu\sum_{p,q,r = 1}^{m,n,d} |\mathcal{X}_{pqr}|\right\}. \label{equ:min_fun}
	 \end{equation}
	 Because the model is convex, we use ADMM directly without any outer iteration.
Upon introducing $(K+1)$ auxiliary variables $\mathcal{U}_1$ and $  \mathcal{U}_0^{(i)}\  i = 1, \dots, K$, we obtain the following augmented Lagrangian function $\mathcal{L}_g(\mathcal{U}_0,\mathcal{U}_1, \mathcal{X})$:
\begin{equation*}
\begin{split}
	 & \sum\limits_{i=1}^K \frac{1}{2} \left\| \mathcal{T}\mathcal{U}_0^{(i)} + b \mathbb{1}- G^{(i)}  \right\|^2_F + \mu\sum_{p,q,r = 1}^{m,n,d} \left|(\mathcal{U}_1)_{pqr}\right| \\  
	 + &  \frac{\beta_0}{2} \sum_{i=1}^{K}
\left\|(\mathcal{A}^{(i)} \ast \mathcal{X}) - \mathcal{U}_0^{(i)}-\eta_0^{(i)}  \right\|^2 + \frac{\beta_1}{2}\|\mathcal{X}- \mathcal{U}_1\|^2 +   I_+(\mathcal{U}_1 ).
\end{split} 
\end{equation*}
The iterative scheme is 
\begin{equation}\label{equ:outer_G}
	\begin{cases}
		\mathcal{U}_{0}^{k+1} &= \arg\min\limits_{\mathcal{U}_{0}} \mathcal{L}_g(\mathcal{U}_0,\mathcal{U}_1^k, \mathcal{X}^k) \\
		\mathcal{U}_1^{k+1} &= \arg\min\limits_{\mathcal{U}_1 \geq 0} \mathcal{L}_g(\mathcal{U}_0^{k+1},\mathcal{U}_1, \mathcal{X}^k) \\
		\mathcal{X}^{k+1} &= \arg\min\limits_{\mathcal{X}}\mathcal{L}_g(\mathcal{U}_0^{k+1},\mathcal{U}_1^{k+1}, \mathcal{X})\\
		\left(\eta_0^{(i)}\right)^{k+1} &= \left(\eta_0^{(i)}\right)^{k} - {\rho \Big(\left(\mathcal{U}_0^{(i)}\right)^{k+1} -\mathcal{A}^{(i)} \ast \mathcal{X}^{k+1}\Big)} \\ &\quad i = 1, 2, \dots K\\
		\eta_1^{k+1} &= \eta_1^{k} - {\rho( \mathcal{U}_1^{k+1} - \mathcal{X}^{k+1})}.
	\end{cases}
\end{equation}

In Stage 2, we still use an alternating scheme to estimate the spectral signatures and  then refine the 3D localizations. However, the inner loop only minimizes a least squares error, which yields the estimated flux in the $i$-th band, $\h f^{(i)}$ as 
\begin{equation*}
	\h f^{(i)} = \left(H^{(i)}\right)^{+} \h g^{(i)}. 
\end{equation*}
The classification algorithm for the Gaussian noise case in Stage 3 is the same as the one for the Poisson noise case, as it is independent of the noise model.

%

 \section{Numerical Experiments}\label{sec:numerical}

Here we test our proposed numerical algorithms for the localization and classification of point sources using multispectral rotating PSFs and compare the results with those obtained using the single band approach. All the numerical experiments are conducted on a standard desktop with Intel i7-6700, 3.4GHz CPU and $\mathrm{MATLAB \ 9.2 \ (R2017a)}. $
 
We use real spectral data provided by Kira Abercombie \cite{NASA_DATA}.
 This test data corresponds to materials which make up various satellites, rocket bodies, etc., from the NASA JSC Space Object Materials Spectral Database. Some of the data was collected from lab tests, and other data was from flown satellites imaged from ground-based telescopes. 

Similar data was also used in \cite{Nagy2015deblurring,zhang2008tensor,zhao2013deblurring}.
We assume here that the space debris in question consist of parts of man-made orbiting objects 
with each debris being unresolved and thus treatable as a point source by our imaging system.
The signatures cover a band of spectra from 400nm to 2500nm over 100 evenly distributed sampling points. Each material type shows a different spectrum based on its composition. Using low-resolution reflectance spectroscopy and comparing the absorption features and overall shape of the spectra, it is possible to determine the material type of man-made orbiting objects. 

\begin{figure}
\centering
\subfloat[M1: Hubble alumninum ]{\includegraphics[width=0.15\textwidth]{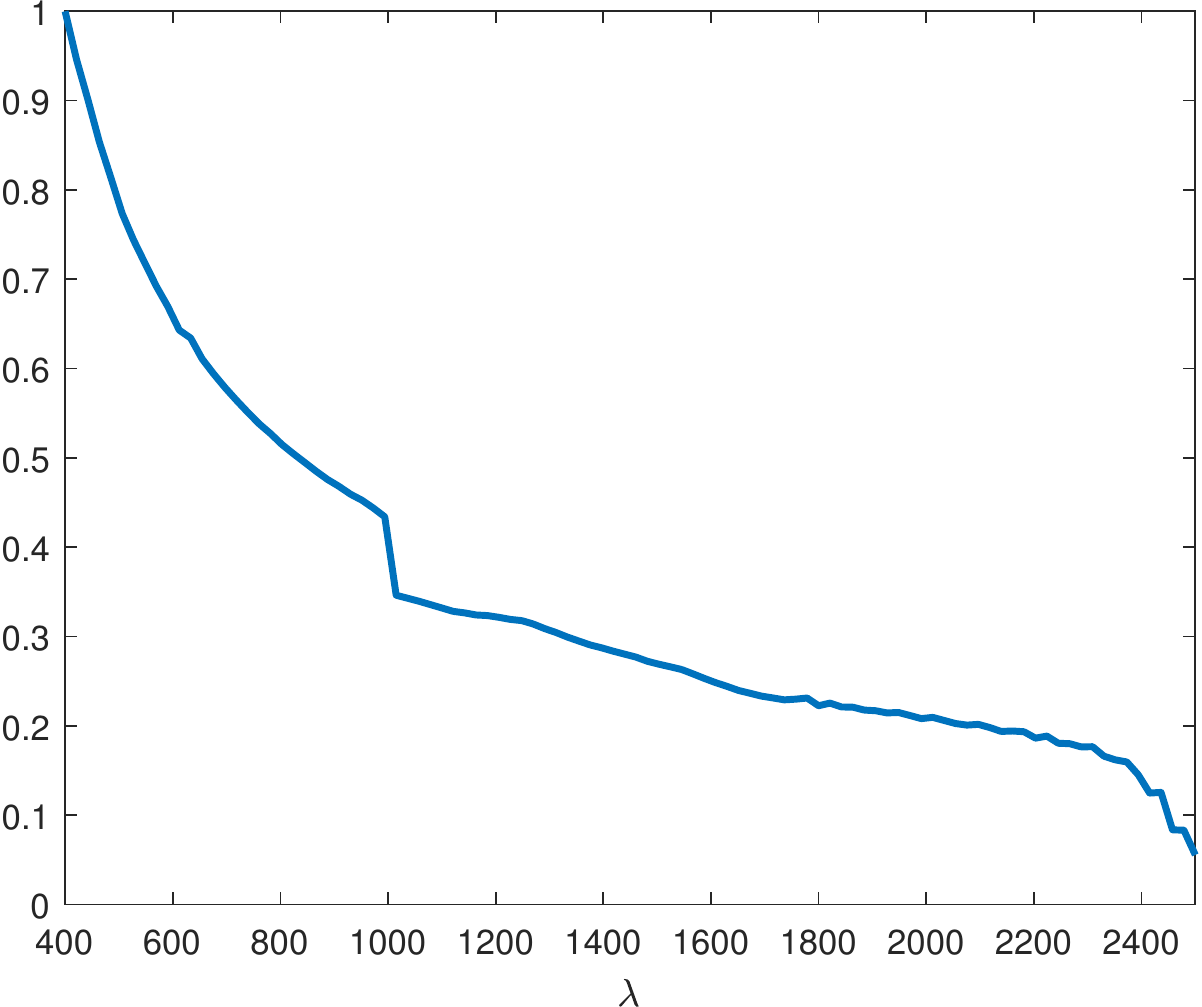}} \hspace{0.01mm}
\subfloat[M2: Hubble glue]{\includegraphics[width=0.15\textwidth]{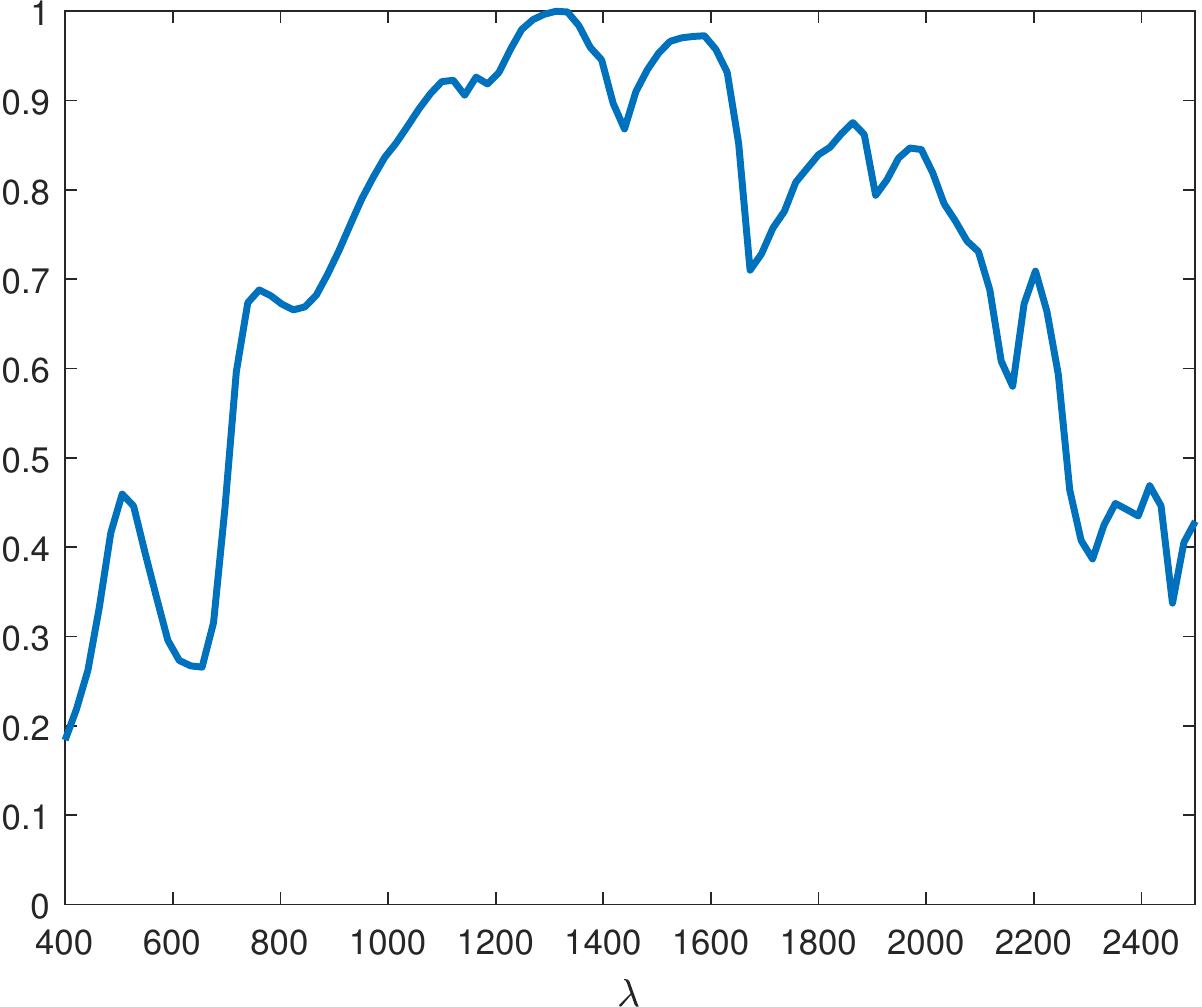}} \hspace{0.01mm}
\subfloat[M3: Hubble solar cell]{\includegraphics[width=0.15\textwidth]{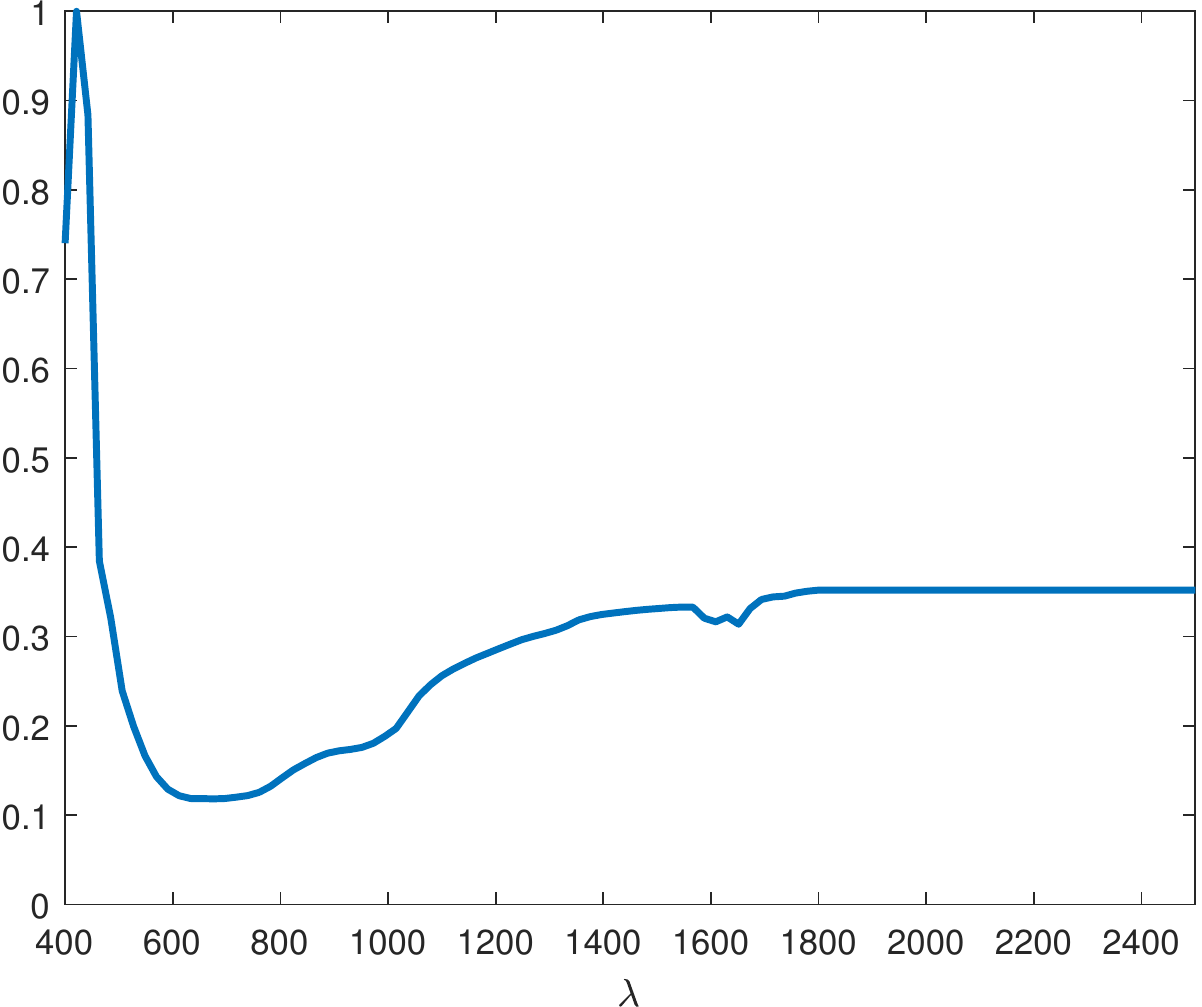}}\\
\subfloat[M4: Black rubber edge]{\includegraphics[width=0.15\textwidth]{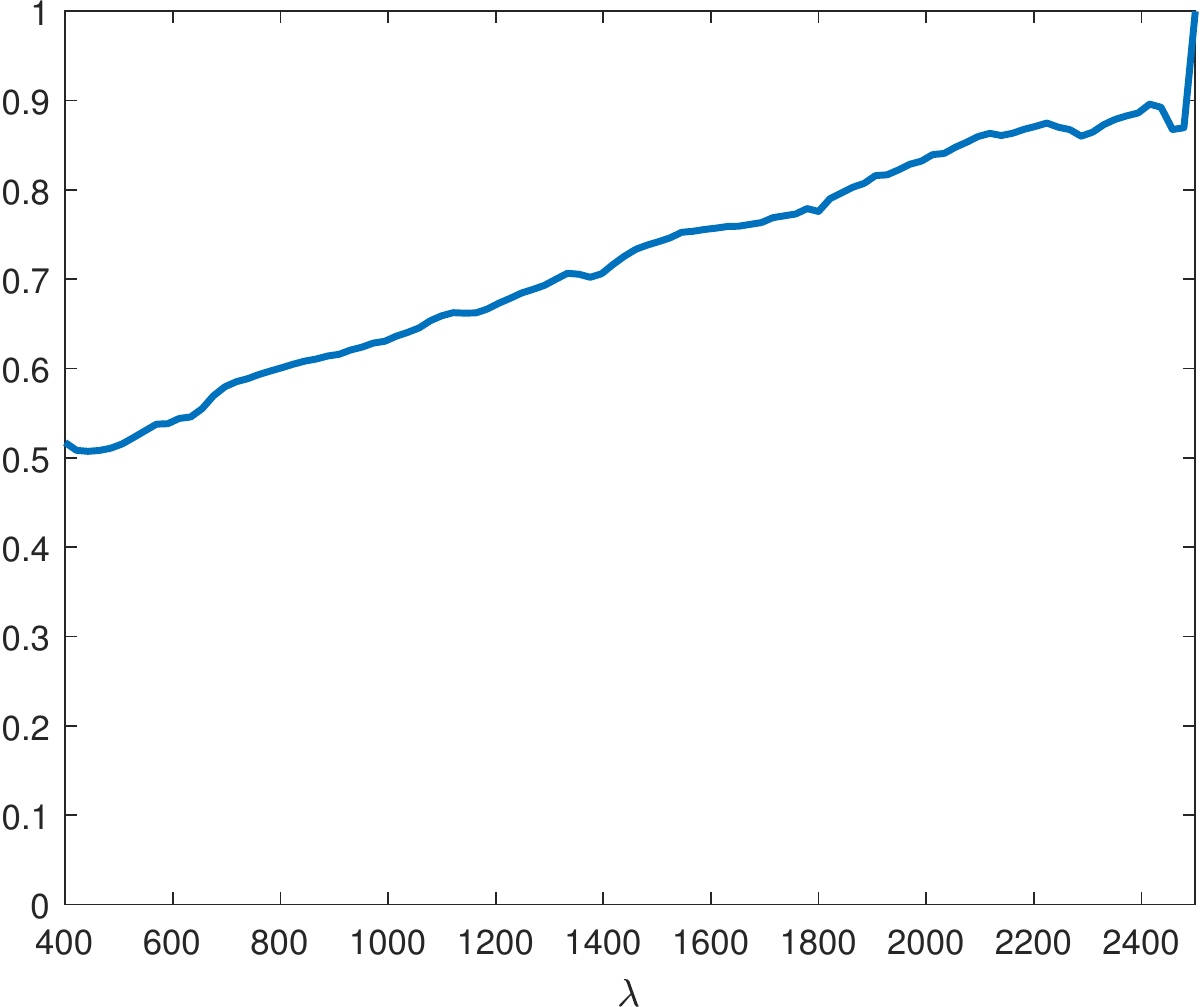}} \hspace{.1mm}
\subfloat[M5: Bolts]{\includegraphics[width=0.15\textwidth]{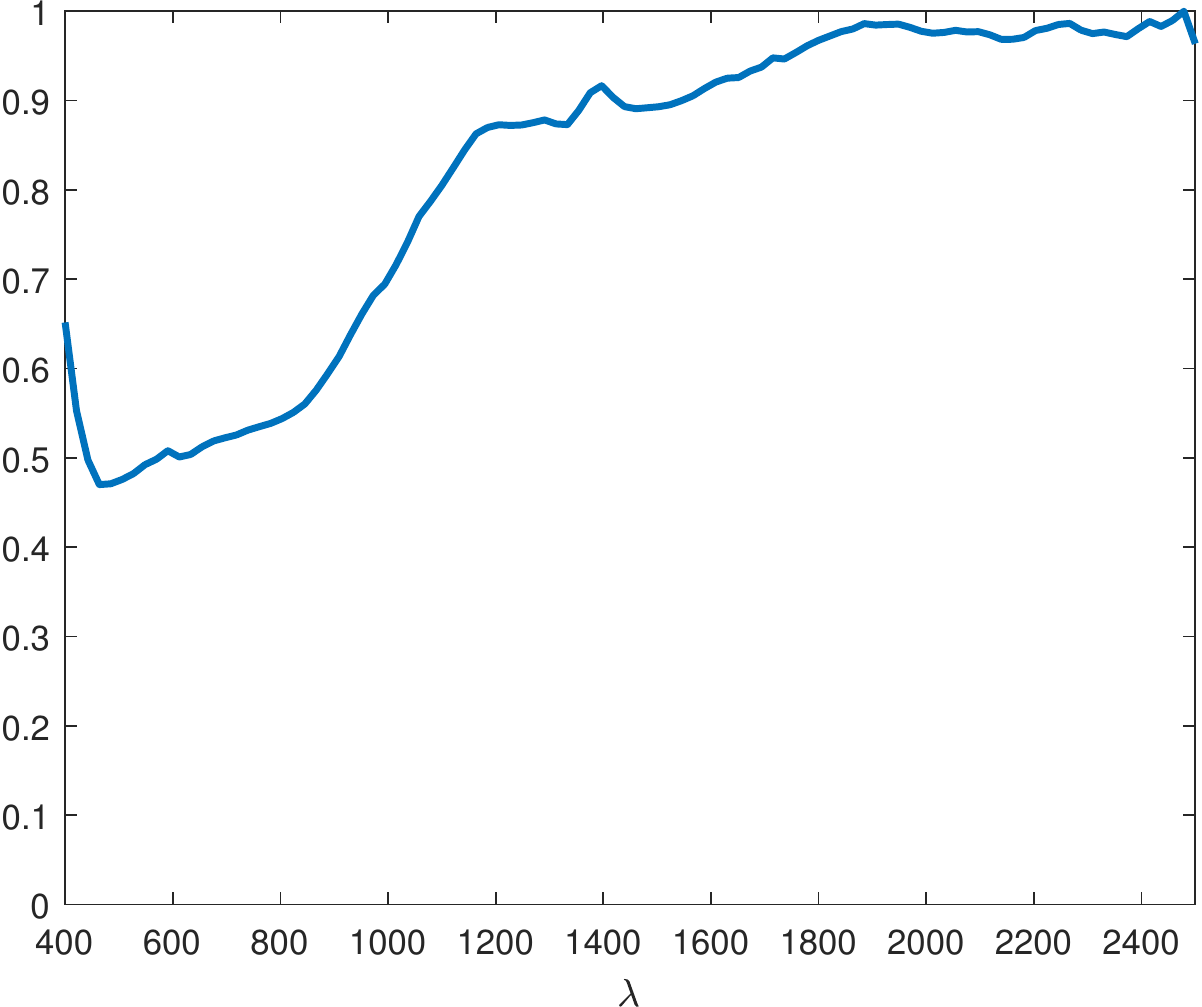}} \hspace{0.01mm}
\caption{Spectral signatures of five materials assigned to the simulated space debris. }
\label{fig:spectral_signatures_without_sampling}
\end{figure}

We assume that each point source contains only one material and test our ideas on 3D localization and classification. We consider five different kinds of materials. The spectral signatures and the names as well as the corresponding false colors for the materials are shown in \Cref{fig:spectral_signatures_without_sampling}.

\begin{figure}
\centering
\includegraphics[width=0.44\textwidth]{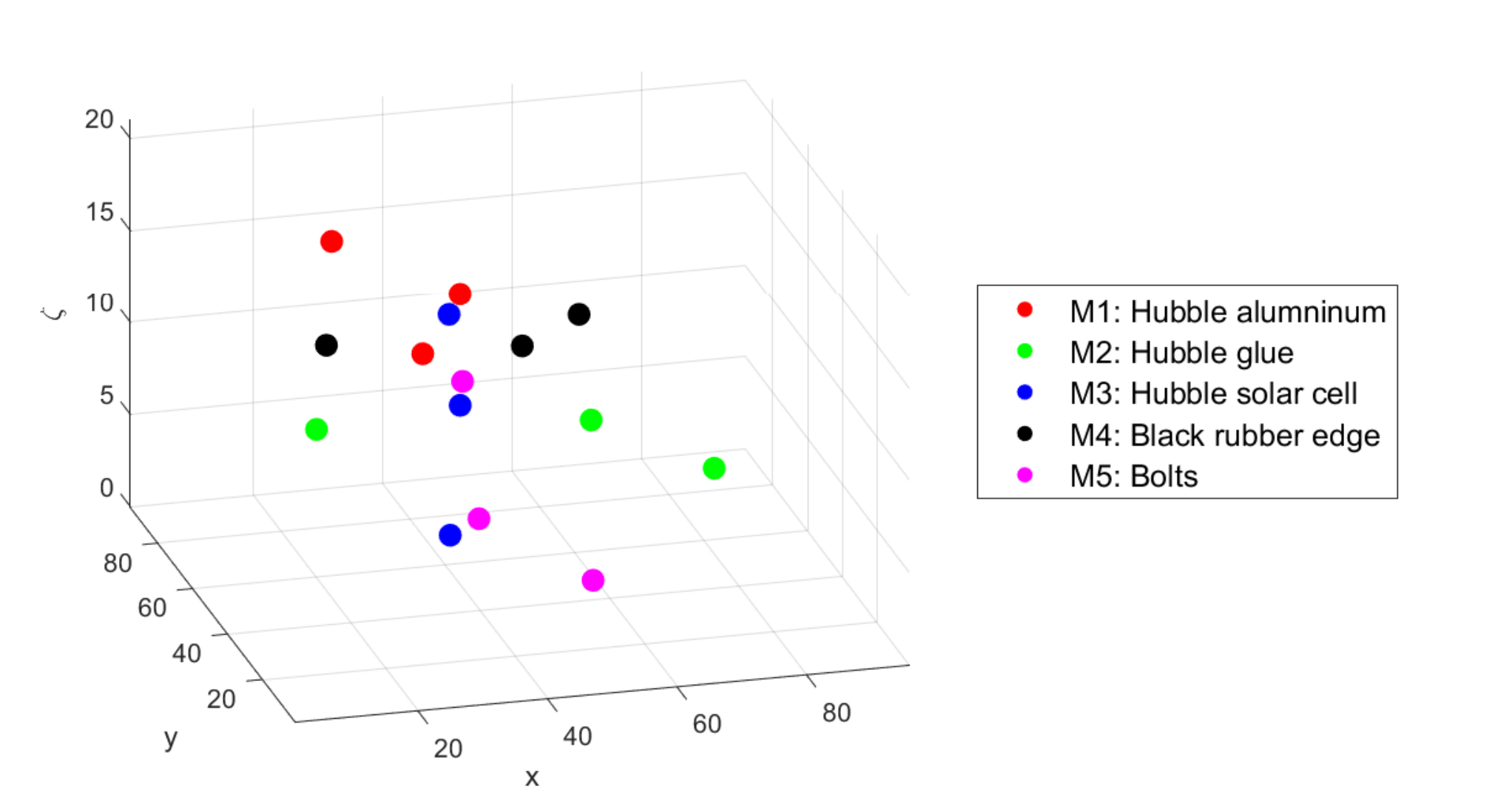}
\caption{Simulated space debris with 5 materials.}
\label{fig:3Dloc}
\end{figure}

 For the Poisson noise case, only the bands with wavelengths between 400 nm and 1000 nm are considered. In the Gaussian noise case, we only consider the higher wavelength bands between 1500nm and 2500nm, over which 
 the photon number per mode 
 is likely to be large compared to 1 and the statistics of the intrinsic photon number more likely to be Gaussian, much like the read noise, which would combine to yield Gaussian statistics for the observed counts at the sensor.

We present results of detailed numerical experiments only for the case of Poisson noise.  In \Cref{subsec:gaussian}, we merely summarize our results for Gaussian noise. Here all 2D simulated observed images are described by 96-by-96 matrices. The number of zones in the spiral phase mask that generates rotating PSF imagery is 7, and the aperture-plane side length is 4 for the Poisson noise case and 2 for the Gaussian noise case. 
Based on \cite{kl_nc}, both the 3D dictionary and our discretized 3D space contain 21 slices in the axial direction, with the corresponding values of the defocus parameter, $\zeta$, distributed uniformly over the range, $[ -21, 21]$. 
Since the ground truth point sources may not be on  grid points, 
we consider a point source to be recovered if its location is estimated correctly to within one pixel in the axial direction and two pixels in the transverse 
directions in our comparisons with ground truth \cite{kl_nc}.  

The fidelity of localization is assessed in terms of the {\bf recall rate}, defined as {\it  the   number of correctly identified true  point sources divided by the number of true  point sources}, and the {\bf precision rate}, defined as {\it  the   number of correctly identified true positive point sources divided by the number of all point sources obtained by the algorithm, which includes the false positives}; see \cite{Book_SR_micro2017}. To quantitatively evaluate the performance of classification, we use the following two widely-used metrics: {\bf overall accuracy (OA)} -  {\it  the percentage of correctly classified point sources among the identified true positive point sources} and {\bf kappa coefficient  (kappa)} -  {\it the percentage of correctly classified pixels corrected by the number of agreements that would be expected purely by chance}  \cite{kappa}.

The original images and observed images for 5 bands are shown in \Cref{fig:org_obs}.  
The corresponding center wavelengths in these five bands are 400.0nm, 548.5nm, 697.0nm, 845.5nm, and 993.9nm, respectively.  The number of photons emitted by each point source in a particular band is taken to be 2000 times the normalized spectral  power of that band. 
The corresponding sequence of the 15 simulated space debris, with 3 different debris for each of the 5 materials, is shown in \Cref{fig:3Dloc}. Each space debris is represented as a point source with a different spectral trace representing a different material.
In \Cref{fig:org_obs}, we plot the corresponding original images in each band for this 15 point sources case, as well as the noisy observed images. Some of the PSF images overlap and we cannot recognize their rotation. The angle of rotation changes with changing wavelength. With longer wavelengths, the PSF images spread out due to greater diffraction. All these factors make the 3D localization of the point sources very challenging.

\newcommand{\figwidth}{.16\textwidth}
\begin{figure*}[tbhp]
	\centering 
	\begin{tabular}{ccccc}
		Band 1 & Band 2 & Band 3 & Band 4 & Band 5\\
		\includegraphics[width=\figwidth]{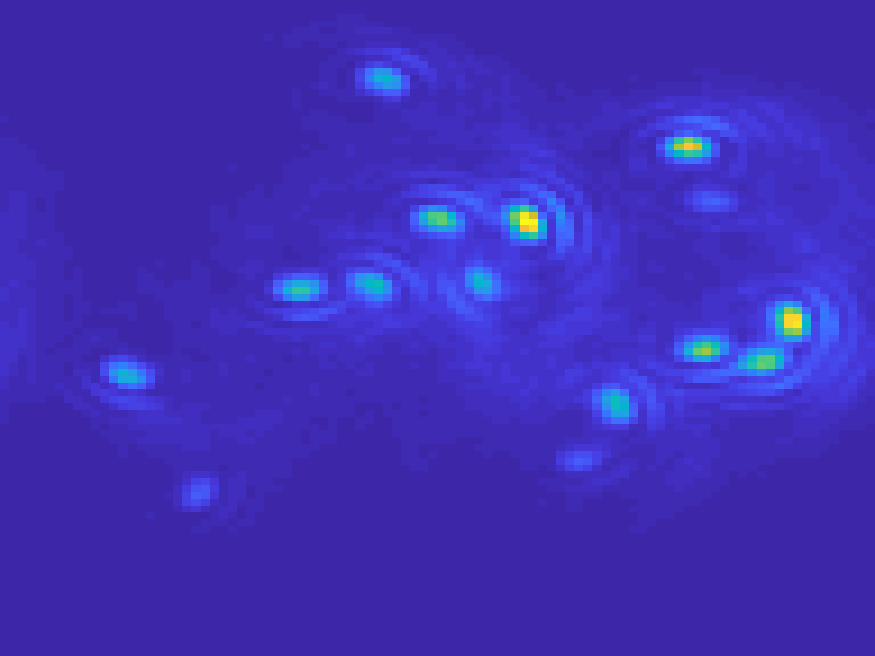}&
		\includegraphics[width=\figwidth]{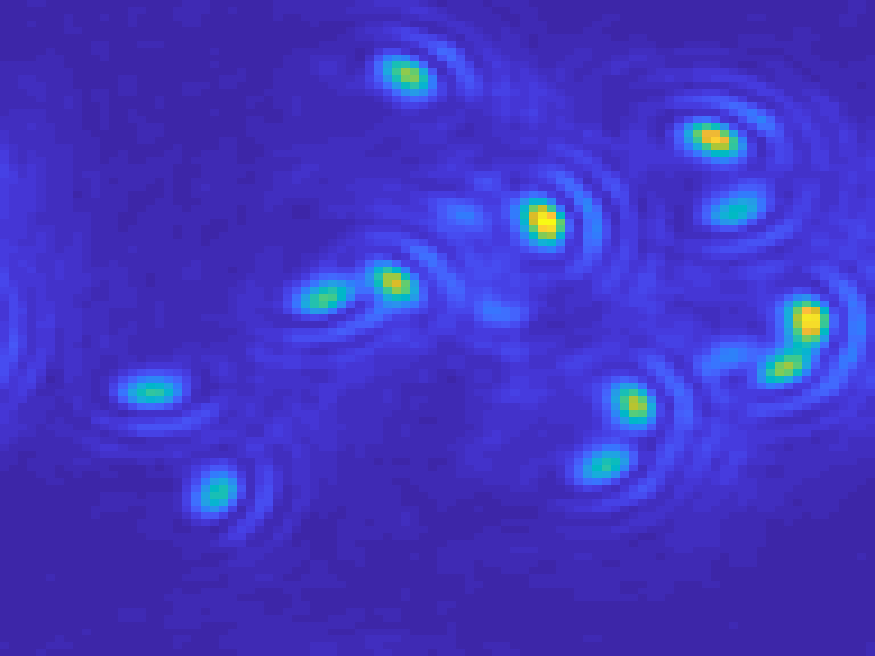}&
		\includegraphics[width=\figwidth]{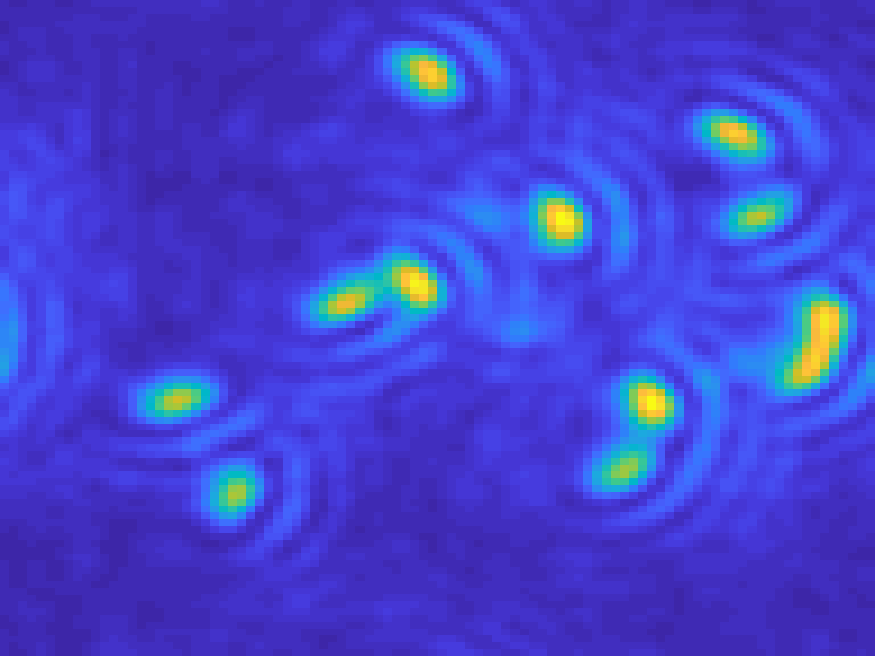} &
		\includegraphics[width=\figwidth]{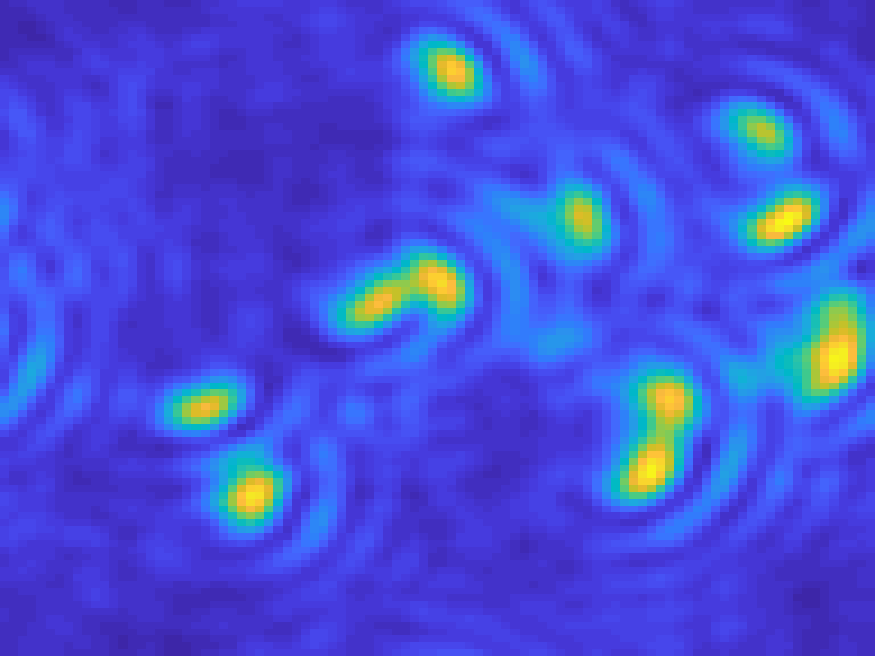} &
		\includegraphics[width=\figwidth]{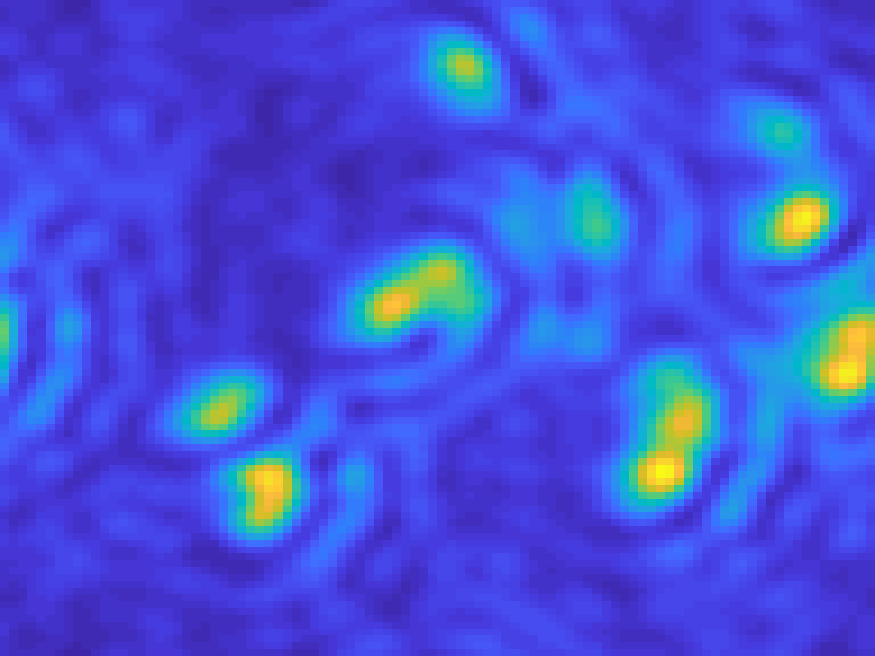}\\
		\includegraphics[width=\figwidth]{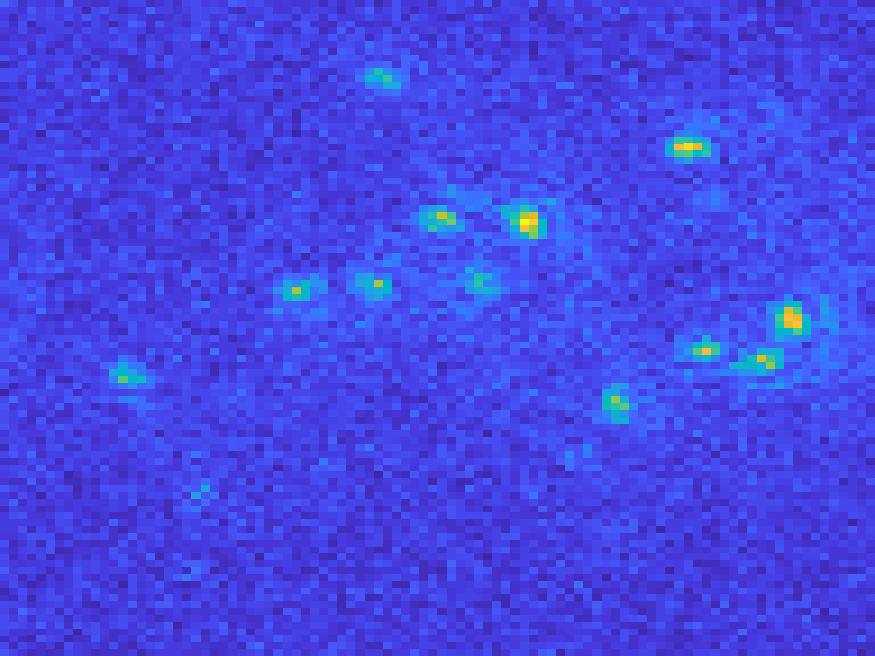}&
		\includegraphics[width=\figwidth]{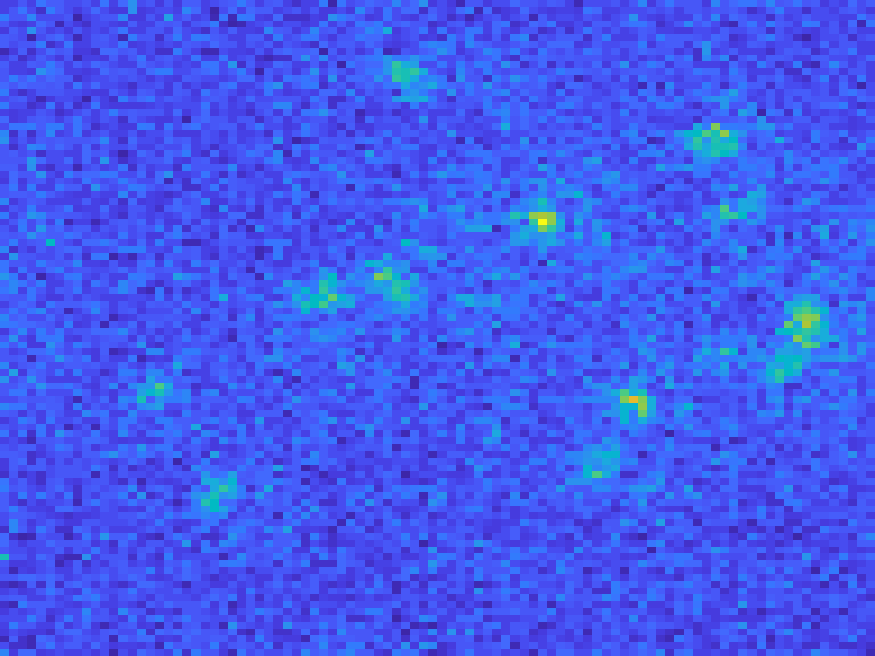}&
		\includegraphics[width=\figwidth]{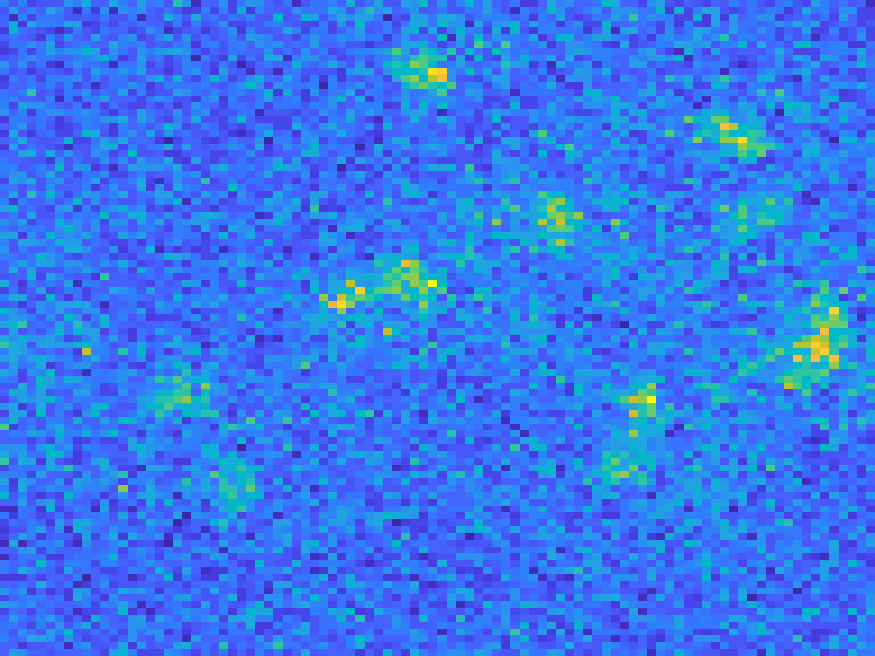}&
		\includegraphics[width=\figwidth]{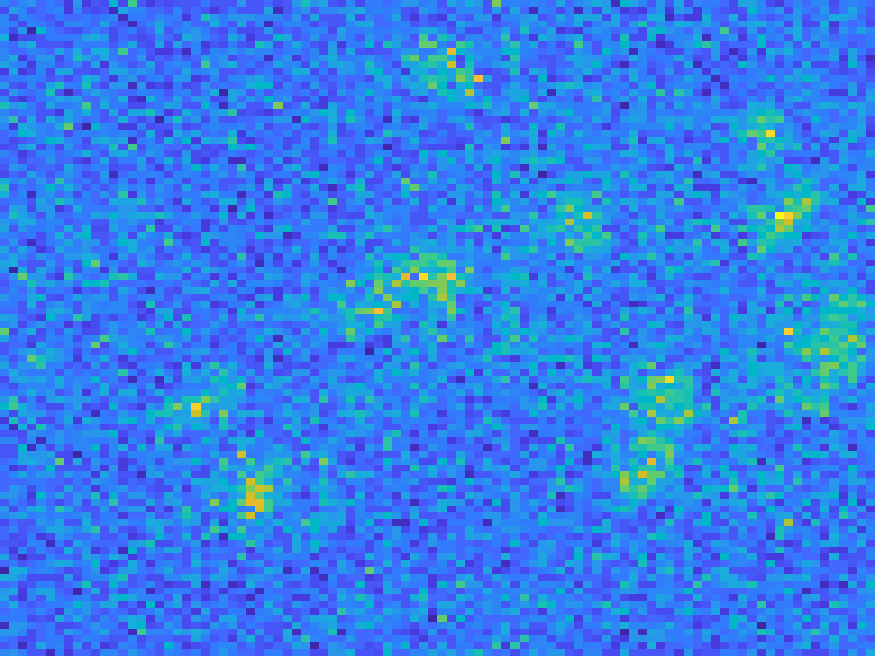}&
		\includegraphics[width=\figwidth]{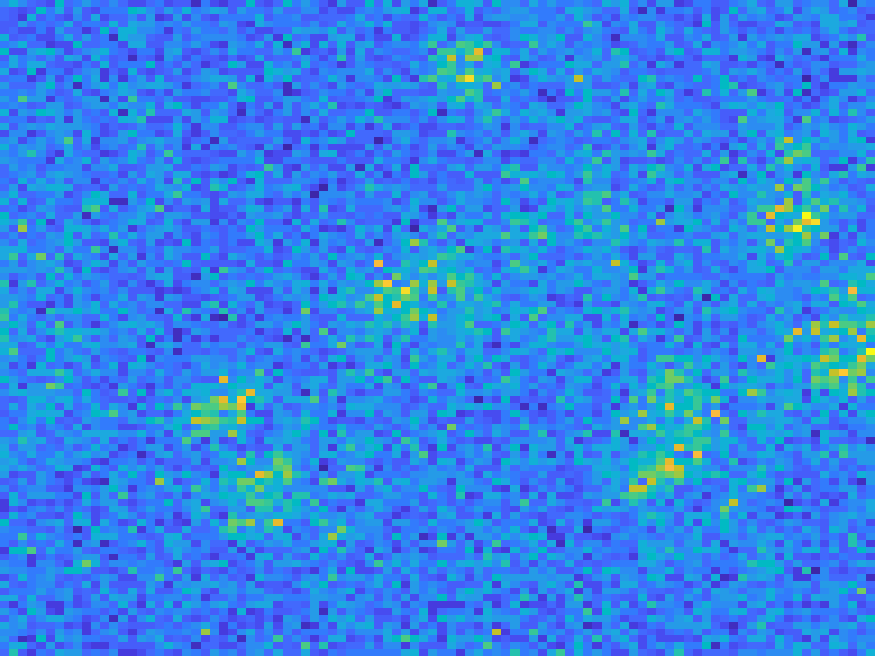}
	\end{tabular}
	\caption{Top: the original image in each band, bottom: the observed image in each band. }
	\label{fig:org_obs}
\end{figure*}

\subsection{Comparisons with the Single Band Approach}
We compare the use of single band images here 
by replacing the algorithm in Stage 1 by the single band approach for localization that we reviewed in Section \ref{sec:single_band}. That is to say, we estimate the 3D locations of point sources via the method in  \cite{kl_nc}  for a single frame $G^{(1)}$ corresponding to the shortest wavelength and then we do the same processing in Stage 2 and 3.  Note that there is no chance to do the classification without multispectral information for these tests. The single band algorithm \cite{kl_nc} can thus achieve only localization. So, the single band result in \Cref{fig:class_loc} is set up  to use a single band in Stage 1 with our proposed stopping criterion (see \Cref{sec:stopcri} for the details) and use multiband information in Stages 2 and 3. 

\subsubsection{Sample Trial}

\begin{figure}[htbp]
\centering
	\includegraphics[width=0.23\textwidth]{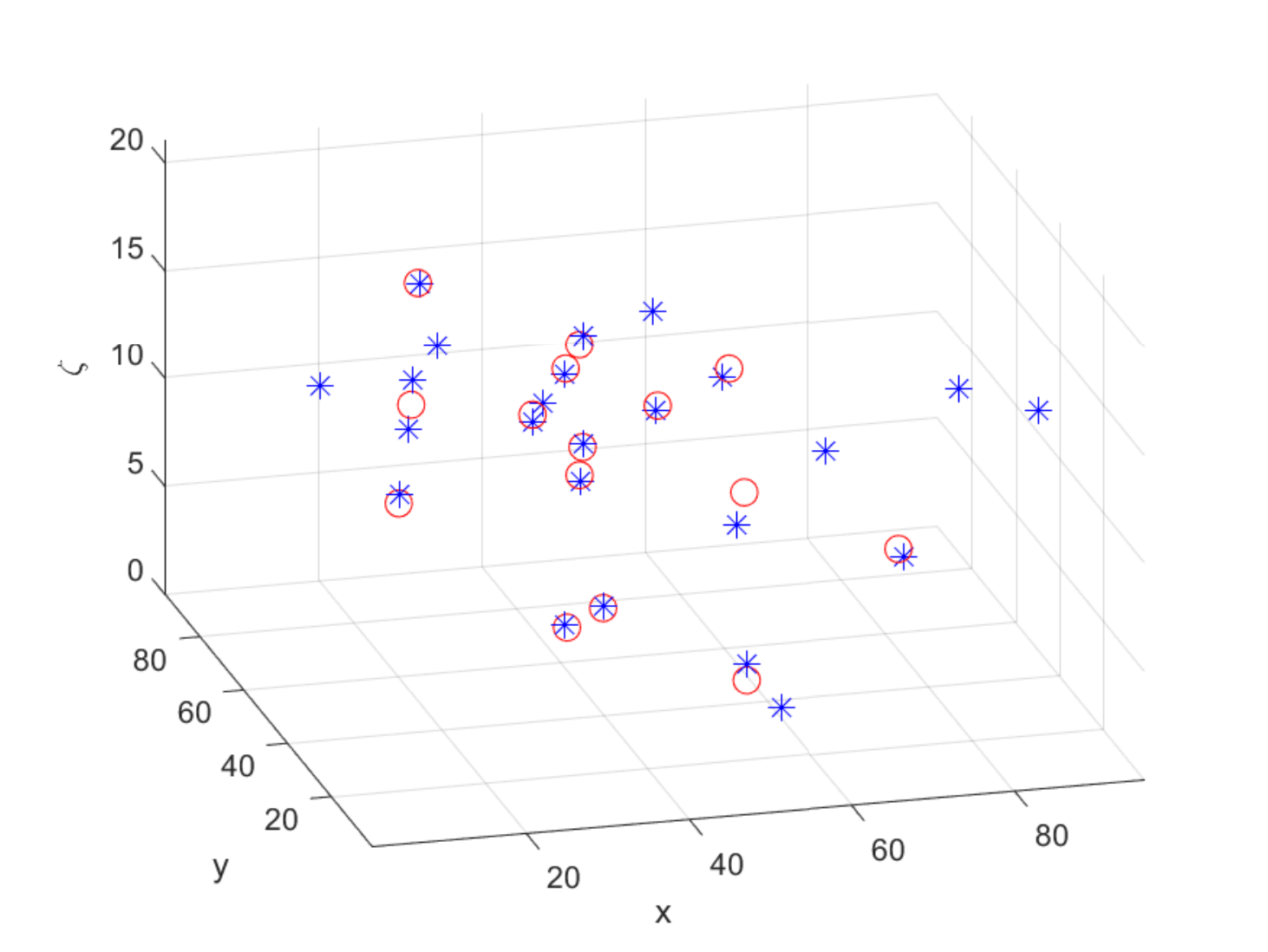} 
	\includegraphics[width=0.23\textwidth]{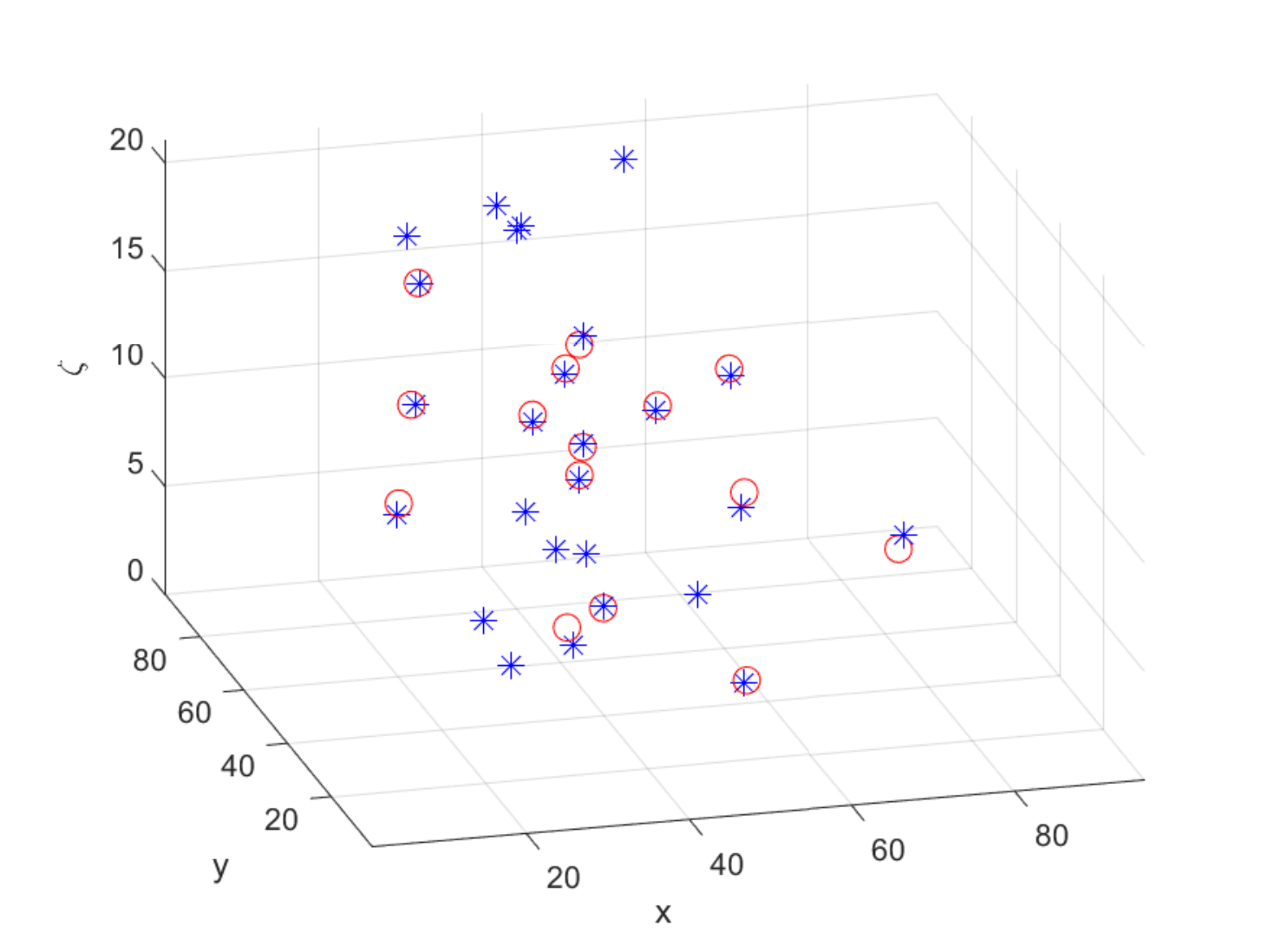}
	\caption{Localization results in Stage 1: ``o''  is ground truth and ``+'' is the estimated point source. Left side: single band. Right: multiple bands.  }
	\label{fig:class_loc_s1}
\end{figure}
\begin{figure}[htbp]
\centering
	\includegraphics[width=0.23\textwidth]{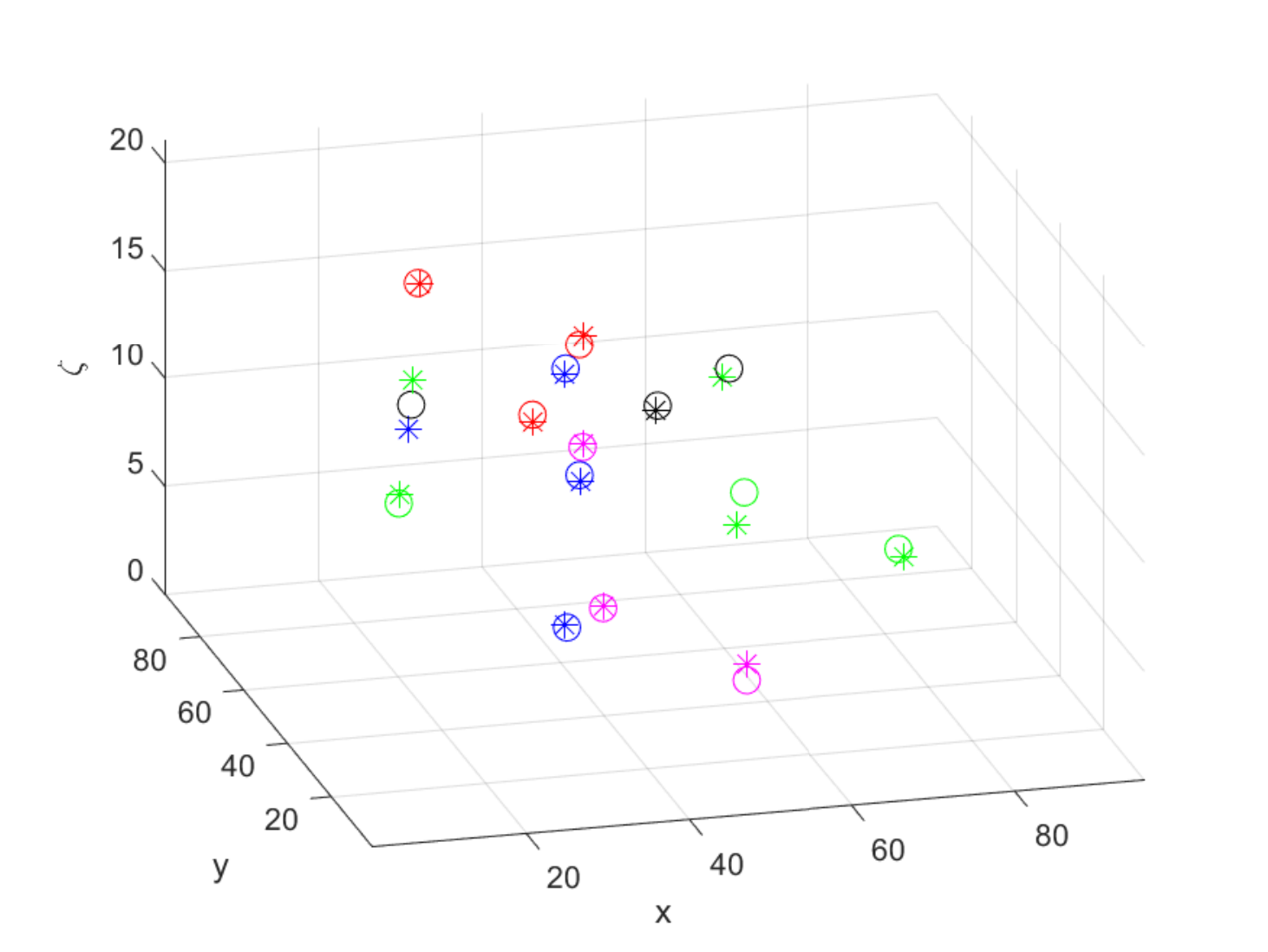} 
	\includegraphics[width=0.23\textwidth]{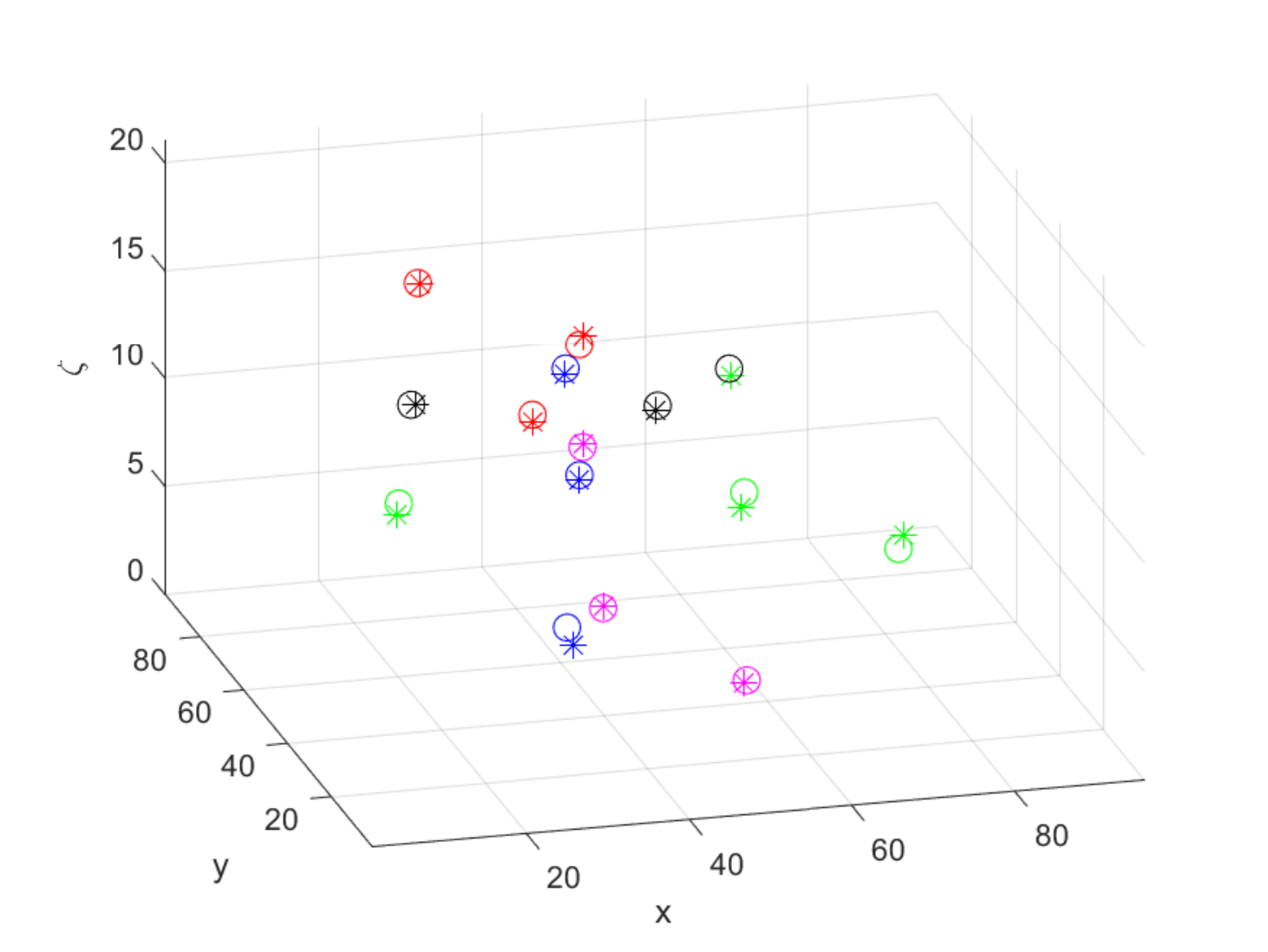}
	\caption{Localization results in Stage 2 and classification results in Stage 3: ``o''  is ground truth and ``+'' is the estimated point source. Different colors represents different materials.  Left side: single band. Right: multiple bands.  }
	\label{fig:class_loc}
\end{figure}

In \Cref{fig:class_loc_s1,fig:class_loc} we show an example of the 15 source case. Here we use information in 5 bands for Stage 2 for both cases and 4 bands in Stage 1 for multispectral  images.
\Cref{fig:class_loc_s1} shows the localization results for  single and multispectral bands. Not only do we recover more estimated point sources using  multispectral  images but we also obtain more identified true positives compared to the single band case. 
There are 14 identified true positives and 10 false positive point sources  in the single band image while multispectral images identifies  15 true positives with 11 false positives.
The recall rate is 100\% and precision rate is 57.69\% for the multispectral  images while it is 93.33\% and 58.33\% for single band images. \Cref{fig:class_loc} shows both classification and localization results for these two cases.
We also see that the single band image has more false positives than our multiple bands case even though we have run the same process Stage 2 to further remove false positives. The recall rate is 93.33\% and precision rate is 87.50\% for single band images, which means there are 16 estimated point sources including 14 identified true positives. In multispectral images,  both recall and precision rates  are 100\%.   As for classification, we misclassify  two point sources  for single band case, while only one point is misclassified in multispectral case. Here the overall average is 85.71\% and kappa is 82.28\% for single band images, while it is 93.33\% and 91.67\% for the multispectral  images. To further explore the performance of classification, we plot the estimated spectra for both  cases in \Cref{fig:est_5all_spectral_signatures}. We observe that their performances are similar except for the fourth material. The estimated spectra in the single band image is further away from the ground truth. Here we misclassify two point sources out of three,   while for our multispectral one we only misclassify one. Note that we plot all the estimated spectra corresponding only to identified true positives with respect to location. For example, in the 2nd material the single band image only has two estimated spectra.

\renewcommand{\figwidth}{.18\textwidth}
\begin{figure*}[htbp]
	\centering 
	\begin{tabular}{ccccc}
		\includegraphics[width=\figwidth]{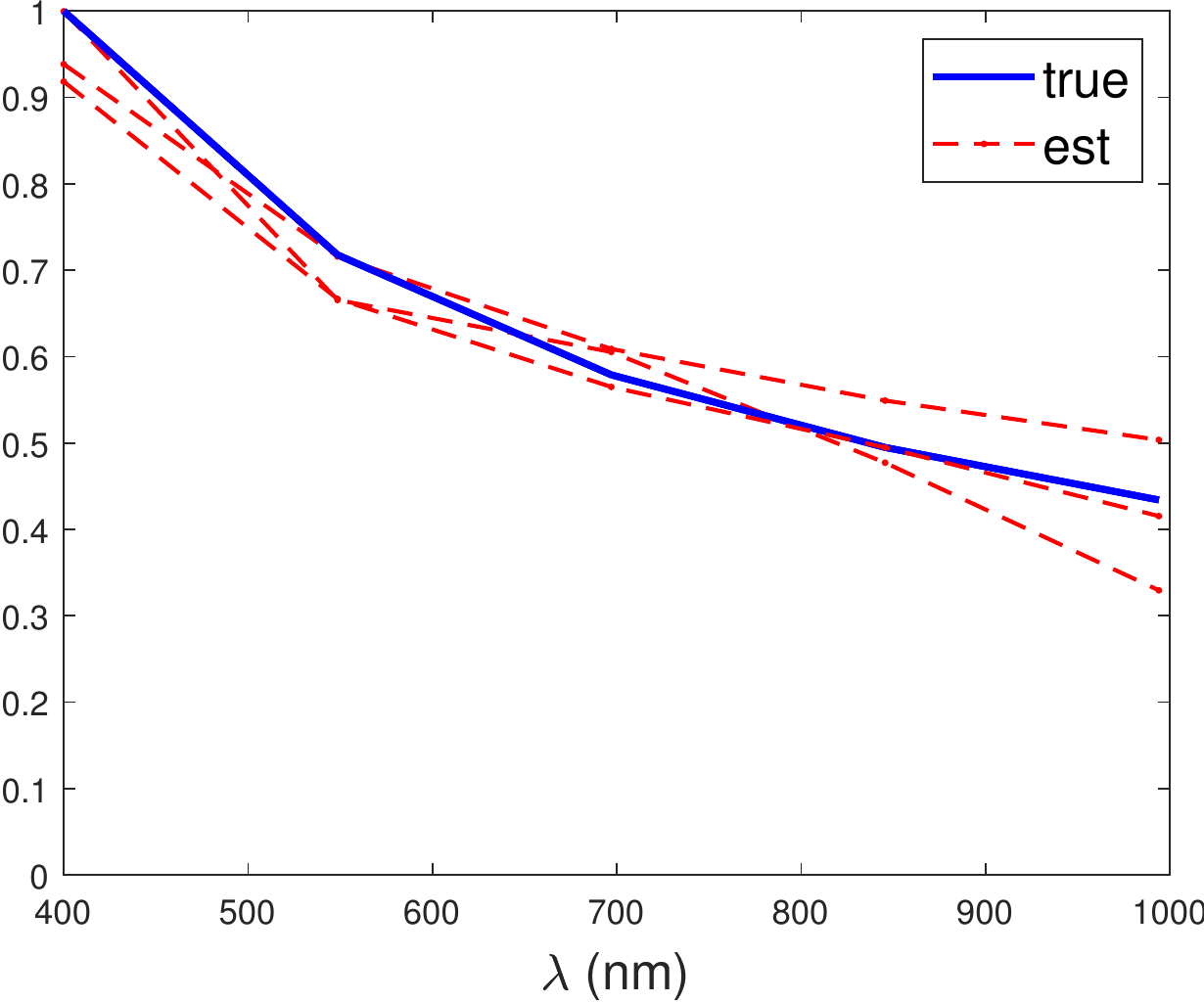}&
		\includegraphics[width=0\figwidth]{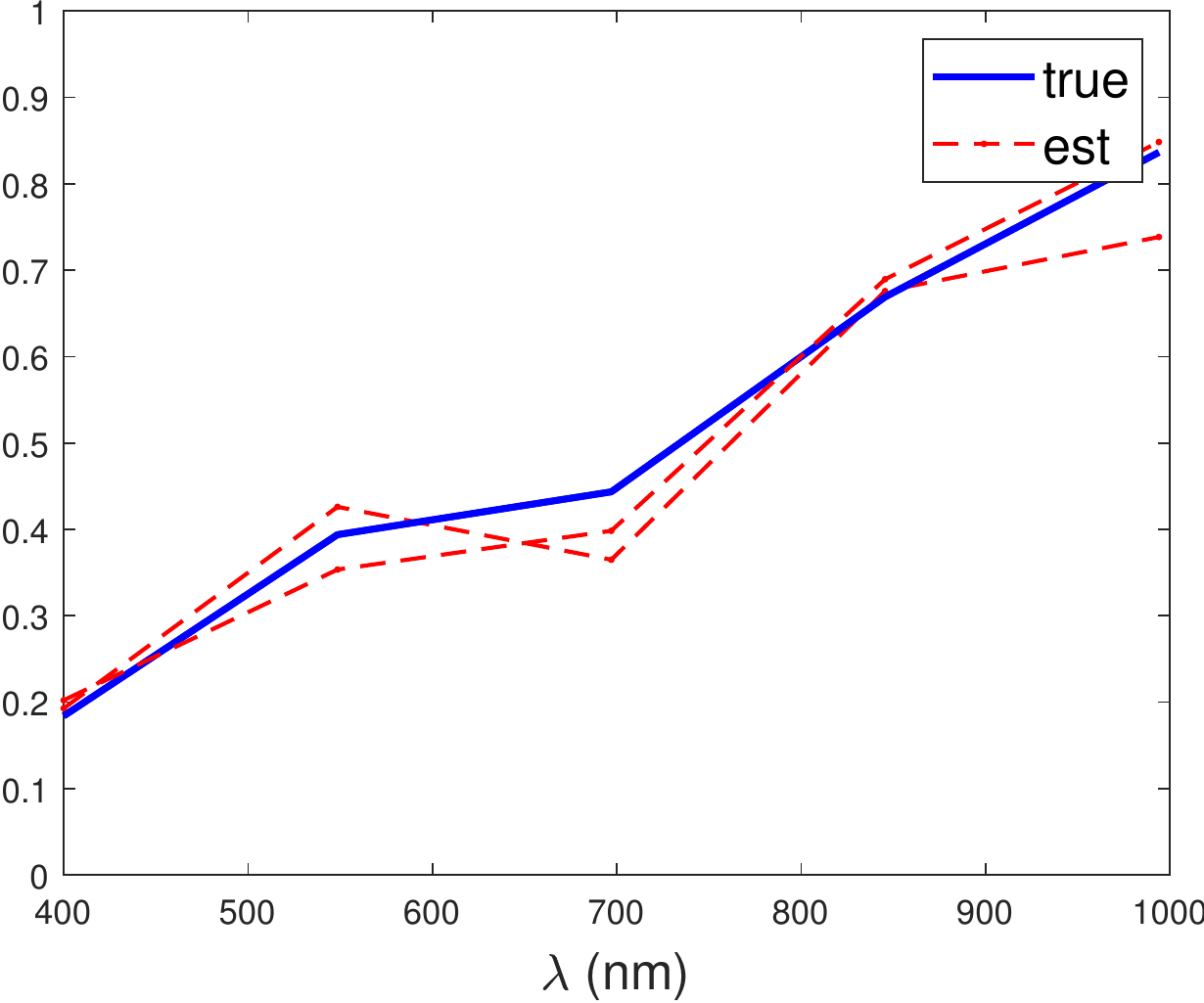}&
		\includegraphics[width=0\figwidth]{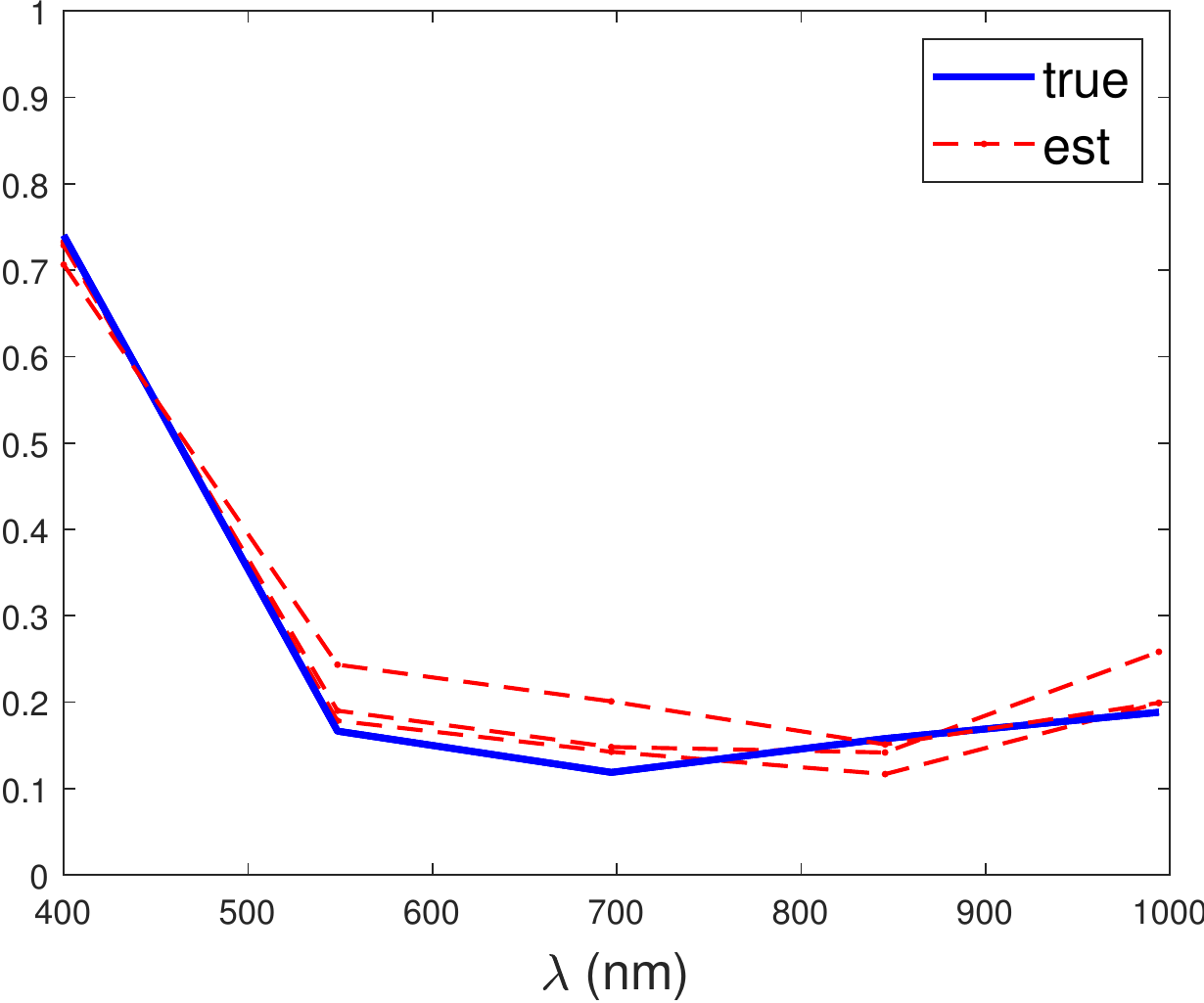} &
		\includegraphics[width=0\figwidth]{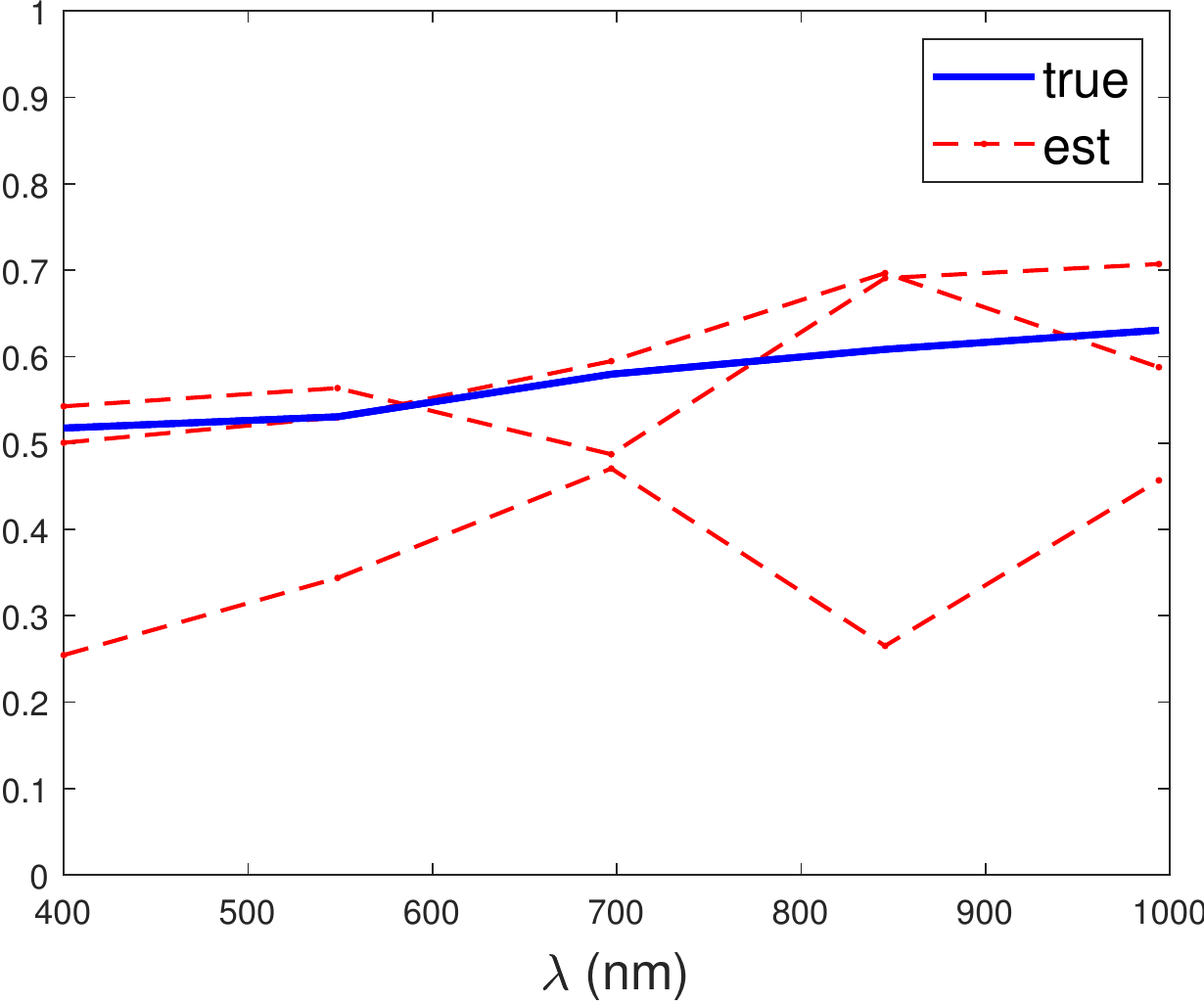} &
		\includegraphics[width=0\figwidth]{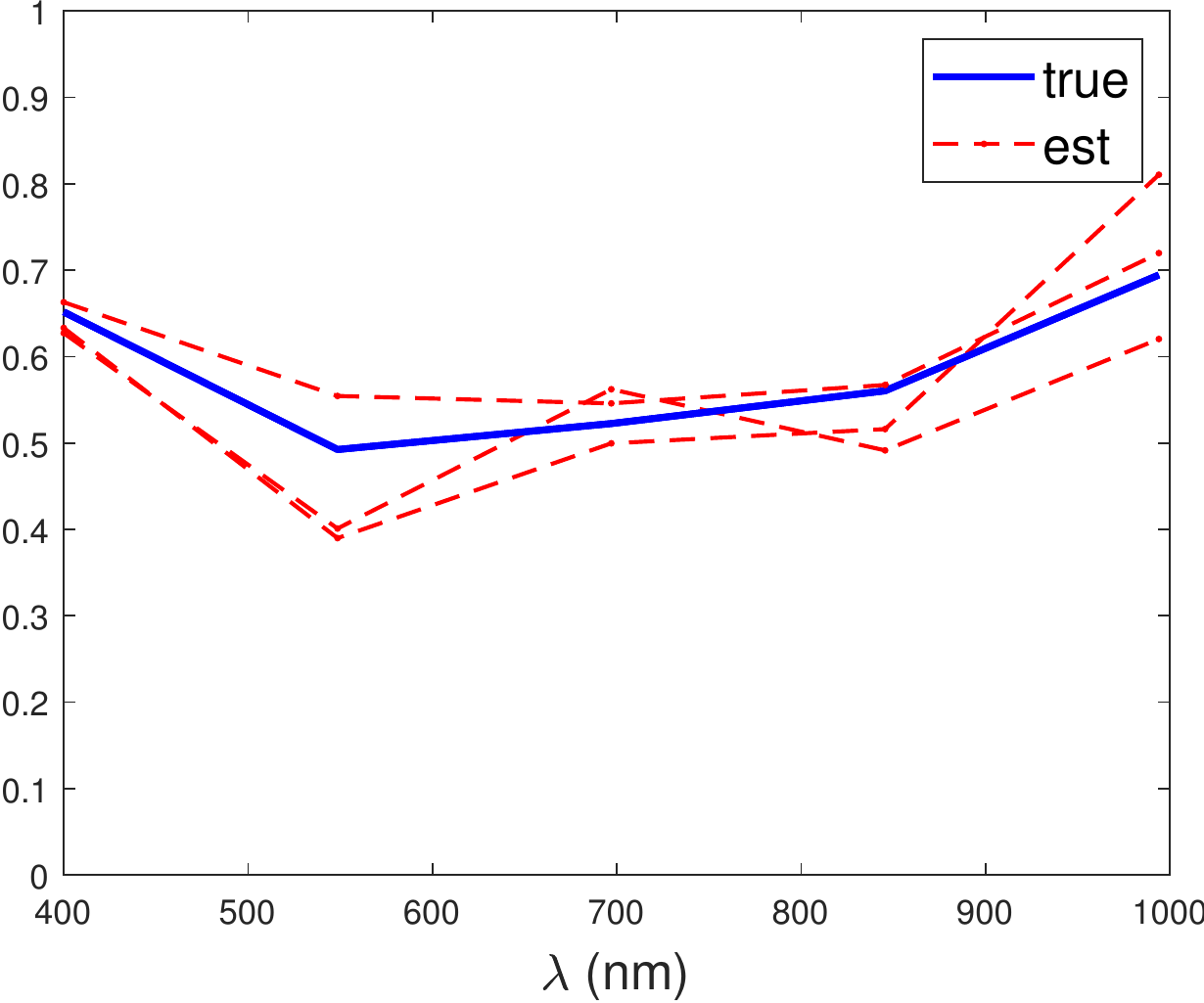}\\
		\includegraphics[width=0\figwidth]{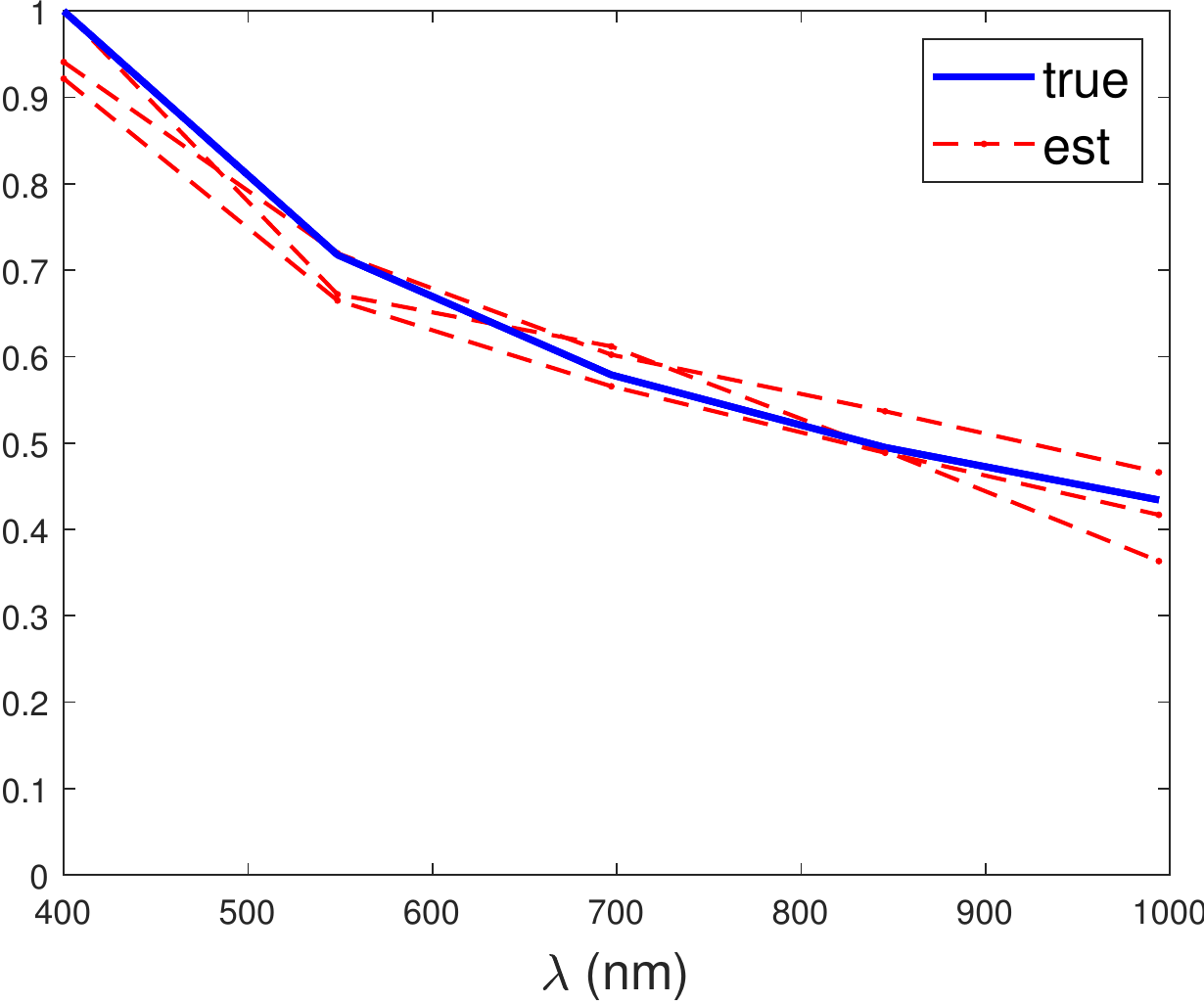}	&
		\includegraphics[width=0\figwidth]{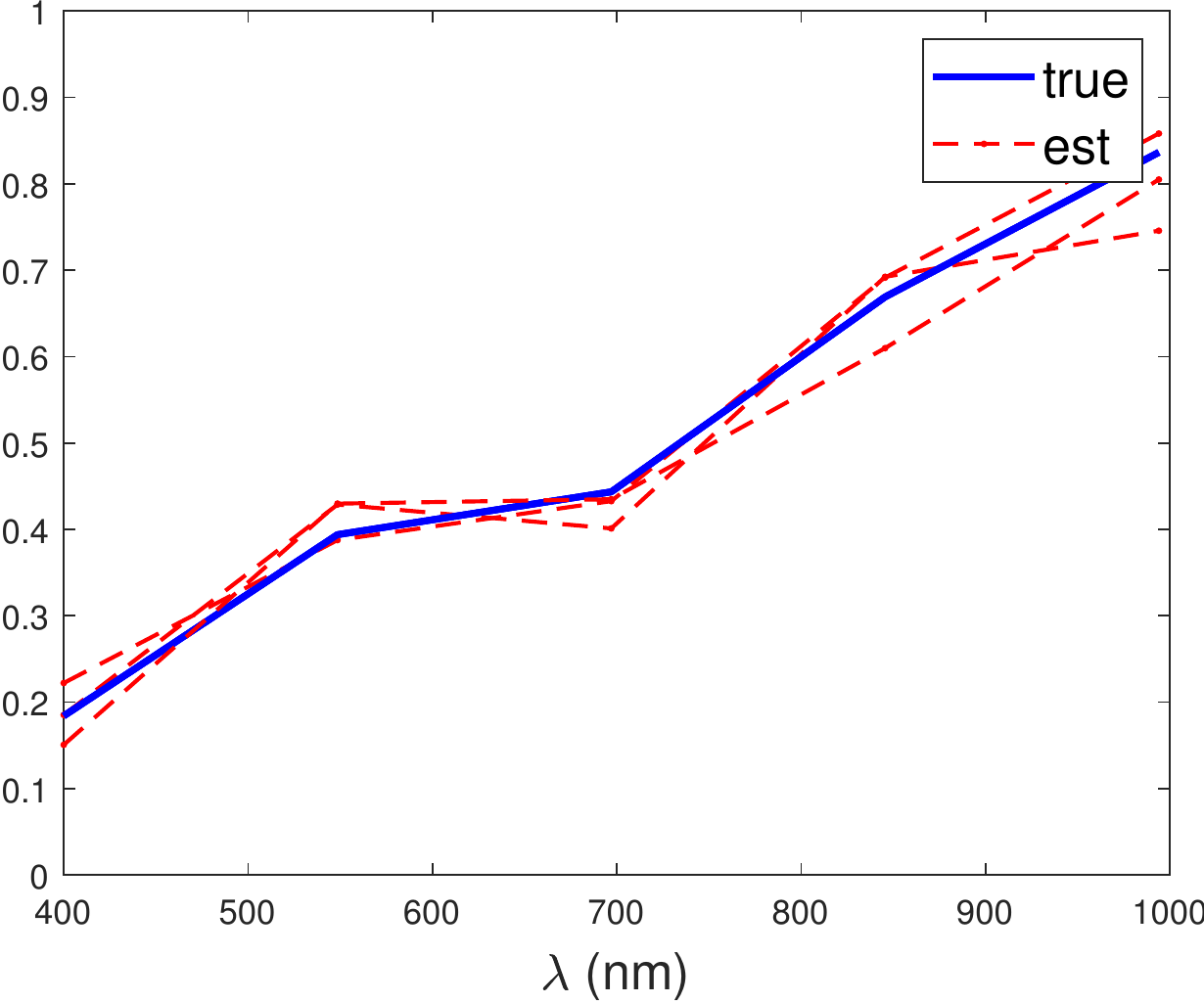}&		
		\includegraphics[width=0\figwidth]{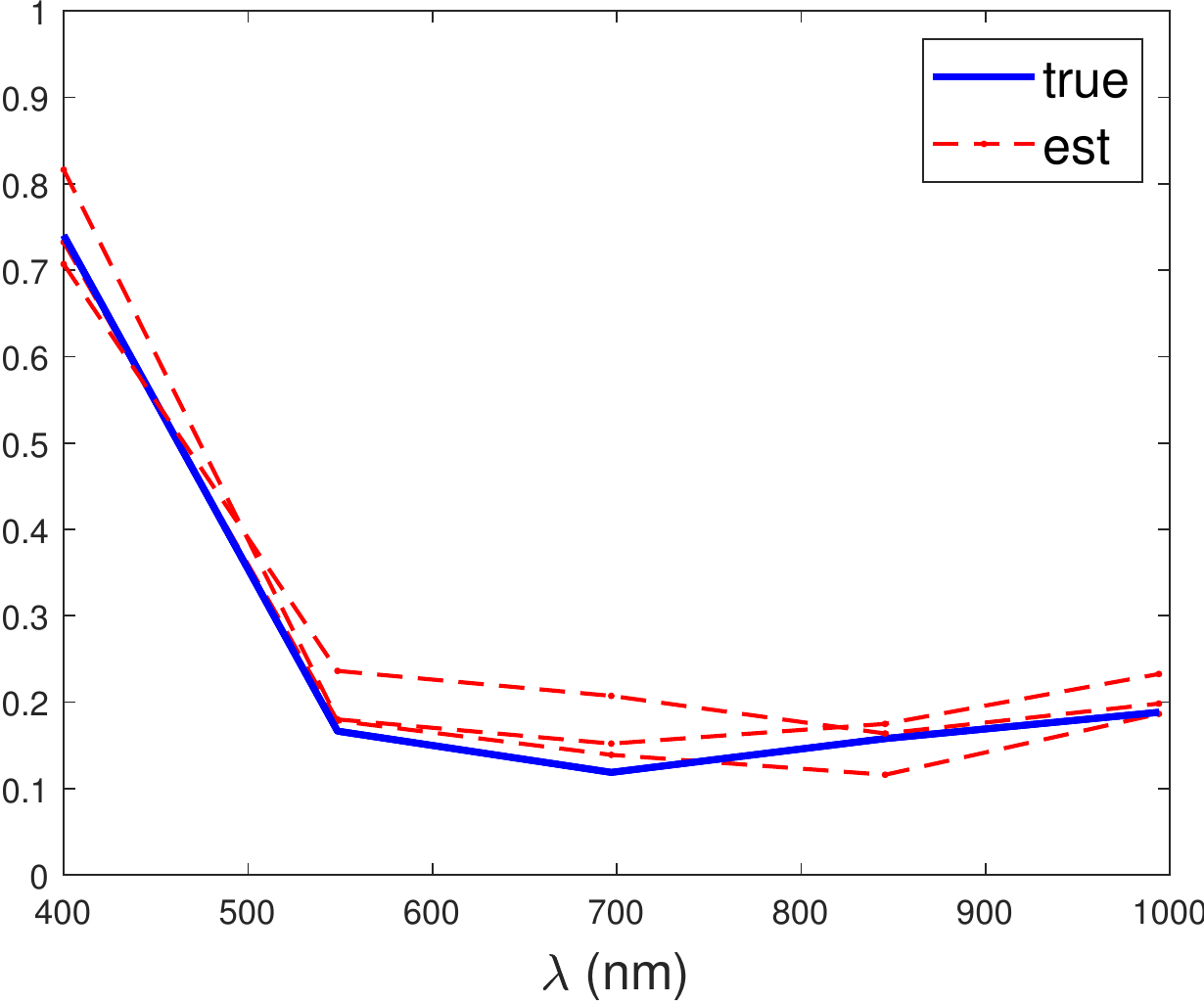}&	
		\includegraphics[width=0\figwidth]{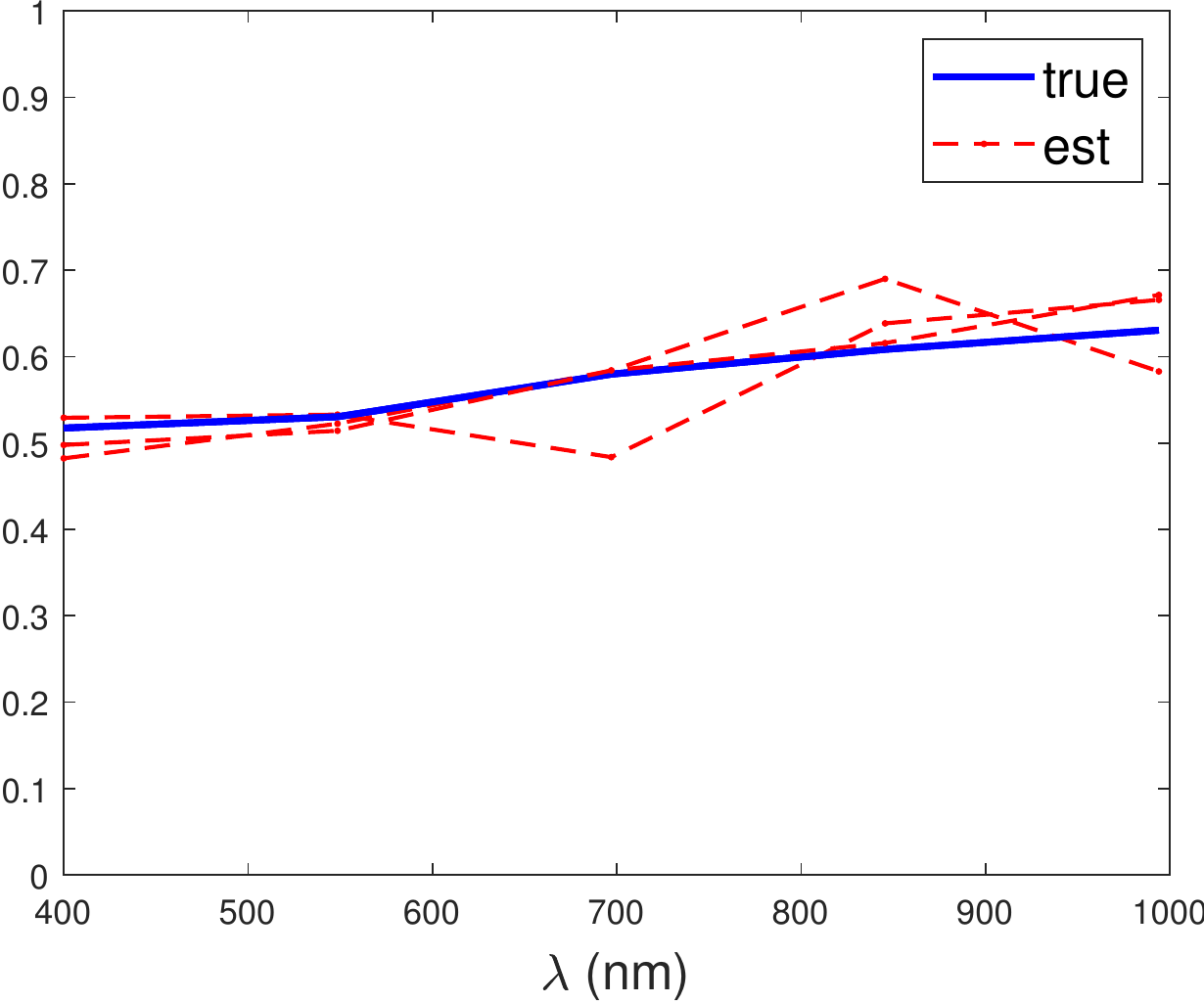}&	
		\includegraphics[width=0\figwidth]{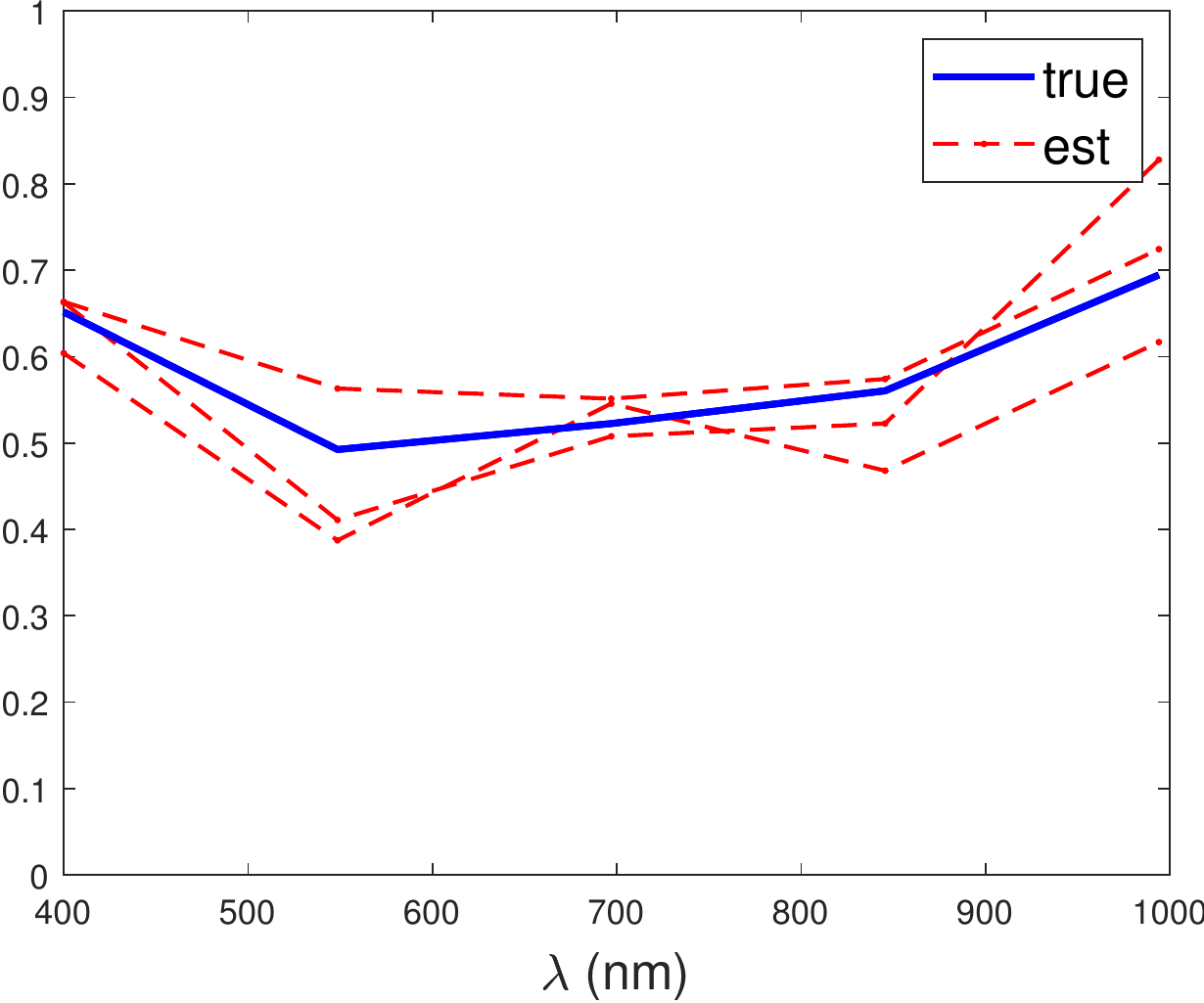}
	\end{tabular}
	\caption{Estimated spectra (\textcolor{red} {red}) and the ground truth spectra (\textcolor{blue} {blue}) for three materials with 5 bands in Stage 2. The first row is for single-band localization method and the second row is for the multispectral one. From left to right corresponds to M1 to M5.  }
	\label{fig:est_5all_spectral_signatures}
\end{figure*}

\subsubsection{Repeated Trials}

We also test for  localization and classification using 50 random tests for each case and compute the average value of the performance metrics; see \Cref{Tab:single_multi}. Here we use the number of bands in Stage 1 and Stage 2 to represent the different conditions we test. 
A missing number in Stage 2 means that we skip Stage 2 and show the localization result from Stage 1. 
In \Cref{Tab:single_multi}, the multispectral images perform much better  than the single band image in both Stages 1 and  2. We also plot the confusion matrix in 
\Cref{Tab:class_confu}. Neither case performs well in the last two classes. In the single band image, 24.67\% point sources in class 4 are wrongly estimated as in class 2. This is because the spectral signature is not estimated well. We know the false positives affect the estimation of flux very seriously and the classification results relies on the performance of localization.   

\begin{table}[htbp]
\begin{center}
\footnotesize
\caption{Comparisons of  3D localization and classification  by single-band and multispectral images. } 
\begin{tabular}{c|cc|cc|cc}
\hline
  Localization & \multicolumn{2}{c|}{No. bands} & \multicolumn{2}{c|}{Localization} & \multicolumn{2}{c}{Classification}  \\ \cline{2-7}
  method & Stage 1 & Stage 2 & Recall & Precision & OA & kappa  \\ \hline
 \multirow{2}{*}{Single}&  1 & - & 87.47\% & 39.22\% & - & - \\
 &1 & 5 & 86.00\% & 83.95\% & 88.09\% & 84.81\%  \\ \hline
  \multirow{2}{*}{Multiple}& 4 & - & 95.20\% & 63.78\% & - & - \\
 & 4 & 5 & 94.80\% & 95.34\% & 93.28\% & 91.57\% \\
         \hline
\end{tabular}\label{Tab:single_multi}
\medskip
\end{center}
\end{table}

\begin{table}[htbp]
\begin{center}
\footnotesize
\caption{Comparisons using class accuracy and a confusion matrix. } 
\begin{tabular}{c|ccccc}
\hline Localization method  &  \multicolumn{5}{c}{Confusion matrix} \\
\hline 
\multirow{5}{*}{Single} & 96.00 &   \, 0.00      &   \,  1.66 &       \,   1.66 &       \,   1.33 \\ 
     & \, 0.00    &   88.00 &       \, 0.00    &     \, 0.00    &      \,    0.67 \\
    &  \, 0.00    &     \, 0.00    &   95.67 &   \,  0.00      &      \,    4.00 \\
    &\,  2.00 &     24.67 &   \,  3.33 &    74.33 &   \,  6.33\\
   & \, 4.67 &   \,  8.00 &   \,  2.00 &  \,  7.00 &    82.33\\
\hline 
\multirow{5}{*}{Multiple} &
96.00 &    \,  0.00    &   \,  0.66 &        \,  3.00 &     \,     0.66 \\
    &  \, 0.00    &   98.67 &    \,   0.00   &      \,    1.00 &      \,    0.67 \\
   &   \, 0.00     &  \,   0.00      &   93.33 &   \,  0.00     &   \,       0.66 \\
   & \, 1.33 &    \, 7.00 &    \, 4.67 &    87.67&    \, 2.67 \\
   & \, 4.00 &    \, 4.67 &  \, 0.00    &   \, 6.33&    86.67\\
\hline
\end{tabular}\label{Tab:class_confu}
\medskip
\end{center}
\end{table}

\subsection{Effects of the Stopping Criterion}\label{sec:stopcri}

In this subsection, we  propose a stopping criterion based on our observations and empirically show its efficiency.
In the numerical test, we observe the relative change of $\mathcal{U}_1$ fails to be a stopping criterion. In the field of inverse problems, some works have focused on the behavior of data-fitting term to adjust the parameter in the optimization \cite{hansen2010discrete}. One of the more famous methods is called the discrepancy principle \cite{dp2012methods,youwei2012parameter,Idivergence2013minimization}  which tunes the regularization parameter until the data-fitting term is close to the noise level. It motivates us to consider the behavior of data-fitting term for our problem. 

\begin{figure}[htbp]
\centering
\begin{tabular}{cc}
\includegraphics[width=0.22\textwidth]{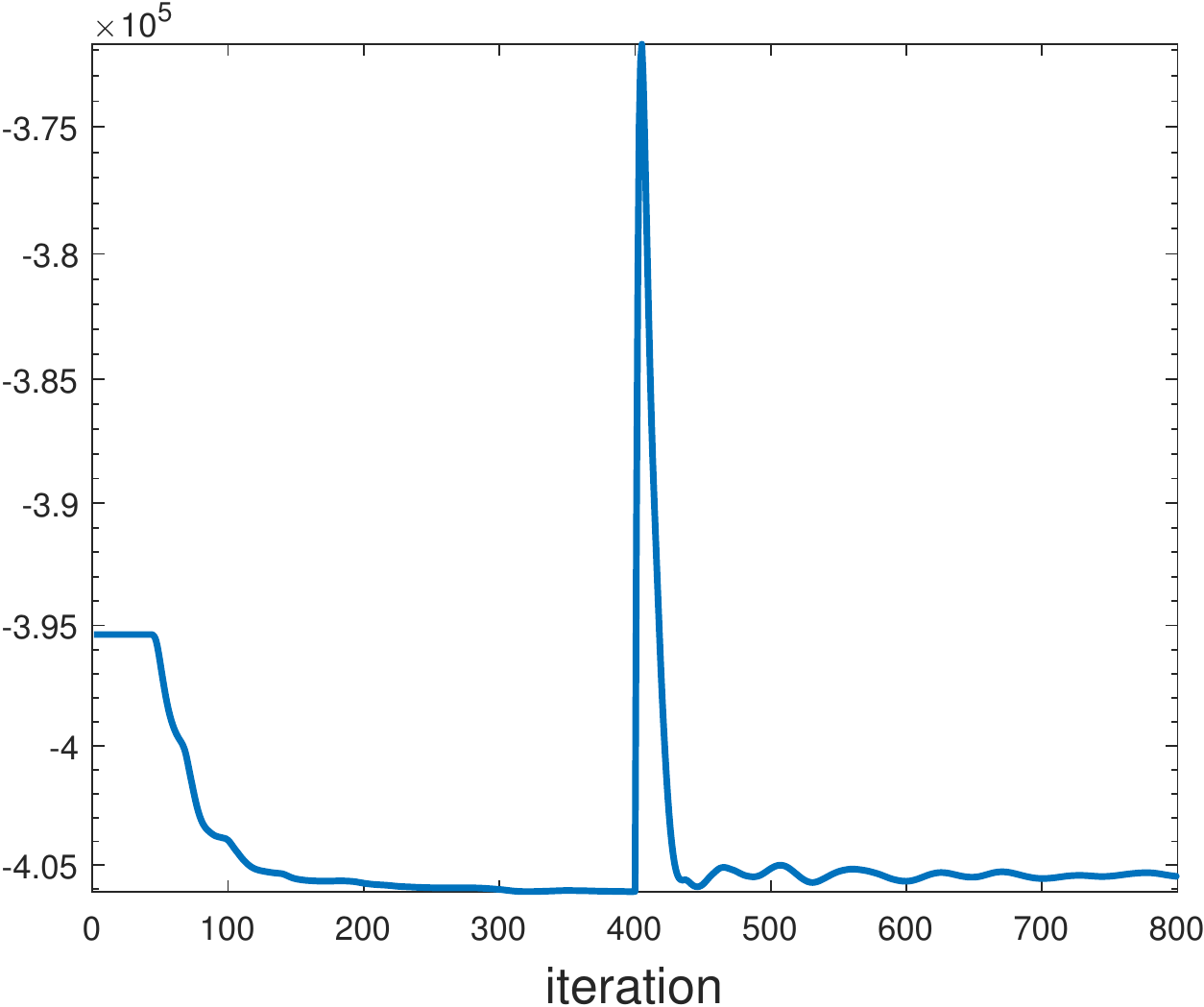} &
\includegraphics[width=0.22\textwidth]{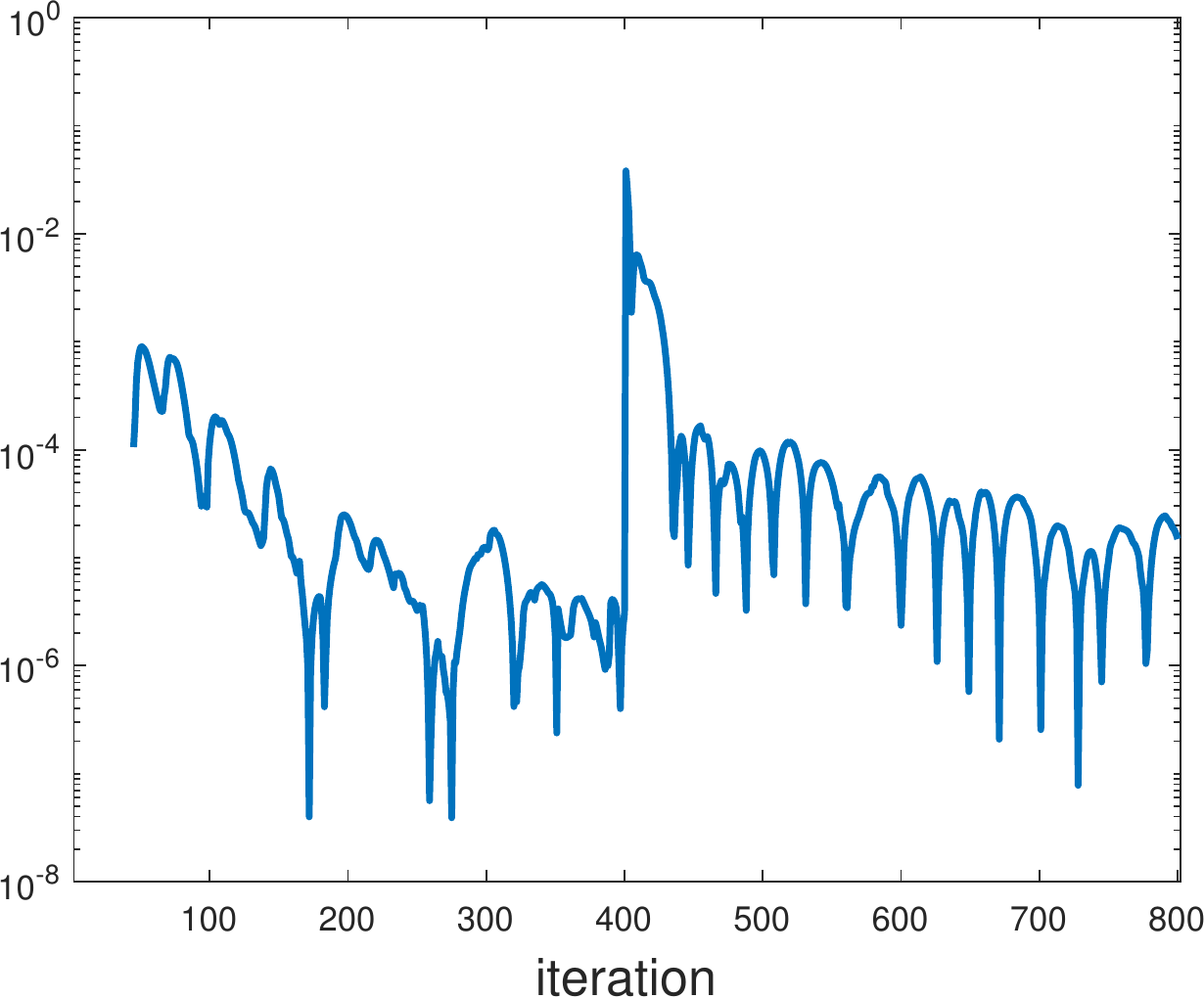}
\end{tabular}
\caption{ $D_m(\mathcal{U}_1^t) $ (left); $\frac{|D_{m}(\mathcal{U}_1^{t})-D_m(\mathcal{U}_1^{t-1})|}{|D_m(\mathcal{U}_1^{t})|}$ (right). }
\label{fig:data_fitting}
\end{figure}

\begin{figure}[htbp]
\centering
\begin{tabular}{cc}
\includegraphics[width=0.22\textwidth]{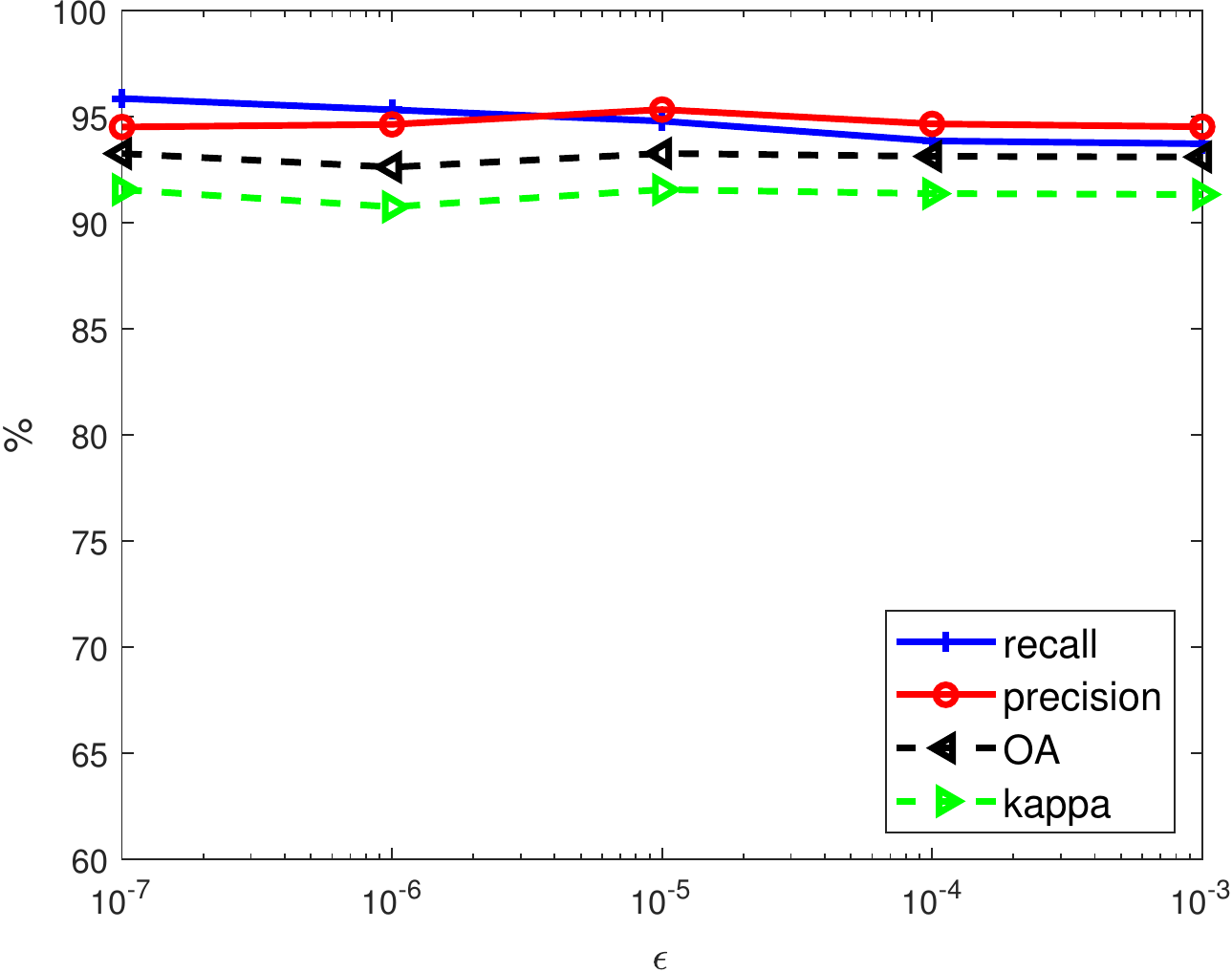} &
\includegraphics[width=0.22\textwidth]{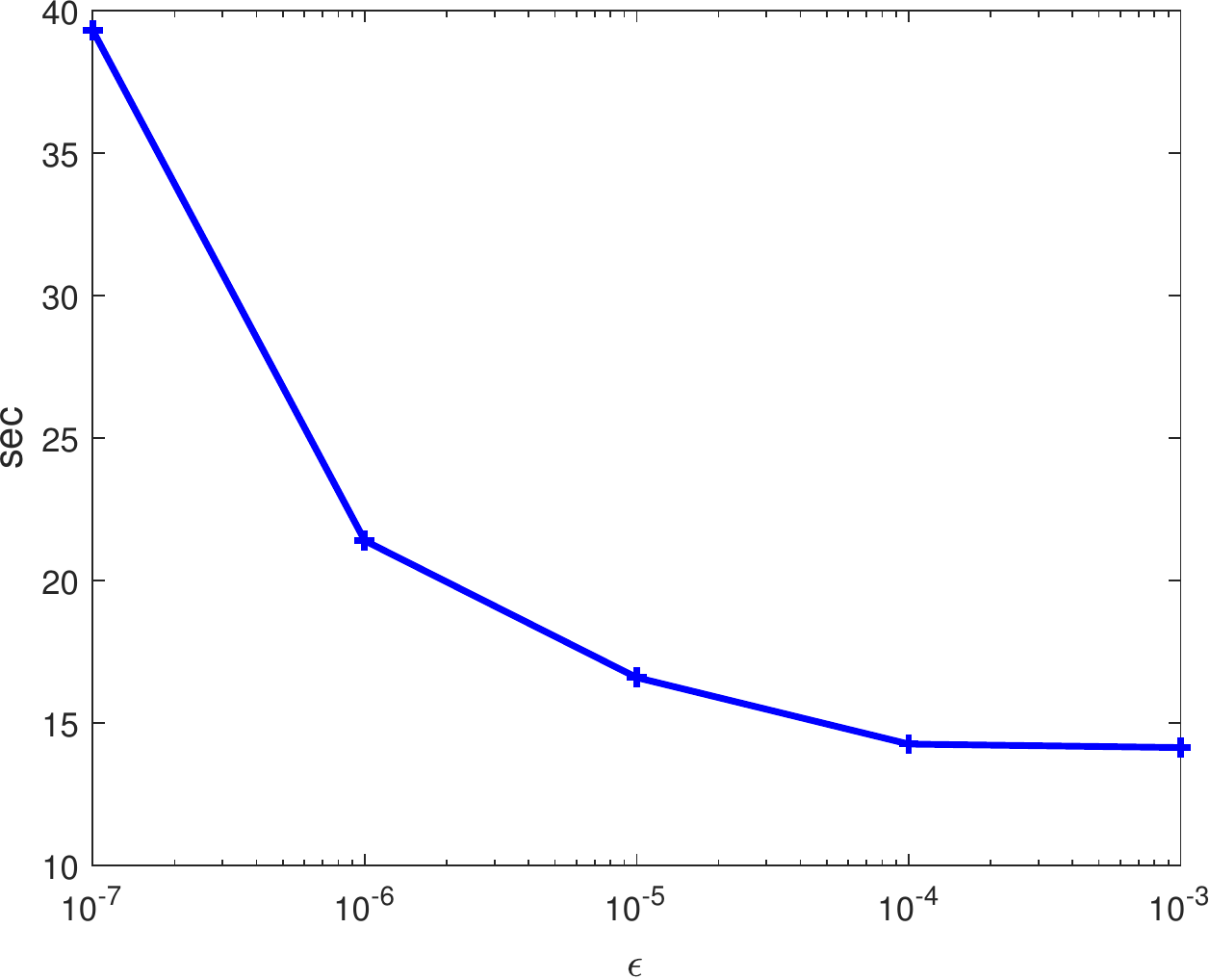}
\end{tabular}
\caption{Comparison of different stopping criterion thresholds: Localization and classification accuracy (left) and computation time (right)}
\label{fig:stop_criterion}
\end{figure}

\Cref{fig:data_fitting}(a) shows the value of the data-fitting term in each iteration. Here we set the maximum number of outer iterations as 2 and
maximum number of inner iterations as 400.
The curve  of the data fitting term is very flat and then significantly drops down for some iterations before becoming flat again. 
Note that the data-fitting term in the Poisson noise case does not reach the noise level nor do we add any prior information on the noise level. Therefore instead of the data-fitting term, 
we turn to its relative change between the previous iteration and the current one, i.e., $\frac{|D_m(\mathcal{U}_1^{t})-D_m(\mathcal{U}_1^{t-1})|}{|D_m(\mathcal{U}_1^{t})|}, $
where 
$$D_m(\mathcal{X}):=\sum\limits_{i=1}^K D_s(\mathcal X, \mathcal A^{(i)}, G^{(i)}). $$
The corresponding plot is shown  in \Cref{fig:data_fitting}(b). Even though the curve fluctuates a lot, it drops down to a sufficiently low level after some iterations. 
Based on this observation, we set up the stopping criterion as  $$\frac{|D_m(\mathcal{U}_1^{t})-D_m(\mathcal{U}_1^{t-1})|}{|D_m(\mathcal{U}_1^{t})|}< \epsilon. $$ We generate 50 random tests for the case of 15 point sources and run our algorithm with different values of $\epsilon$. In \Cref{fig:stop_criterion}, we see that the recall, precision and OA are similar with different values of $\epsilon$, but the computational time is significantly reduced by setting up $\epsilon$ to be not too small. In the following tests, we set $\epsilon=10^{-5}$.

\subsection{Effects of Thresholds in Stage 2}\label{sec:thr_s2}

In this subsection, we discuss the behaviors of algorithm when changing the threshold in Stage 2, i.e., using different values of $\gamma$.  
In the space object spectral data \cite{NASApaper} that we use for testing purposes in 5 bands (see the blue curves in \Cref{fig:est_5all_spectral_signatures}),  the minimal summation of spectra over the maximal summation is 0.425, i.e., $$\min_{k \neq l}  \frac{\h 1^T \h f^\ast_k}{\h 1^T \h f^\ast_l }=0.425. $$ According to \eqref{equ:gamma}, we need to choose $\gamma$ to be smaller than 0.425. This is to  avoid the possible over-estimation in spectra.
Here we choose different values of $\gamma$ between 0.1 and 0.5 and evaluate the accuracy in both localization and classification. In \Cref{Tab:thred}, we observe that the performance is similar when $\gamma <0.4 $ while the recall rate decrease greatly when $\gamma \geq 0.4$. These illustrate the robustness of algorithm when $\gamma$ satisfies \eqref{equ:gamma}. In the numerical tests, we set $\gamma = 0.2$ for both Poisson and Gaussian noise cases. 
\begin{table}[htbp]
\begin{center}
\footnotesize
\caption{Comparisons of 3D localization and classification accuracies with changing  threshold. } 
\begin{tabular}{c|cc|cc}
\hline \multirow{2}{*}{$\gamma$}  &  \multicolumn{2}{c|}{Localization} & \multicolumn{2}{c}{Classification}  \\ \cline{2-5}
& Recall & Precision & OA & kappa \\ \hline
  0.1   &  94.80\% &  90.45\%  &  93.28\%  & 91.57\%    \\ \hline
  0.2   & 94.80\%  &  95.34\%  &  93.28\%  & 91.57\%    \\ \hline
  0.3   & 91.33\%  &  96.61\%  &  91.55\%  & 89.17\%     \\ \hline
  0.4   & 82.00\%  &  97.06\%  &  89.93\%  & 86.17\%    \\ \hline
  0.5   & 72.53\%  &  97.82\%  &  85.08\%  & 79.48\% \\
\hline 
\end{tabular}\label{Tab:thred}
\medskip
\end{center}
\end{table}

\subsection{Effects of the Number of Bands Used}

In the first two stages described earlier, we used only a few bands. We now consider a general multi-spectral problem and ask the question --- how many  bands do we need to use in stages 1 and 2 to produce acceptable localization and classification results? 

\begin{figure}[htbp]
\centering
\begin{tabular}{cc}
\includegraphics[width=0.22\textwidth]{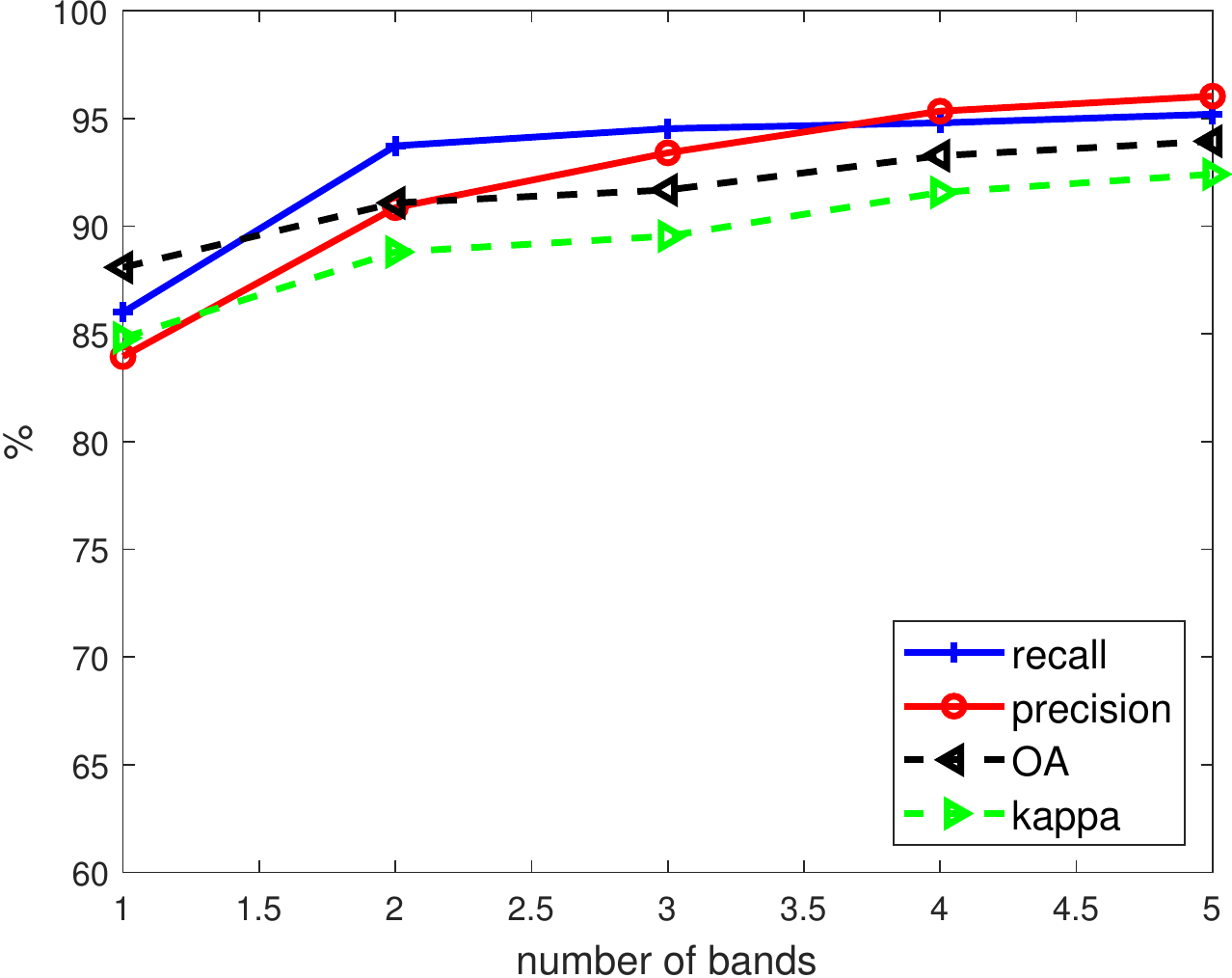} & 
\includegraphics[width=0.22\textwidth]{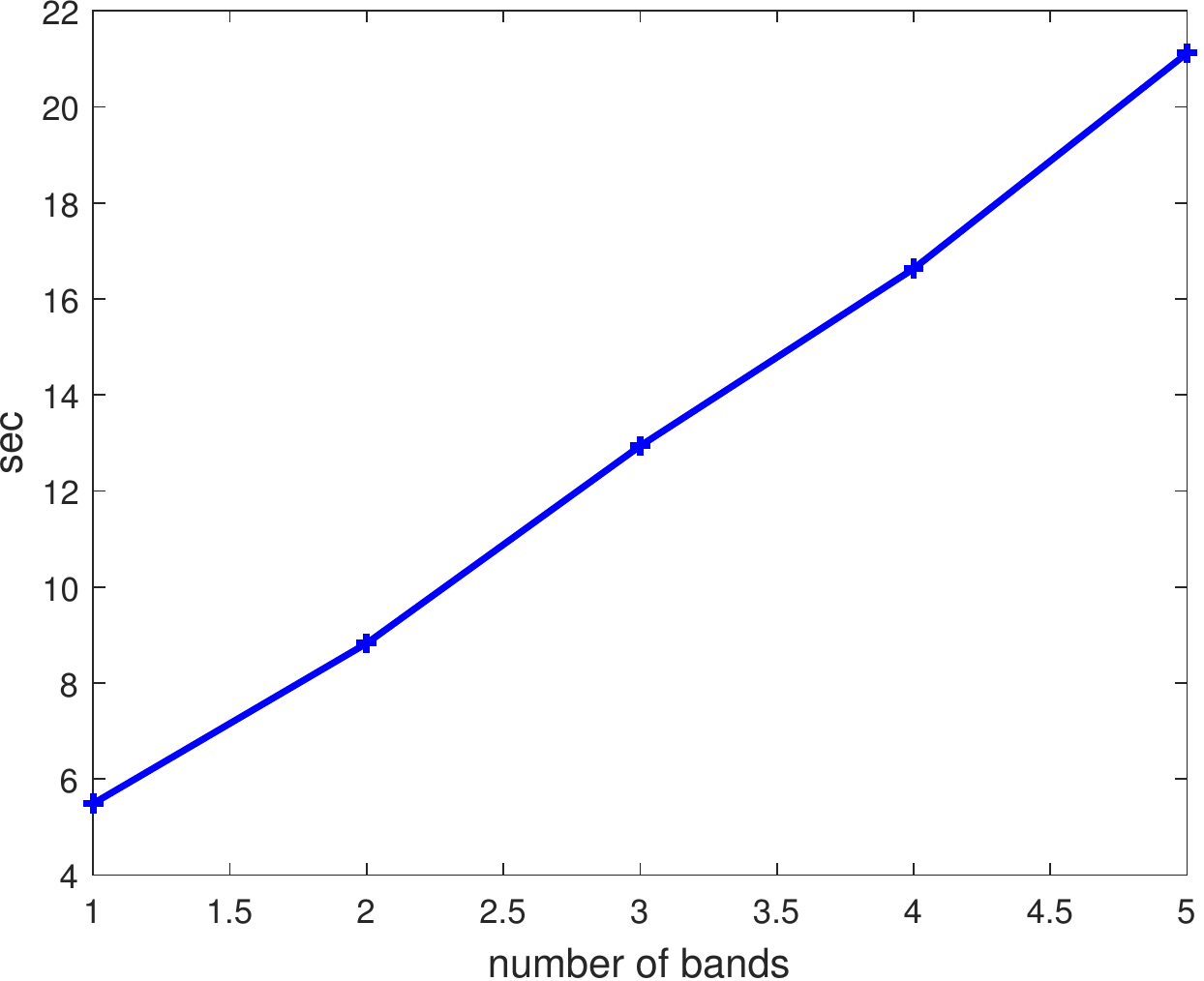}
\end{tabular}
\caption{Effects of the number of bands used in Stage 1: the localization and classification accuracy (left) and computational times (right)}
\label{fig:Effects_wavelengths_S1}
\end{figure}

We still consider 5 different materials and 15 point sources. Each material is assigned to 3 point sources, as before.
 In the first test, we fix the number of bands in Stage 2 as 5 and try various numbers of bands in Stage 1. The 5 bands are consistent with those in the previous subsection. 
 When the number of bands in Stage 1 is 1, we use Band 1 for Stage 1, similarly,  we use both Band 1 and Band 2 when the number of bands in Stage 1 is 2. 
 Here, we randomly generate 50 tests and compute the average of our performance metrics ---  recall rate, precision rate, overall accuracy (oa), and kappa as well as the computational times in Stage 1, which are shown in \Cref{fig:Effects_wavelengths_S1}. We observe that both localization and classification results get better as the number of bands increases. The improvement is especially significant when increasing the number of bands from 1 to 2. The curves gradually become flat, especially after 4 bands.  The computational time in Stage 1 increases almost linearly. All these tests use our proposed stopping criterion. The computational time is around 17 seconds, for even a single band image. 


\begin{figure}[htbp]
\centering
\begin{tabular}{cc}
\includegraphics[width=0.22\textwidth]{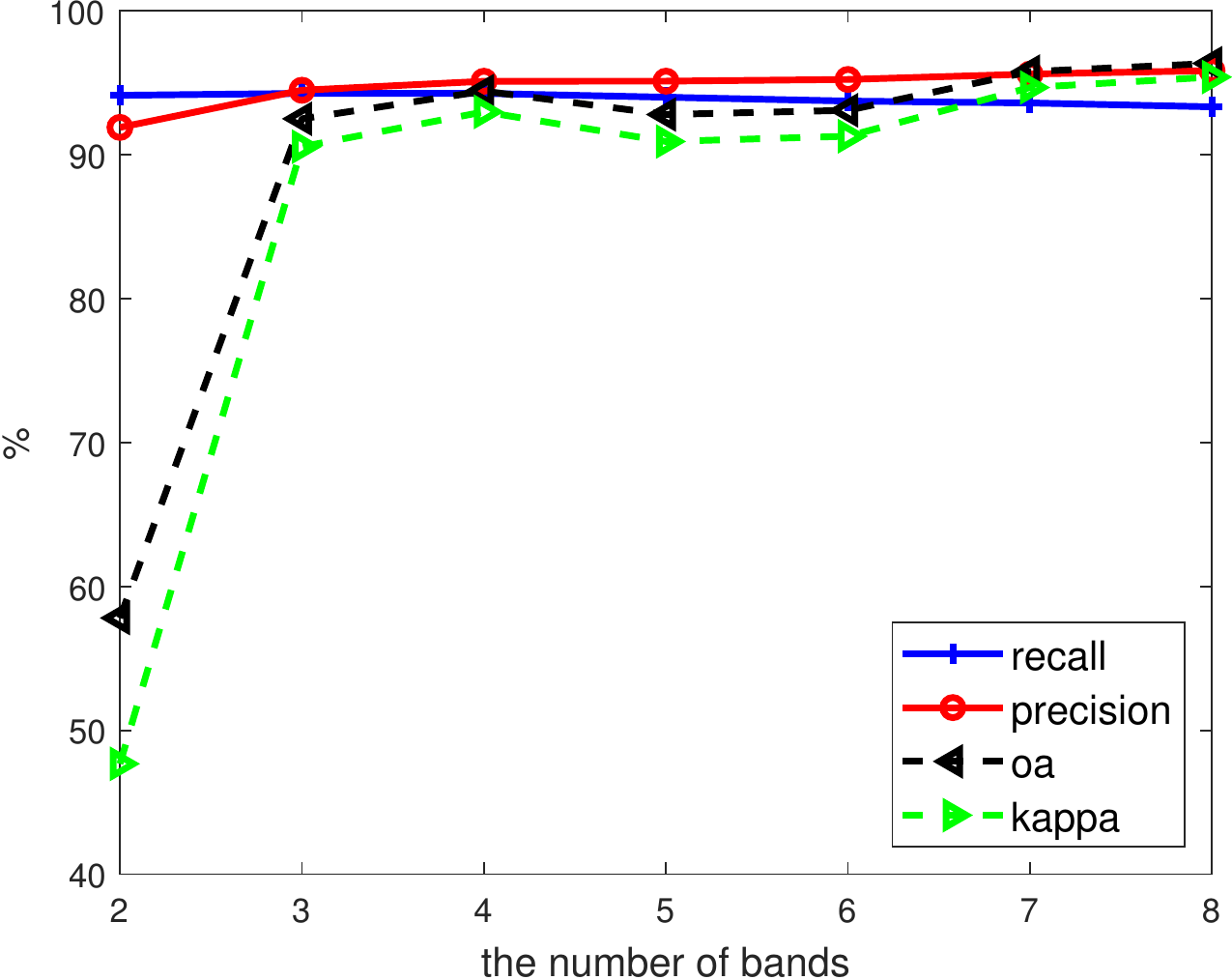} & 
\includegraphics[width=0.22\textwidth]{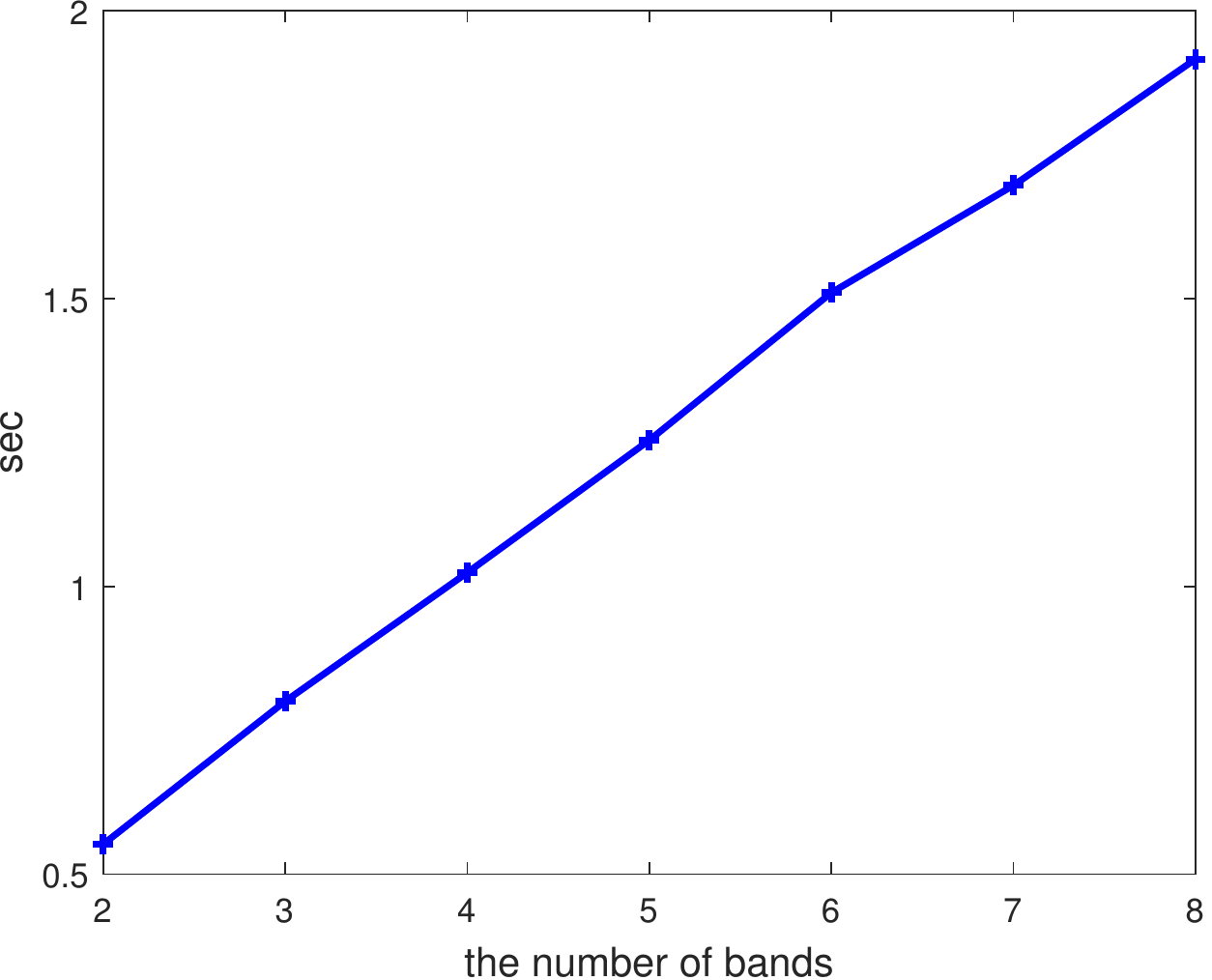}
\end{tabular}
\caption{Effects of the number of bands used in Stage 2: the localization and classification accuracy (left) and computational times (right)}
\label{fig:Effects_wavelengths_S2.}
\end{figure}

In the second test, we fix the number of bands in Stage 1 as 4, and  try different numbers of bands in Stage 2 from 2 to 8. The corresponding eight wavelengths are 400.00nm,  484.84nm,  569.69nm  654.54nm,  739.39nm, 824.24nm, 909.09nm, and 993.93nm, respectively. The fixed four bands in Stage 2 use the first four wavelength's information.  The average recall rate, precision rate, overall accuracy (oa), and kappa as well as computational time in Stage 2 using 50 random tests are shown in \Cref{fig:Effects_wavelengths_S2.}. We observe that using at least  4 bands leads to acceptable results. Using two bands in Stage 2 results in a value of kappa less than 50\%. In addition, with more spectral information, the precision rate increases as we remove more false positives, but the recall rate decreases. This means we have also removed some true positives. The computational time in Stage 2 increases linearly but it is much less than  in Stage 1, when we compare \Cref{fig:Effects_wavelengths_S1}(b) with \Cref{fig:Effects_wavelengths_S2.} (b).

\subsection{Effects of Regularization}

\begin{figure}[htbp]
\centering
	\includegraphics[width=0.23\textwidth]{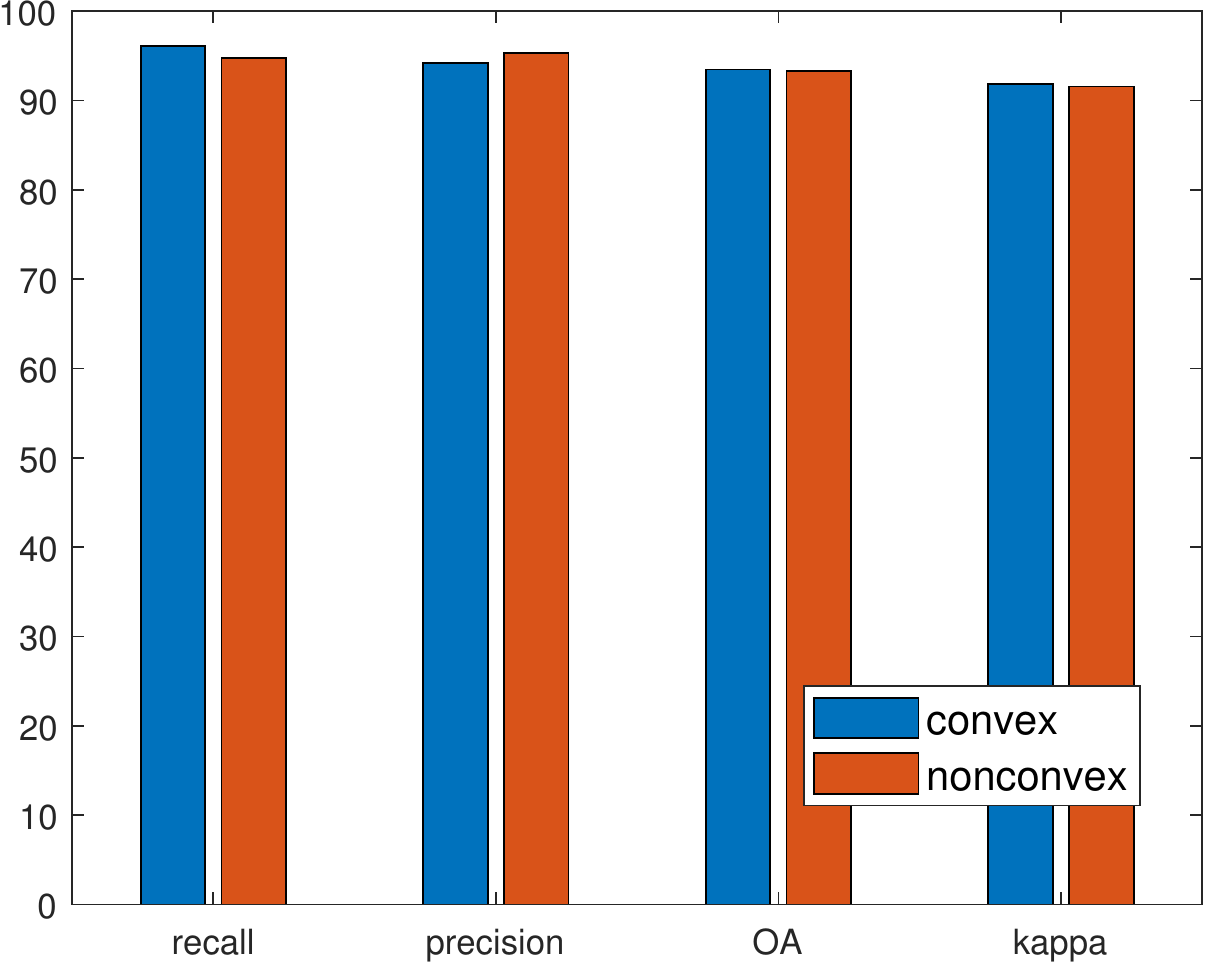} 
	\caption{Comparison of the results between convex regularization and nonconvex regularization.  }
	\label{fig:bar_l1}
\end{figure}

In this subsection, we discuss effect of regularization in the optimization model used for Stage 1. Since we propose Stage 2 to remove false positives, it is worth our attention to consider whether the convex regularization is good enough for classification. 
Here we test the $\ell_1$ regularization term and compare it with the results of our non-convex approach and similarly use, as before, a set of 50 random tests for each case. The minimization model for the $\ell_1$ regularization in the multispectral image is 
\begin{equation}
	 	\min\limits_{\mathcal{X}\geq 0 }\left\{  \sum\limits_{i=1}^K D_s(\mathcal X, \mathcal A^{(i)}, G^{(i)}) + \mu\sum_{p,q,r = 1}^{m,n,d} |\mathcal{X}_{pqr}|\right\}. \label{equ:min_fun_muti_l1}
	 \end{equation}
The problem is the same as  the $\mathcal{X}$-subproblem in \eqref{equ:outer_mult}, except that weights $w^k_{pqr}$ have all been set equal to a single parameter, $\mu$. Therefore, we use the same algorithm to solve \eqref{equ:min_fun_muti_l1}. 
In \Cref{fig:bar_l1},  localization and classification results are shown. Both  models have very similar performance. However, the convex model only takes 10.77 seconds while the nonconvex one needs 16.72 seconds, on average. The performance also illustrates the efficiency of the  three-stage method. The second stage improves the localization and loosens the requirement of sparsity in the first stage.

\subsection{Gaussian Noise Case}\label{subsec:gaussian}
In the last subsection, we summarize our results for the Gaussian noise case.  \Cref{fig:org_obs_g} shows a sequence of 2D original images as well as the observed images in 5 bands corresponding to the wavelengths, 1530.0nm, 1736.4nm, 1969.7nm, 2203.0nm, and 2436.4nm. 
Since the wavelengths corresponding to the selected bands are different from the ones in the Poisson noise case, the effect of diffraction is larger for the Gaussian noise case. To keep the image blurring comparable across the two different wavelength ranges corresponding to the two different noise models, we assume the aperture diameter to be twice as large in the Gaussian noise case. Correspondingly, the side lengths are only half as great as the ones in the Poisson noise case.
In \Cref{fig:org_obs_g}, the diffraction effect is serious even in band 1. Comparing the results to those in \Cref{fig:org_obs} for the Poisson noise case, we see that the background noise is much stronger since Gaussian noise is additive and independent of the signal.  

\renewcommand{\figwidth}{.18\textwidth}
\begin{figure*}
	\centering 
	\begin{tabular}{ccccc}
		Band 1 & Band 2 & Band 3 & Band 4 & Band 5\\
		\includegraphics[width=\figwidth]{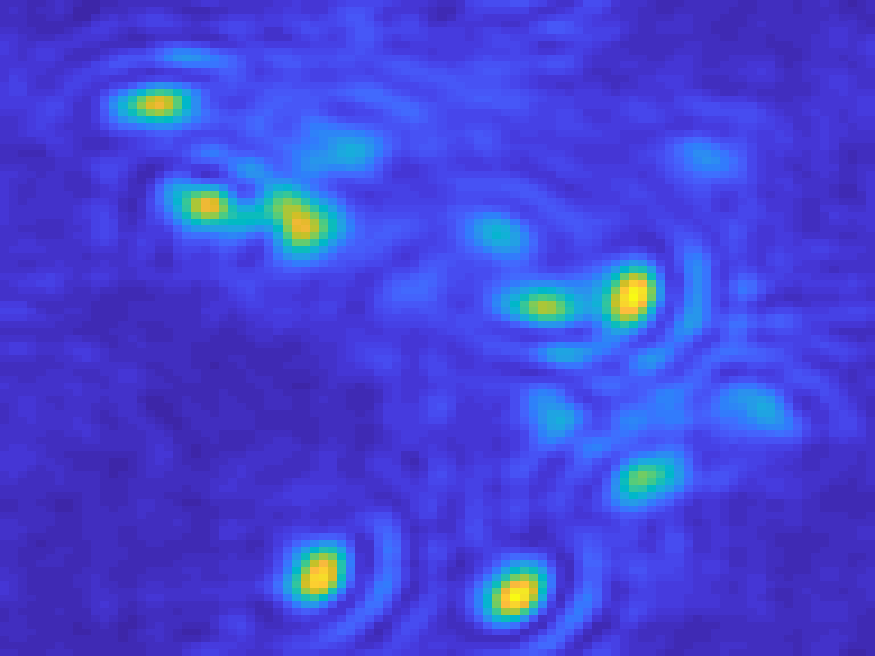}&
		\includegraphics[width=\figwidth]{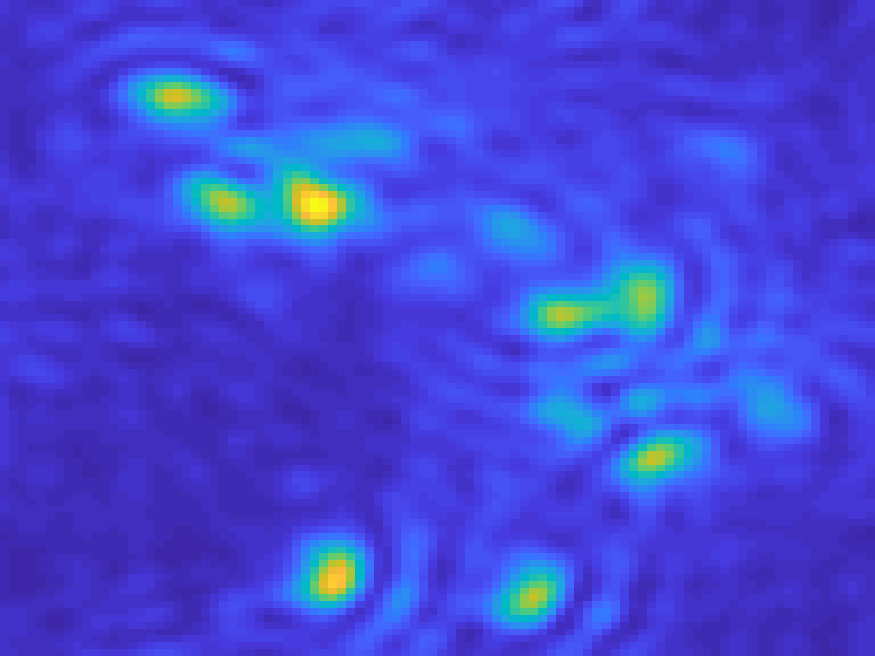}&
		\includegraphics[width=\figwidth]{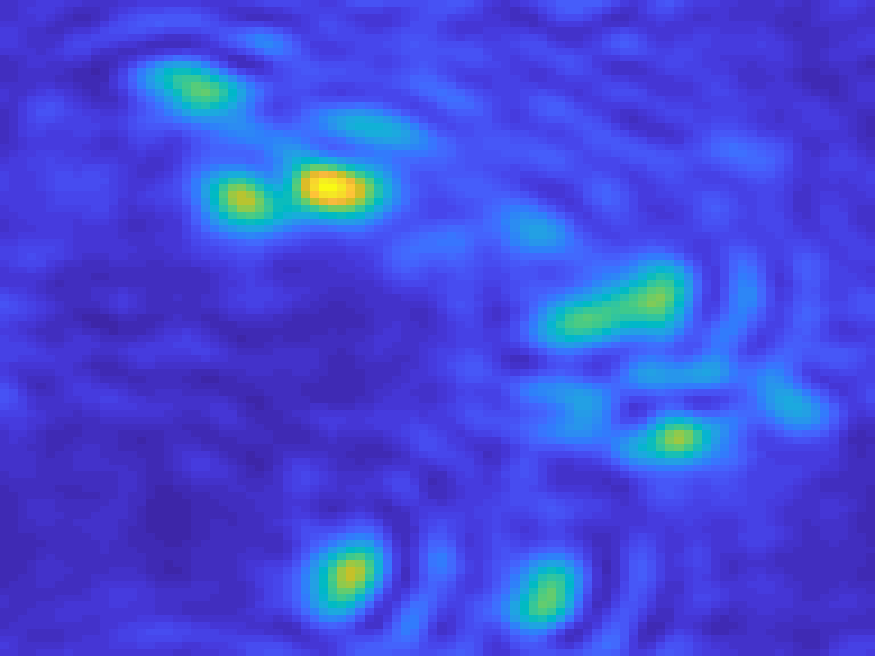} &
		\includegraphics[width=\figwidth]{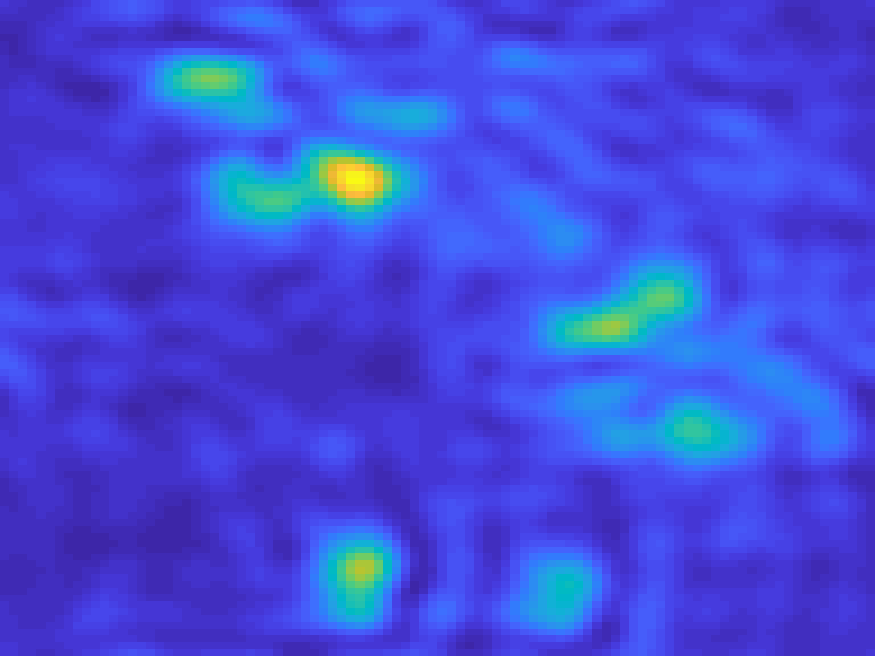} &
		\includegraphics[width=\figwidth]{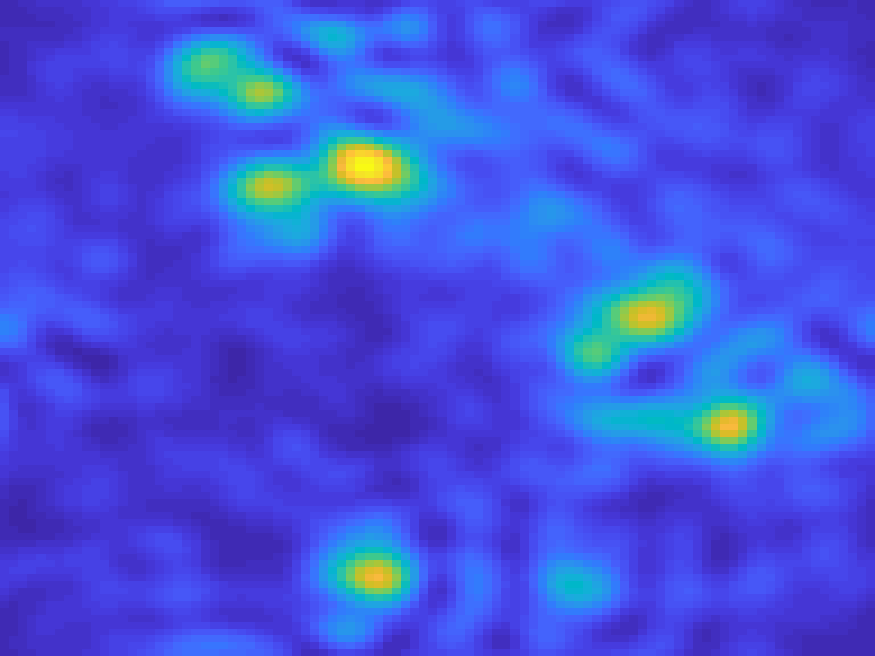}\\
		\includegraphics[width=\figwidth]{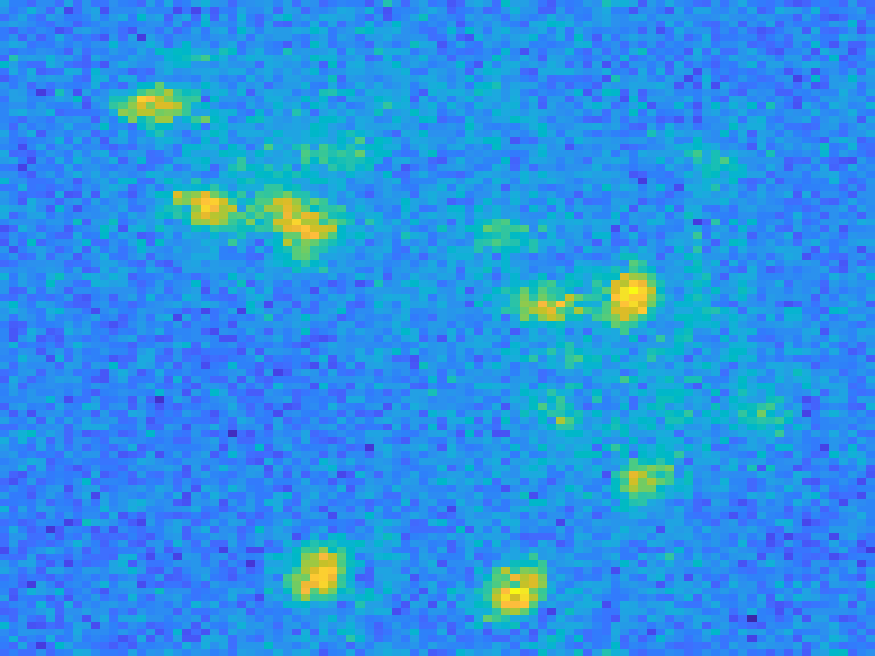}&
		\includegraphics[width=\figwidth]{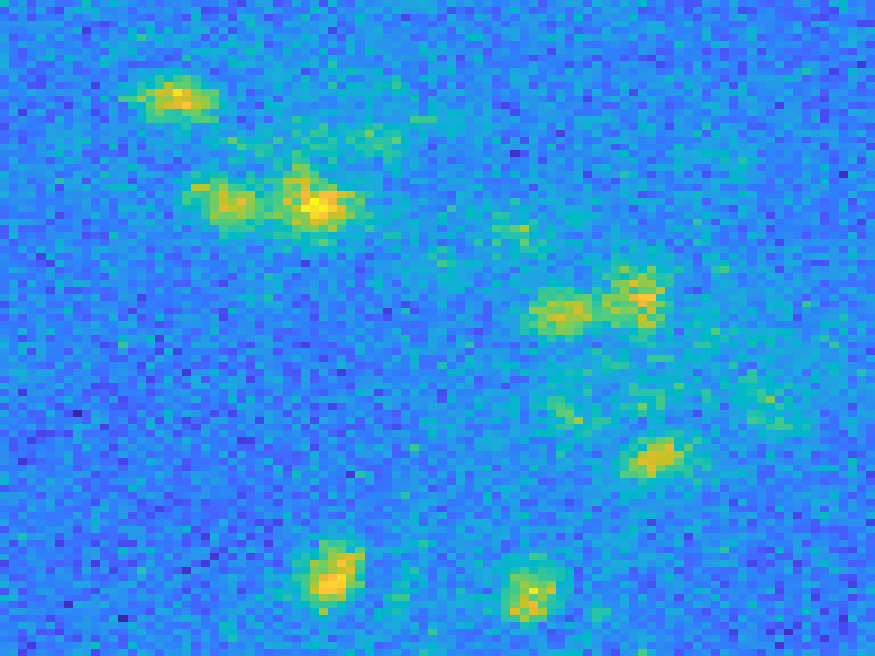}&
		\includegraphics[width=\figwidth]{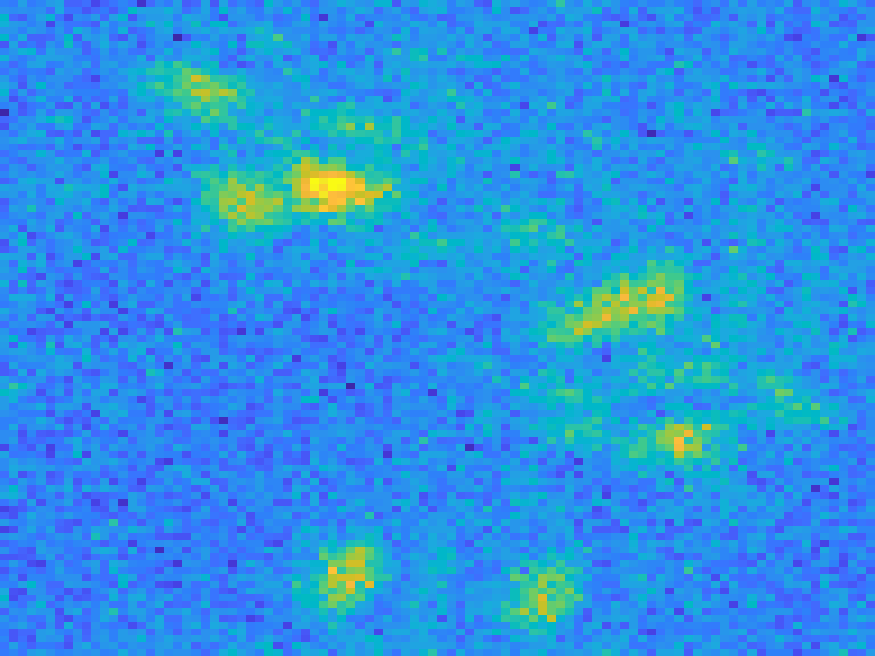}&
		\includegraphics[width=\figwidth]{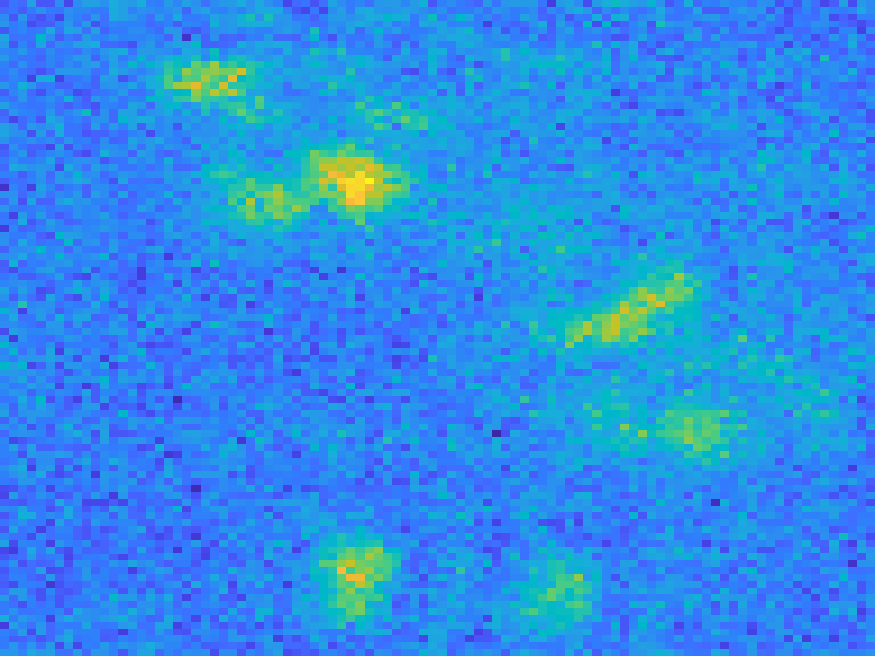}&
		\includegraphics[width=\figwidth]{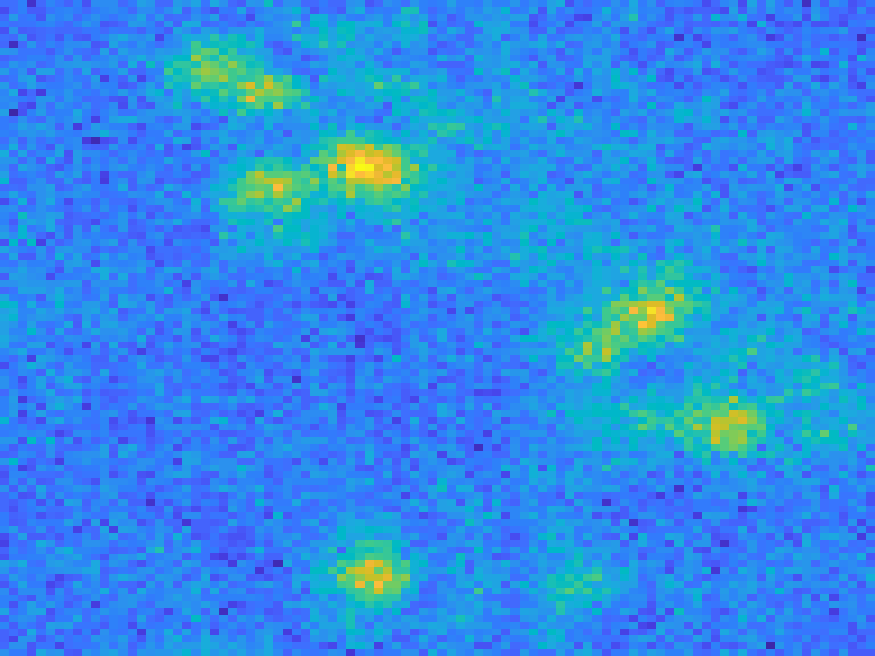}
	\end{tabular}
	\caption{Gaussian noise case. Top: the original image in each band, bottom: the observed image in each band.  }
	\label{fig:org_obs_g}
\end{figure*}

\begin{figure}[htbp]
\centering
\subfloat{\includegraphics[width=0.22\textwidth]{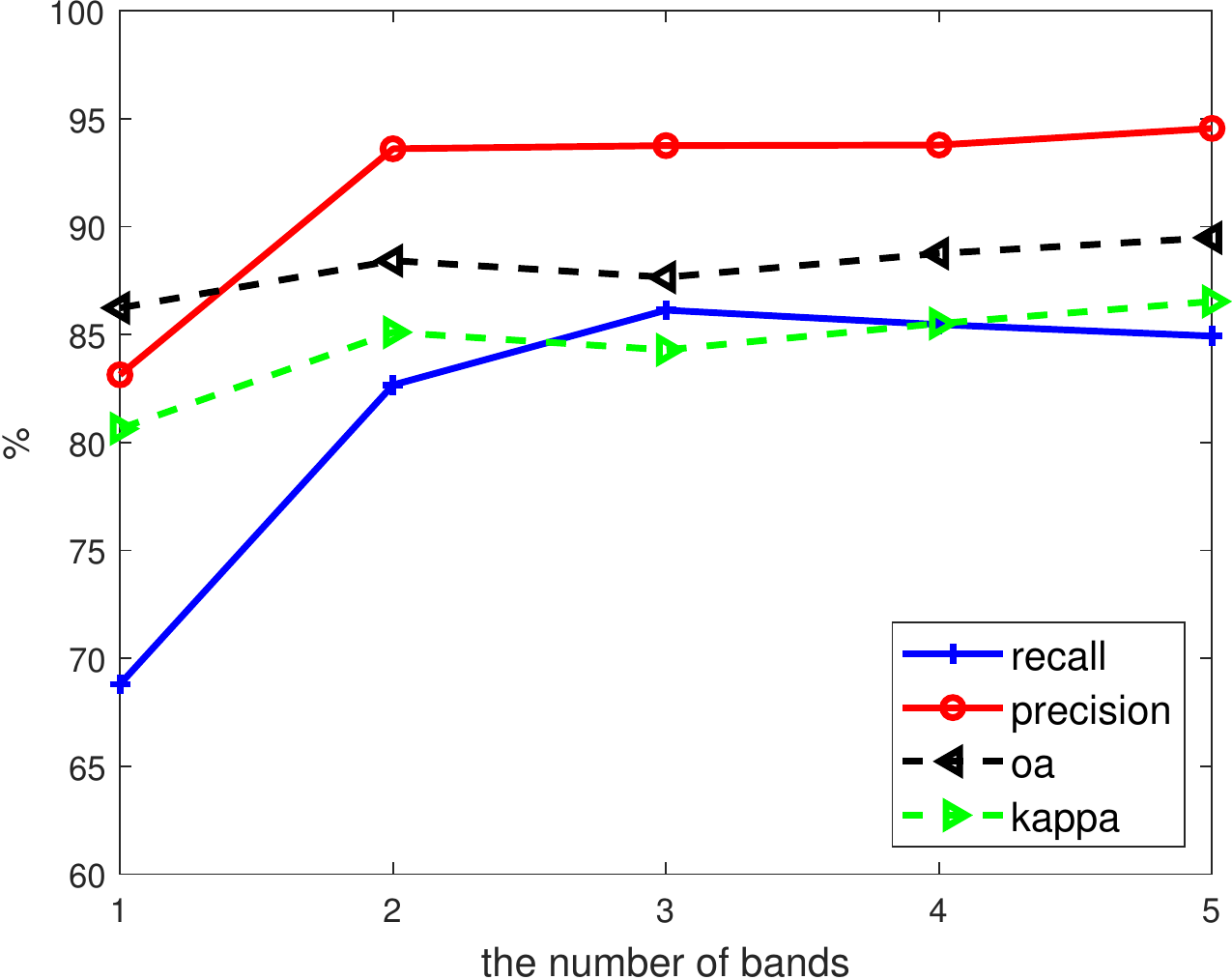}} \hspace{0.01mm}
\subfloat{\includegraphics[width=0.22\textwidth]{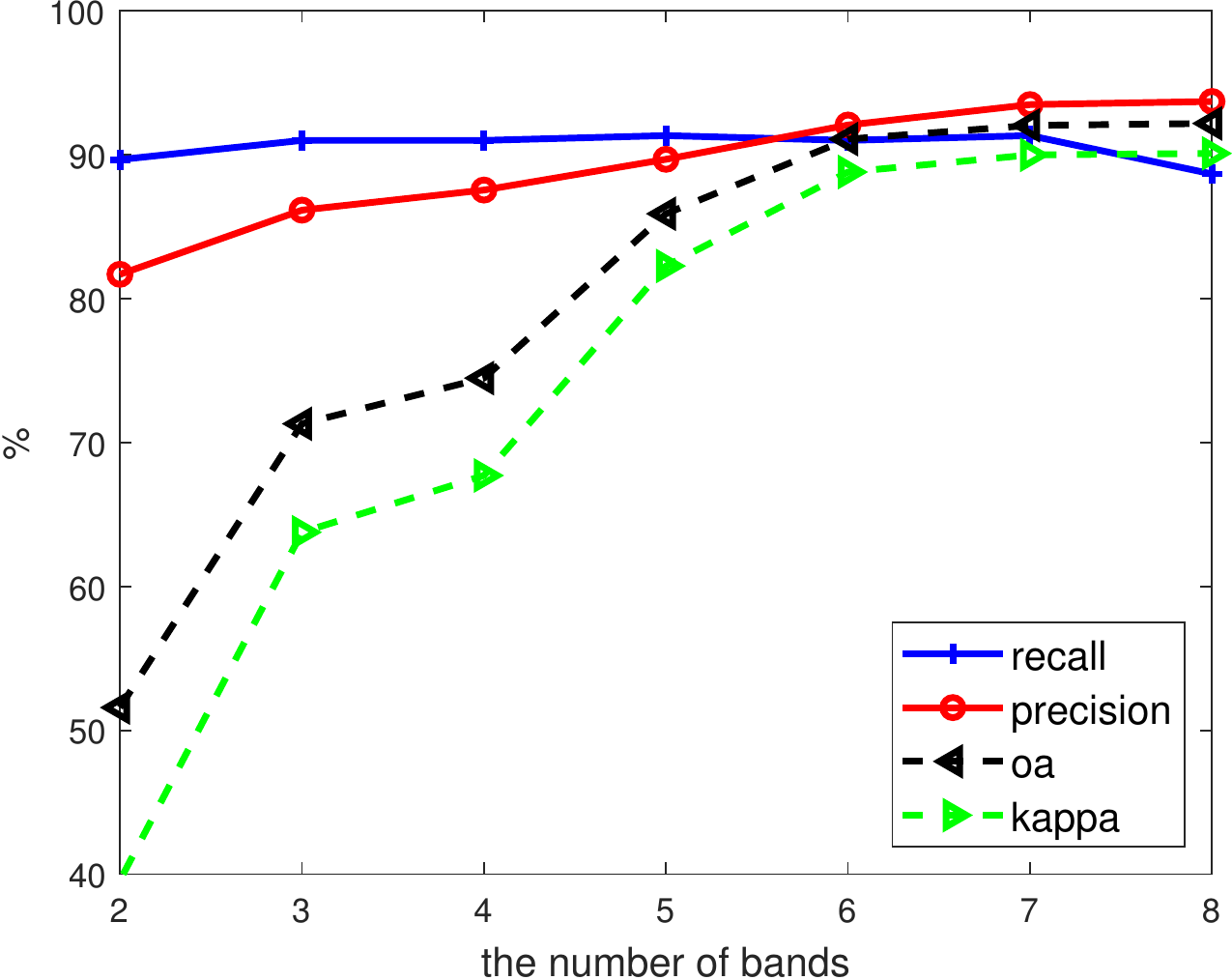}} 
\caption{Gaussian noise case: effects of the number of bands used in Stage 1 (left) and Stage 2 (right)}
\label{fig:Effects_wavelengths_S1S2_g.}
\end{figure}

We do the same experiments with different numbers of bands in Stages 1 and 2, and plot the performance in \Cref{fig:Effects_wavelengths_S1S2_g.}. The corresponding wavelengths in \Cref{fig:Effects_wavelengths_S1S2_g.}(a) are the five ones as in above discussion. The eight wavelengths for \Cref{fig:Effects_wavelengths_S1S2_g.} are 1503.0nm,     1630.3nm,    1757.6nm,   1884.8nm,    2012.1nm,    2139.4nm, 2266.7nm, and 2393.9nm, respectively. In \Cref{fig:Effects_wavelengths_S1S2_g.}(a), we observe that the performance metrics are lower than for the Poisson noise case, but the trend is the same --- there is significant improvement from single band to multispectral images. In \Cref{fig:Effects_wavelengths_S1S2_g.}(b), we observe that the classification results do not improve much with an increase in the number of bands used only when that number becomes larger than 6. This is a basis for the poor performance in \Cref{fig:Effects_wavelengths_S1S2_g.}(a). In the Gaussian noise case, we need to use more spectral images in Stage 2.

\section{Conclusions and Future Work}\label{sec:conclusions}
In this paper we have proposed a three-stage method for employing multispectral RPSF images to localize and classify space debris. With image data from multiple bands, not only the localization results based on the use of a single band can be improved, but also classification can be better addressed for identifying the material components. We have extended the single band's nonconvex optimization model and adapted it for the multiple band case in the first stage of the proposed three-stage approach. In Stage 2, we have estimated the spectral signatures using each band separately. An alternative scheme was proposed which can further remove the false positives and achieve better estimation for both locations and spectral signatures. Stage 3 was devoted to classification from the estimated spectral signatures. This method can be adapted to different noise models, as we have showed here for Poisson and Gaussian models, and different regularization terms. The numerical results were tested for the case of a pure material for each space debris component. We illustrated the efficiency and stability in the numerical results and illustrated that much better accuracy can be obtained by using  multiple bands.  In the future,
we will study the more involved joint localization-spectral unmixing problem in which each piece of space debris may contain multiple material components. In addition, spectral image based recovery of extended space debris of different shapes, sizes, and material compositions will be of interest as a future direction for our work reported here.

\section*{Funding}
The research was supported by the US Air Force Office of Scientific Research under grant FA9550-15-1-0286.

\end{document}